\documentclass[reprint,amsmath,amssymb,aps]{revtex4-2}
\usepackage{amsmath,amssymb}
\usepackage{graphicx}
\usepackage{float}
\usepackage{subfigure}
\usepackage{verbatim}

\usepackage{bbold}
\usepackage{amsfonts}
\usepackage{dcolumn,tabularx}
\usepackage{bm}
\usepackage{color}
\usepackage[colorlinks,citecolor=blue]{hyperref}
\usepackage{braket}
\allowdisplaybreaks[4]

\usepackage{soul}
\newcommand{\be}{\begin{equation}}
	\newcommand{\ee}{\end{equation}}
\newcommand{\bq}{\begin{eqnarray}}
	\newcommand{\eq}{\end{eqnarray}}

\begin{document}

\title{Photonic simulation of Majorana-based Jones polynomials}

\author{Jia-Kun Li}
\affiliation{CAS Key Laboratory of Quantum Information, University of Science and Technology of China, Hefei 230026, People's Republic of China}
\affiliation{CAS Center For Excellence in Quantum Information and Quantum Physics, University of Science and Technology of China, Hefei 230026, People's Republic of China}

\author{Kai Sun}
\email{ksun678@ustc.edu.cn}
\affiliation{CAS Key Laboratory of Quantum Information, University of Science and Technology of China, Hefei 230026, People's Republic of China}
\affiliation{CAS Center For Excellence in Quantum Information and Quantum Physics, University of Science and Technology of China, Hefei 230026, People's Republic of China}

\author{Ze-Yan Hao}
\affiliation{CAS Key Laboratory of Quantum Information, University of Science and Technology of China, Hefei 230026, People's Republic of China}
\affiliation{CAS Center For Excellence in Quantum Information and Quantum Physics, University of Science and Technology of China, Hefei 230026, People's Republic of China}

\author{Jia-He Liang}
\affiliation{CAS Key Laboratory of Quantum Information, University of Science and Technology of China, Hefei 230026, People's Republic of China}
\affiliation{CAS Center For Excellence in Quantum Information and Quantum Physics, University of Science and Technology of China, Hefei 230026, People's Republic of China}

\author{Si-Jing Tao}
\affiliation{CAS Key Laboratory of Quantum Information, University of Science and Technology of China, Hefei 230026, People's Republic of China}
\affiliation{CAS Center For Excellence in Quantum Information and Quantum Physics, University of Science and Technology of China, Hefei 230026, People's Republic of China}

\author{Jiannis K. Pachos}
\affiliation{School of Physics and Astronomy, University of Leeds, Leeds LS2 9JT, United Kingdom}

\author{Jin-Shi Xu}
\email{jsxu@ustc.edu.cn}
\affiliation{CAS Key Laboratory of Quantum Information, University of Science and Technology of China, Hefei 230026, People's Republic of China}
\affiliation{CAS Center For Excellence in Quantum Information and Quantum Physics, University of Science and Technology of China, Hefei 230026, People's Republic of China}

\author{Yong-Jian Han}
\email{smhan@ustc.edu.cn}
\affiliation{CAS Key Laboratory of Quantum Information, University of Science and Technology of China, Hefei 230026, People's Republic of China}
\affiliation{CAS Center For Excellence in Quantum Information and Quantum Physics, University of Science and Technology of China, Hefei 230026, People's Republic of China}

\author{Chuan-Feng Li}
\email{cfli@ustc.edu.cn}
\affiliation{CAS Key Laboratory of Quantum Information, University of Science and Technology of China, Hefei 230026, People's Republic of China}
\affiliation{CAS Center For Excellence in Quantum Information and Quantum Physics, University of Science and Technology of China, Hefei 230026, People's Republic of China}

\author{Guang-Can Guo}
\affiliation{CAS Key Laboratory of Quantum Information, University of Science and Technology of China, Hefei 230026, People's Republic of China}
\affiliation{CAS Center For Excellence in Quantum Information and Quantum Physics, University of Science and Technology of China, Hefei 230026, People's Republic of China}

\begin{abstract}
Jones polynomials were introduced as a tool to distinguish between topologically different links. Recently, they emerged as the central building block of topological quantum computation: by braiding non-Abelian anyons it is possible to realise quantum algorithms through the computation of Jones polynomials. So far, it has been a formidable task to evaluate Jones polynomials through the control and manipulation of non-Abelian anyons. In this study, a photonic quantum system  employing two-photon correlations and non-dissipative imaginary-time evolution is utilized to simulate two inequivalent braiding operations of Majorana zero modes. The resulting amplitudes are shown to be mathematically equivalent to Jones polynomials at a particular value of their parameter. The high-fidelity of our optical platform allows us to distinguish between a wide range of links, such as Hopf links, Solomon links, Trefoil knots, Figure Eight knots and Borromean rings, through determining their corresponding Jones polynomials. Our photonic quantum simulator represents a significant step towards executing fault-tolerant quantum algorithms based on topological quantum encoding and manipulation.     

\end{abstract}

\maketitle

\noindent
\textbf{Introduction}

Link or knot invariants \cite{Alexander1928Topological,1986Examples,1994Temperley} serve as a powerful tool to determine whether or not two knots are topologically equivalent. In 1984, Vaughan Jones identified a novel type of knot invariant, known as the Jones polynomial \cite{Jones1985A}. While being conceptually simple, these polynomials are sufficiently powerful in distinguishing between most topologically inequivalent knots. Currently, there is much interest in determining Jones polynomials as they are important in the field of low-dimensional topology and have applications in various disciplines such as DNA biology \cite{doi:10.1073/pnas.96.23.12974,2000A} and condensed matter physics \cite{PhysRevE.85.036306,PhysRevLett.124.186402}. Unfortunately, even approximating the value of Jones polynomials at certain non-lattice roots of unity falls within the \#P-hard complexity class, with the most efficient classical algorithms requiring an exponential amount of resources \cite{2009How}.


Surprisingly, it has been demonstrated that quantum states of non-Abelian anyons can be used to efficiently evaluate Jones polynomials 
\cite{1977On,PhysRevLett.49.957}. Non-Abelian anyons have exotic statistics manifested in the evolution  of their quantum state when two of them are exchanged or ``braided'' \cite{RevModPhys.80.1083,doi:10.1126/science.aaz5601}. These topological evolutions can be related to Jones polynomials in the following way.
Consider a set of identical non-Abelian anyons corresponding to the SU(2)$_k$ Chern-Simons theory with $k$ a positive integer \cite{WOS:A1989T934900001}. We pair-create these anyons from the vacuum and we move them on the plane so they can span arbitrary $(2+1)$-dimensional paths with their worldlines. Closed worldlines that form a desired link, $L$, are constructed by demanding that the final anyons are also pairwise combined to the vacuum state. The quantum amplitude of such an anyonic evolution is proportional to the Jones polynomial of $L$ evaluated at the polynomial parameter $t = e^{\frac{2\pi i}{(k+2)}}$. 
The Jones polynomials can be considered as computational primitives for topological quantum computation. Universality is obtained for anyons that correspond to SU(2)$_k$ Chern-Simons theory with $k\neq2,4$. Anyons with $k=2,4$ generate the Pauli group, which can become universal with the addition of a simple dynamical gate \cite{doi:10.1126/sciadv.aat6533}. The fault-tolerance of topological quantum computation is inherited by the topological invariance  of the Jones polynomials against continuous deformations of the worldlines that do not change the topology of the corresponding link. 

\begin{figure*}[ht!]
\centering
\includegraphics[width=2\columnwidth]{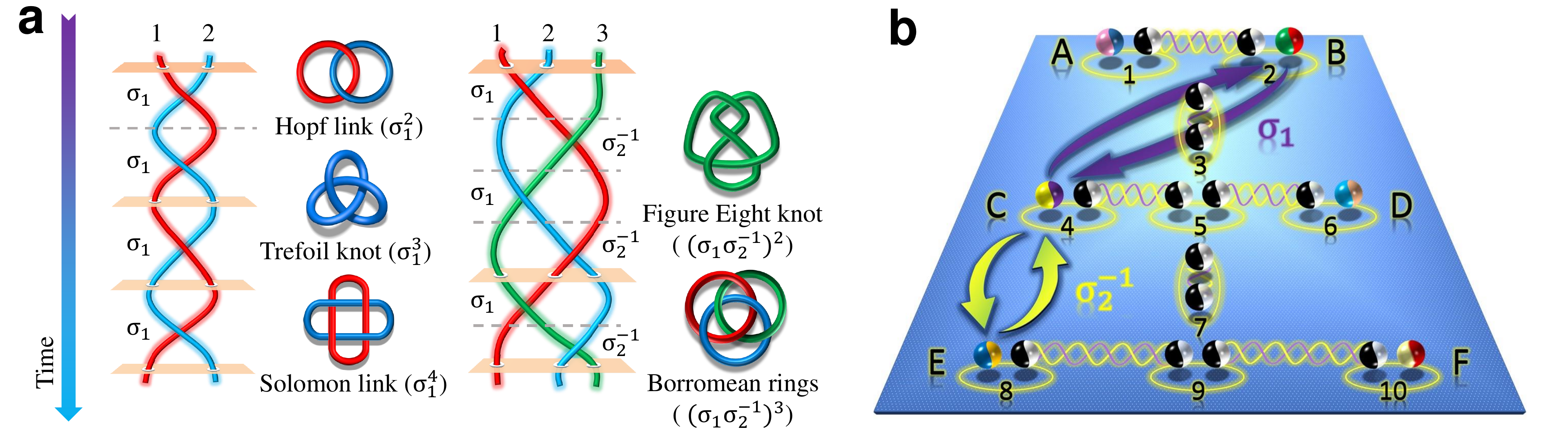}
\caption{Theoretical framework. \textbf{a}. The braiding generators and the relations with specific links. Different intertwining of strings constitute different braid words, which represent specific links when the opposite ends of the strands are connected together. Here, five common links are presented, Hopf link (generated by the sequence $\sigma^{2}_{1}$ ), Trefoil knot (generated by the sequence $\sigma^{3}_{1}$), Solomon link (generated by the sequence $\sigma^{4}_{1}$), Figure Eight knot (generated by the sequence $(\sigma_{1}\sigma^{-1}_{2})^{2}$) and Borromean rings (generated by the sequence $(\sigma_{1}\sigma^{-1}_{2})^{3}$). \textbf{b}. The realisation of braiding generators in the Kitaev chain. The chain consists of ten fermions with six Majorana zero modes (MZMs) identified from A to F. Each sphere represents a Majorana fermion, and a pair of Majorana fermions marked by yellow circles constitutes a normal fermion. The clockwise braiding of MZMs B and C implements braiding operation $\sigma_{1}$, and the anticlockwise braiding of MZMs C and E implements braiding operation $\sigma^{-1}_{2}$. The MZMs B, C and E correspond to string 1, 2 and 3 in \textbf{a}, respectively. }
\label{fig:1}
\end{figure*}

The SU(2)$_k$ non-Abelian anyons are expected to emerge in strongly interacting systems such as fractional quantum Hall liquids \cite{Moore1991NonabelionsIT,PhysRevLett.104.040502,2013Experimental}. Unfortunately, we are not yet in position to experimentally generate and manipulate SU(2)$_k$ non-Abelian anyons. As an alternative, significant effort is dedicated in the experimental realisation of Majorana zero modes (MZMs) \cite{MR1951039,PhysRevLett.98.010506,doi:10.1126/science.aax0274}. Even though the computational power of MZMs is equivalent to SU(2)$_2$ anyons and thus are not universal they are the most plausible candidate for experimentally realising non-Abelian statistics. Still, the practical realisation of specific quantum algorithms on this platform, such as the calculation of Jones polynomials, is yet to be achieved as the braiding operations using MZMs in an experiment remains a challenging and unresolved issue. As an alternative, quantum simulations of MZMs offer a platform to test the viability of these non-Abelian anyons, investigate their properties and determine the resources required for their full realisation.

In this study, we calculate Jones polynomial based on the braiding operation of Majorana zero modes. We experimentally simulate the realisation and manipulation of MZMs through our all-optical multi-state interferometers. By employing this photonic quantum simulator that harnesses two-photon correlations and non-dissipative imaginary-time evolution, we implement a sequence of Hamiltonians that induce specific braiding evolutions of MZMs. This approach enables us to perform two distinct braiding operations and apply them to compute the Jones polynomials for five typical links: the Hopf link, Trefoil knot, the Solomon link, Figure Eight knot and Borromean rings \cite{1926}, as shown in Fig. \ref{fig:1}{\bf a}. The obtained quantum amplitudes demonstrate the topological non-equivalence of these links. Our research demonstrates the feasibility of simulating non-Abelian braiding operations and the subsequent evaluation non-trivial Jones polynomials, setting the stage for further exploration of non-Abelian anyons. Importantly, the topological encoding of quantum information in non-Abelian anyons and its manipulation by braiding them with each other guaranties the highly desired fault-tolerance of the algorithm, throughout its execution. Moreover, it showcases that our photonic quantum simulator, even with its limited scalability, facilitates the experimental realisation of complex topological quantum algorithms.
\\

\noindent
\textbf{Results}\\        
\textbf{Majorana-based Jones polynomials.} The Jones polynomial $V_L(t)$ of a link $L$, with variable $t$, is related to the quantum amplitude of braided SU(2)$_k$ anyons by
\begin{equation}
	V_L(t) = (-\sqrt[4]{t})^{-3w(L)} d^{n-1} \langle \phi_0|U(t)|\phi_0\rangle,
	\label{eq:JonesAmp}
\end{equation}
where $t = e^{2\pi i \over (k+2)}$, determines the type of employed anyons, $d = -\sqrt{t}-1/\sqrt{t}$ is the quantum dimension of the anyons and $w(L)$ is the writhe of the link $L$. State $|\phi_0\rangle$ and $\langle\phi_0|$ describe to vacuum anyonic pair creation and annihilation, respectively. We denote by $|\phi_0\rangle = |0_{1}...0_{n}\rangle_{l}$ the pair creation of $n$ pairs of anyons from the vacuum. The unitary $U(t)$ describes the braiding evolution involving one anyon from each pair. The amplitude $\langle \phi_0|U(t)|\phi_0\rangle$ corresponds to the Markov trace of the braiding operations \cite{pachos_2012} that span the link $L$ with the anyonic worldlines, as shown in Fig. \ref{fig:1}{\bf a} \cite{2009A}.

Strictly speaking the MZMs are not described by a Chern-Simons gauge theory. Nevertheless, the elementary braiding evolutions of Majorana fermions differ from those of SU(2)$_2$ anyons by an overall constant phase factor and they both have quantum dimension $d=\sqrt{2}$. Hence, we can mathematically relate the expectation values, $\bra{\phi_{0}}U_M\ket{{\phi_{0}}}$, of arbitrary Majorana evolutions $U_M$ with the Jones polynomial $V_L(i)$ corresponding to the Chern-Simons theory with $k=2$, in the following way:
$V_L(i) = e^{{9i\pi \over 8}{(w(L)-\chi(L))}}  {2}^{n-1 \over 2} \langle \phi_0|U_M|\phi_0\rangle$,
where $\chi(L)=\chi_{\rm ac}(L)-\chi_{\rm c}(L)$ with $\chi_{\rm ac}(L)$ and $\chi_{\rm c}(L)$ the number of anti-clockwise and clockwise exchanges of the link $L$, respectively, and $U_M$ describes the Majorana braiding evolutions that span the link with their worldlines. For the links $L$ considered here obtained from the Markov trace of braiding evolutions $U_M$ we always have $\chi(L)=w(L)$, resulting in the simple expression
\begin{equation}
V_L(i) = {2}^{n-1 \over 2} \langle \phi_0|U_M|\phi_0\rangle.
\label{eq:JonesMajies}
\end{equation}
Note that in the conventional algorithm given in \cite{2009A}, Hadamard test is required to measure the amplitude, i.e. the $Re\bra{\phi_{0}}U_M\ket{{\phi_{0}}}$ and $Im\bra{\phi_{0}}U_M\ket{{\phi_{0}}}$. However, for the SU(2)$_2$ anyons the Jones polynomial $V_L (i)$ is always a real number (see Supplemental Material for the details). Moreover, to simplify the complexity of our photonic experiment we restrict to the calculation of $| \langle \phi_0|U_M|\phi_0\rangle|$, which gives us the value $|V_L(i)|$ of the Jones polynomial up to an overall sign. While this reduces the predictive power of the Jones polynomials, our experiment can still distinguish between several of the targeted links, as we shall see in the following.
\\

We want to realise a set of braiding operations that correspond to a wide family of knots and links, while they are still feasible to implement with our photonic simulator. For that we consider three pairs of Majorana zero modes positioned at the endpoints of three Kitaev chains \cite{Kitaev_2001}. Moreover, we choose the size of the chains to be large enough so MZMs are never brought on the same site during the braiding evolution, which would expose the topologically encoded information to local potentials. To achieve fault-tolerance against local potentials at all times during the performance of our algorithm we employ ten fermion sites, as shown in Fig. \ref{fig:1}{\bf b}. Each Tai Chi-like dual-colored sphere represents a Majorana fermion, and two adjacent spheres together constitute a normal fermion.  Here, canonical operators $c_{j}$ and  $c_{j}^{\dagger}$  can be used for describing these fermions, with position $j$=1,...,10. The first chain is composed of $j$=1, 2; the second chain is composed of $j$=4, 5, 6 and the third chain is composed of $j$=8, 9, 10. Sites $j$=3 and $j$=7 corresponds to the connections between chains that facilitate the braiding.  The Hamiltonian of the initial state in the Kitaev chains can be written as
\begin{equation}
	\begin{split}
		&H_{M_0}=i(\gamma_{1b} \gamma_{2a}+\gamma_{4b} \gamma_{5a}+\gamma_{5b} \gamma_{6a}+\gamma_{8b} \gamma_{9a}+\gamma_{9b}\gamma_{10a}\\&+\gamma_{3a}\gamma_{3b}+\gamma_{7a}\gamma_{7b}),
 \label{eqn:0}
 \end{split}
\end{equation} 
where $\gamma_{ja}=c_{j}+c_{j}^{\dagger}$ and $\gamma_{jb}=i(c_{j}^{\dagger}-c_{j})$. The Majorana operators $\gamma_{m}$ satisfies the relations $\gamma_{m}^{\dagger}=\gamma_{m}$ and $\gamma_{l} \gamma_{m} +\gamma_{m} \gamma_{l} =2\delta_{lm}$ for $l,m=1a,1b,...10a,10b$. Operators $\gamma_{1a}$, $\gamma_{2b}$, $\gamma_{4a}$, $\gamma_{6b}$, $\gamma_{8a}$ and $\gamma_{10b}$ are not in the initial Hamiltonian $H_{M_0}$, thus $[H_{M_0},\gamma_{j}]=0$ for $j=1a, 2b, 4a, 6b, 8a, 10b$. Hence, these Majorana modes have zero energy, corresponding to six endpoint MZMs, denoted from A to F in Fig. \ref{fig:1}{\bf b}.

After completing the two sequences of Hamiltonians of braiding operation  $\sigma_1$ and $\sigma_2^{-1}$ (see Methods for the details), the theoretically resulting braiding evolutions in the logical basis are given by $L_{\sigma_{1}}= \frac{1}{\sqrt{2}}(\mathbb{1} \otimes \mathbb{1}+i\sigma_{x}\otimes\sigma_{x})\otimes\mathbb{1}$ and $L_{\sigma_{2}^{-1}}=\frac{1}{\sqrt{2}} \mathbb{1} \otimes (\mathbb{1} \otimes \mathbb{1}-i\sigma_{y}\otimes\sigma_{x})$. In $L_{\sigma_{1}}$($L_{\sigma_{2}^{-1}}$), the last (first) $\mathbb{1}$ is introduced due to the chain that is not involved in the braiding. According to \eqref{eq:JonesMajies} we can now utilize the combination of the braiding operations $\sigma_1$ and $\sigma_2^{-1}$ to realise the braiding matrix $U_M$ that corresponds to the desired knots and links shown in Fig. \ref{fig:1}{\bf a}.

\noindent
\textbf{Encoding the quantum states based on MZMs.}  
To experimentally implementing the braiding evolution of MZMs, it is necessary to firstly convert the fermionic Hamiltonians into spin Hamiltonians using the Jordan-Wigner (JW) transformation \cite{RN138}. The resulting spin system exhibits an equivalent spectrum to the original fermion system along the braiding evolution, giving rise to the same non-Abelian Berry phases in their corresponding basis of eigenstates \cite{doi:10.1126/sciadv.aat6533}. This enables the simulation of a fermion superconducting system in terms of spin-like degrees of freedom that can be directly encoded in our photonic platform. Methods provide the transformed spin Hamiltonians.  

Next, we encode the initial logical state $|\phi_0\rangle=|0_{1}0_{2}0_{3}\rangle_{l}$ of the three chains in terms of spin states. For convenience, we express the spin states of the system in terms of the eigenvectors $\{|x\rangle,|\bar{x}\rangle\}$, $\{|y\rangle,|\bar{y}\rangle\}$ and $\{|z\rangle,|\bar{z}\rangle\}$ of the corresponding Pauli operators $\sigma^x$, $\sigma^y$ and $\sigma^z$, with eigenvalues $\{1,-1\}$, respectively. Then the transformation is given by: $|0_{1}\rangle_{l}=(|x_{1}x_{2}\rangle+|\bar{x}_{1}\bar{x}_{2}\rangle)/\sqrt{2}$, $|1_{1}\rangle_{l}=(|x_{1}x_{2}\rangle-|\bar{x}_{1}\bar{x}_{2}\rangle)/\sqrt{2}$, $|0_{2}\rangle_{l}=(|x_{4}x_{5}x_{6}\rangle-|\bar{x}_{4}\bar{x}_{5}\bar{x}_{6}\rangle)/\sqrt{2}$, $|1_{2}\rangle_{l}=(|x_{4}x_{5}x_{6}\rangle+|\bar{x}_{4}\bar{x}_{5}\bar{x}_{6}\rangle)/\sqrt{2}$, $|0_{3}\rangle_{l}=(|x_{8}x_{9}x_{10}\rangle-|\bar{x}_{8}\bar{x}_{9}\bar{x}_{10}\rangle)/\sqrt{2}$ and $|1_{3}\rangle_{l}=(|x_{8}x_{9}x_{10}\rangle+|\bar{x}_{8}\bar{x}_{9}\bar{x}_{10}\rangle)/\sqrt{2}$. Thus, the initial logical state $|\phi_0\rangle$ is represented in terms of spins $\ket{\phi_{0}'} = (|x_{1}x_{2}\rangle+|\bar{x}_{1}\bar{x}_{2}\rangle)/\sqrt{2} \otimes (|x_{4}x_{5}x_{6}\rangle-|\bar{x}_{4}\bar{x}_{5}\bar{x}_{6}\rangle)/\sqrt{2} \otimes (|x_{8}x_{9}x_{10}\rangle-|\bar{x}_{8}\bar{x}_{9}\bar{x}_{10}\rangle)/\sqrt{2}$.

\begin{figure*}[ht!]
	\centering
	\includegraphics[width=2\columnwidth]{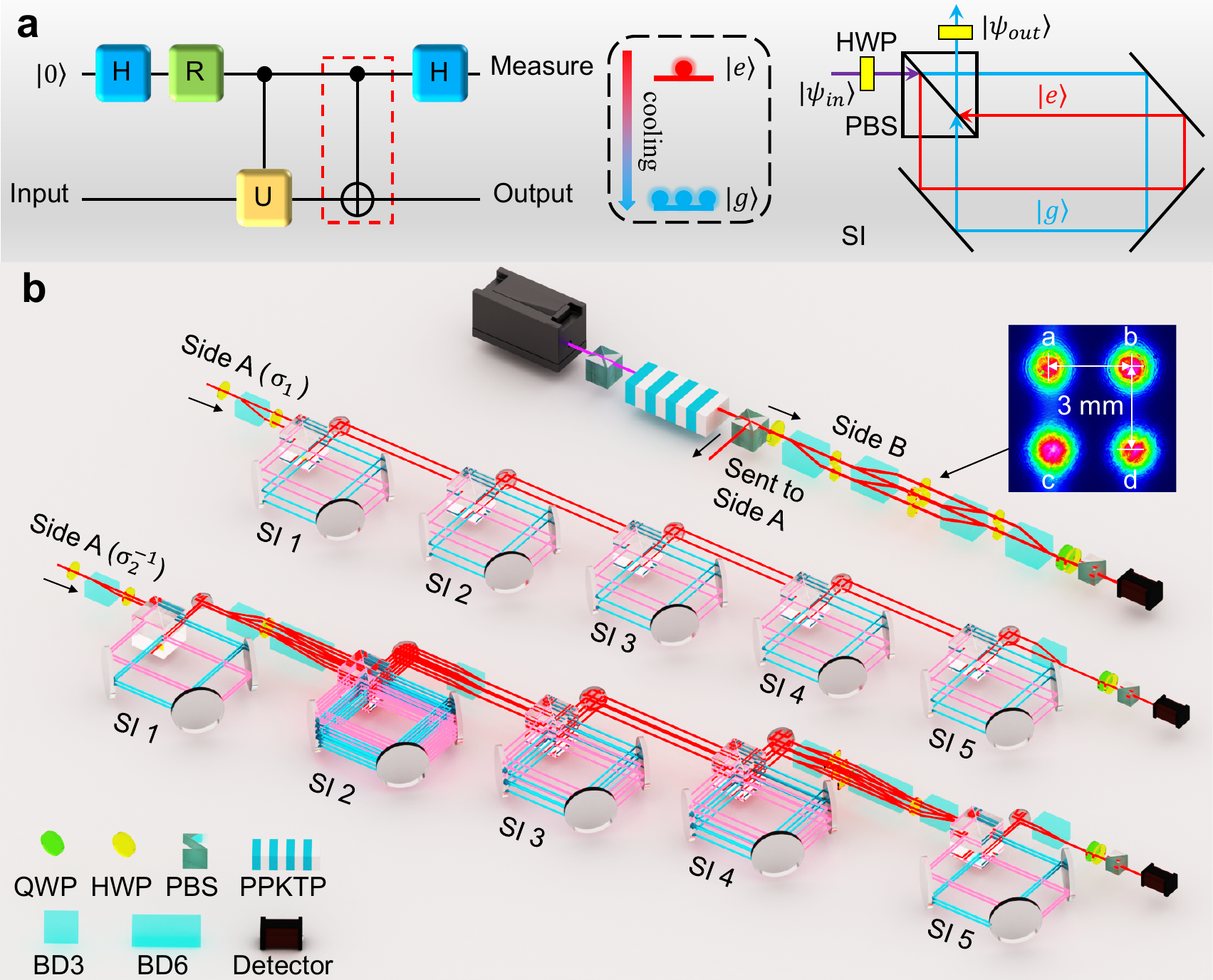}
	\caption{Experimental setup. \textbf{a}. The traditional quantum cooling circuit is composed of two Hadamard gates $(H)$, a local phase gate $(R)$ and a controlled unitary gate $(U)$. Here, a controlled-NOT (CNOT) gate (marked by the red dashed line) is newly introduced to realise non-dissipative imaginary-time evolution (ITE), and the Sagnac interferometer (SI) is implemented to  perform this evolution. \textbf{b}. Correlated photon pairs are generated by pumping a type-\uppercase\expandafter{\romannumeral+2} PPKTP crystal. The generated photon pairs are then transmitted to side A and side B, respectively.  The combination of HWPs (some of which are omitted) and BDs realise the manipulation of these modes thus achieving the corresponding state evolution. Throughout the process of evolution, there have consistently been four distinct spatial patterns on side B. Its coincidence with photons on side A will be implemented to encode the quantum states during the evolution. SIs are employed to achieve non-dissipative ITE at specific steps during braiding operation $\sigma_{1}$ and $\sigma^{-1}_{2}$.  }
	\label{fig:2}
\end{figure*}

Rather than using single photons to encode the desired spin states we improve the efficiency of our photonic system by employing two-photon correlations \cite{PhysRevLett.67.661,doi:10.1126/science.1193515,RevModPhys.81.865}. Methods provide the details of two-photon correlations encoding. Fig. \ref{fig:2} shows our experimental setup. A 404 nm continuous wave diode laser is used to pump a type-\uppercase\expandafter{\romannumeral+2} periodically poled  KTiOPO$_{4}$ (PPKTP) to generate the two correlated photons through spontaneous parametric down conversion \cite{Eisaman2011InvitedRA}. One of the two photons is sent to side A, and the other photon is sent to side B. On each side, spatial modes of the photon are manipulated via beam displacers (BDs), which are birefringent crystals capable of spatially separating light beams based on their horizontal or vertical polarization. Here, two types of beam displacers are introduced: BD3 with beams separated by 3 mm
and BD6 with beams separated by 6 mm.  Additionally, half-wave plates (HWPs) and quarter-wave plates (QWPs) are employed to regulate the polarization of the photon, i.e., exchanging the basis between Pauli operators $\sigma_{x}$, $\sigma_{y}$ and $\sigma_{z}$.  
 
 \begin{figure*}[htbp]
	\centering
	\includegraphics[width=2\columnwidth]{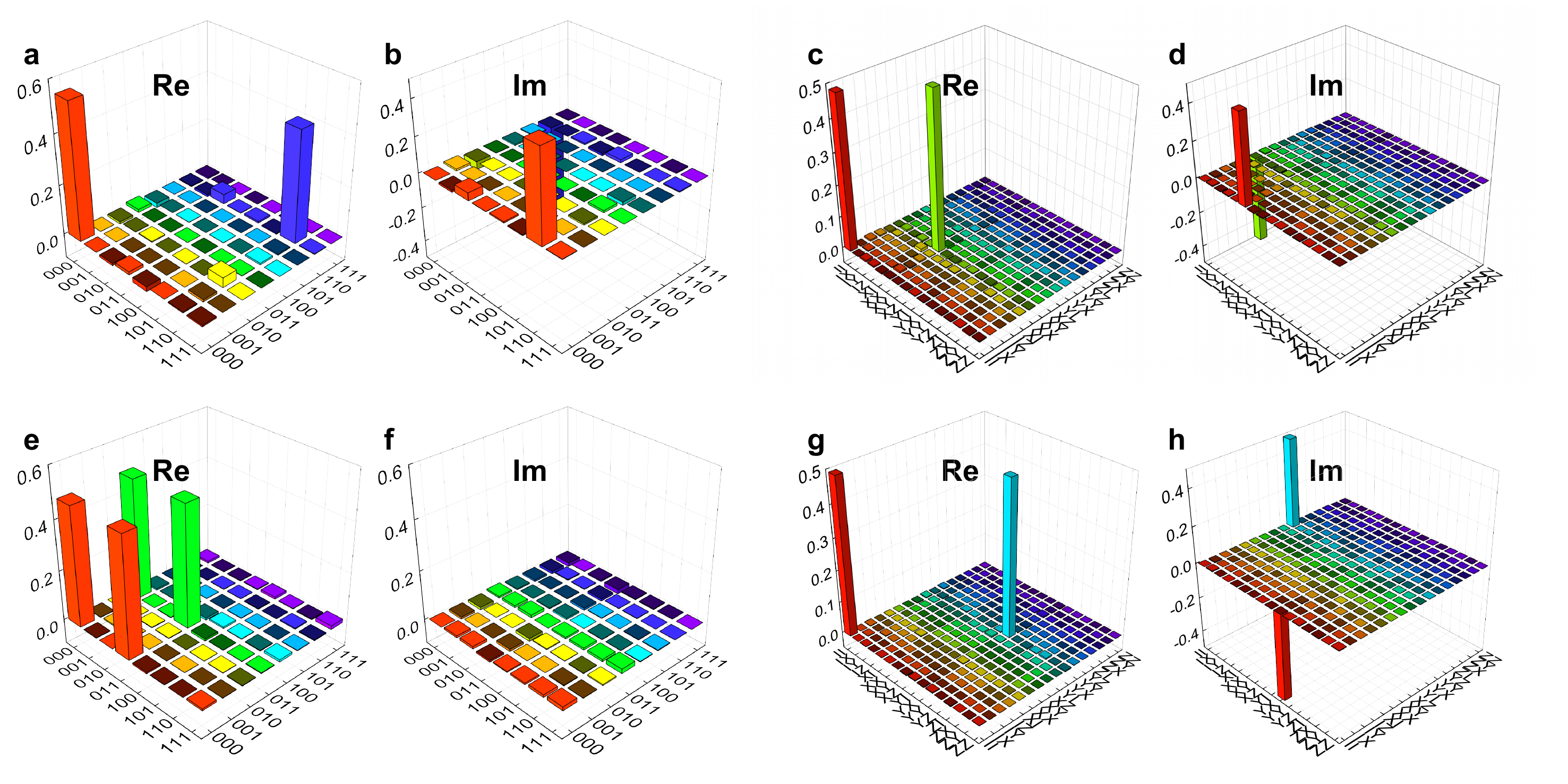}
	\caption{Experimental results for the reconstructed density matrix for braiding operation $\sigma_{1}$ and $\sigma^{-1}_{2}$. $\textbf{a}$. Real (Re) and  $\textbf{b}$. imaginary (Im) parts of final state $|\phi^{\sigma_{1}}_f\rangle=(|000\rangle_{l}+i|110\rangle_{l})/\sqrt{2}$. $\textbf{c}$. Real (Re) and  $\textbf{d}$. imaginary (Im) parts of the process density matrix $(II+iXX)/\sqrt{2}$ of braiding operation $\sigma_{1}$. 
		$\textbf{e}$. Real (Re) and  $\textbf{f}$. imaginary (Im) parts of final state $|\phi^{\sigma^{-1}_{2}}_f\rangle=(|000\rangle_{l}+|011\rangle_{l})/\sqrt{2}$. $\textbf{g}$. Real (Re) and  $\textbf{h}$. imaginary (Im) parts of the process density matrix $(II-iYX)/\sqrt{2}$ of braiding operation $\sigma^{-1}_{2}$.
		The measurement operator $I,X,Y$ and $Z$ represent the identity, $\sigma^x$, $\sigma^y$ and $\sigma^z$, respectively.}
		\label{fig:3}
\end{figure*}

After completing the initial state preparation, we need to perform the braiding evolution according to their corresponding sequences of Hamiltonians. The braiding can be implemented through a series of imaginary-time evolution (ITE) operators (see Supplemental Material for the details), and the whole process for $\sigma_{1}$ and $\sigma_{2}^{-1}$ can be simplified as $e^{-H_{0}t}e^{-H_{3}t}e^{-H_{2}t}e^{-H_{1}t}\ket{\phi_{0}}=e^{-\sigma_{3}^{z}}e^{\sigma_{4}^{x}\sigma_{5}^{x}}e^{-\sigma_{4}^{z}}e^{-\sigma_{3}^{x}\sigma_{4}^{y}}e^{\sigma_{2}^{x}\sigma_{3}^{x}}\ket{\phi_{0}}$ and $e^{-H_{0}t}e^{-H'_{1}t}\\e^{-H'_{2}t}e^{-H'_{3}t}e^{-H'_{4}t}e^{-H'_{5}t}\ket{\phi_{0}}=e^{\sigma_{4}^{x}\sigma_{5}^{x}}e^{-\sigma_{7}^{z}}e^{\sigma_{8}^{x}\sigma_{9}^{x}}\\e^{-\sigma_{8}^{z}}e^{\sigma_{9}^{x}\sigma_{10}^{x}}e^{-\sigma_{7}^{x}\sigma_{8}^{y}}e^{\sigma_{5}^{y}\sigma_{6}^{z}\sigma_{7}^{x}}e^{-\sigma_{4}^{z}}\ket{\phi_{0}}$, respectively. Previously \cite{RN135,doi:10.1126/sciadv.aat6533}, we selected an extended evolution time $t$ in order to diminish the influence of the high-energy excited states (labeled as $\ket{e}$), thereby allowing only the ground states (labeled as $\ket{g}$) with the lowest energy to persist following each evolution step. 

Here, to avoid the dissipation of the excited states, we design a new scheme inspired by the concept of quantum cooling \cite{Xu_2014}. The quantum circuit is shown in Fig. \ref{fig:2}\textbf{a}. By inserting a control-not (CNOT) gate in the traditional quantum cooling circuit, the non-dissipative ITE is achieved. 
This can be realised using a designed polarization-dependent Sagnac interferometer (SI). Methods provide the details of the realisation of the CNOT gate. This process emulates quantum cooling by effectively reducing the excited states to their ground states, resulting in zero photon loss during the ITE evolution. The ratio of reflected to transmitted vertical polarization photons exceeds 500:1, ensuring the accuracy of the non-dissipative ITE evolution. This represents a significant advancement compared to our previous experimental approaches, where approximately half of the photons were discarded in each step of the basis transformation (after $n$ steps, $n=1,2,3...$ the total number of photons will decay to $1/2^{n}$ of the initial photon numbers) and thus making it challenging to extend the evolution to multiple steps. This enhancement significantly improves our ability to conduct experiments involving multiple evolutionary steps, thereby facilitating the experimental simulation of complex physical processes.

For braiding operation $\sigma_{1}$, to realise the non-dissipative evolution according to its spin Hamiltonians, we built five cascaded SIs on side A while keeping the device on side B stationary. For $\sigma^{-1}_{2}$, due to the more complex Hamiltonians,  multiple BD3 and BD6 were used alongside the cascaded SIs on side A, which is more experimentally challenging compared with $\sigma_{1}$. (The experimental details for these two evolutions can be found in Supplemental Material).

Upon completion of the specific evolution operation, we need to   reconstruct the quantum states during the evolution thus ensuring the effectiveness.   The states have  $2^{3}$ basis from $\ket{000}_{meas}$ to $\ket{111}_{meas}$, which corresponds to a three-qubit quantum state. We thus project the final state to 64-measurement basis composed of $\ket{0}_{meas},\ket{1}_{meas},\ket{+}=(\ket{0}_{meas}+\ket{1}_{meas})/\sqrt{2},\ket{R}=(\ket{0}_{meas}-i\ket{1}_{meas})/\sqrt{2}$ to reconstruct the density matrix of the quantum state such as $|\phi_{0}'\rangle$ according to the standard three-qubit quantum state tomography \cite{PhysRevLett.74.4101}.  
Specifically, it requires combining the beams from sides A/B into a single beam.
This process involves the use of BDs and HWPs to recombine all the equal-optical path spatial modes on both sides, thus transforming the spatial modes encoded information to polarization information. The relative phase between different states in the interferometer is adjusted by inserting multiple fused silica plates (4×20×0.15 mm, with reflectivity less than 0.2\%, not depicted in Fig. \ref{fig:2}). The analysis of the polarization information is accomplished through the combination of a QWP, an HWP, and a polarization beam splitter (PBS). (More experimental details for implementing  quantum state tomography can be found in Supplemental Material). \\

\noindent
\textbf{Realisation of braiding operations.} We now demonstrate that our experimental setup can faithfully realise the desired braiding evolutions of MZMs. For braiding operation $\sigma_{1}$, we perform quantum state tomography on the final state that undergoes the corresponding Hamiltonian transformations. Theoretically, the $\sigma_1$ braiding process transforms the logical Hilbert space with the unitary operation $L_{\sigma_{1}}=(\mathbb{1} \otimes \mathbb{1}+i\sigma_{x}\otimes\sigma_{x})\otimes \mathbb{1}/\sqrt{2}$. As a result the initial state $|\phi_0\rangle=|000\rangle_{l}$ should evolve to the final state $|\phi^{\sigma_{1}}_f\rangle=(|000\rangle_{l}+i|110\rangle_{l})/\sqrt{2}$. The experimentally reconstructed density matrix is shown in Fig. \ref{fig:3}{\bf a} (real part) and \textbf{b} (imaginary part). These measurement results show that we can experimentally produce the final quantum state, $|\phi^{\sigma_{1}}_f\rangle$, successfully, with a fidelity of 0.992 $\pm$ 0.001. We also perform quantum state tomography for all the eigenstates of the Hamiltonians during the braiding process. The average fidelity of these states is 0.991 $\pm$ 0.005, indicating the validity and efficacy of each step during the braiding evolution process. 

In addition to the reconstruction of the quantum states, we also perform quantum process tomography \cite{PhysRevLett.93.080502}  to characterize the braiding operation $\sigma_{1}$. We employ the experimentally intuitive basis $I,X,Y$ and $Z$ representing the identity, $\sigma^x$, $\sigma^y$ and $\sigma^z$, respectively. During the $L_{\sigma_{1}}$ braiding evolution the third chain does not participate so we only need to present the reconstruction of the operator $(II+iXX)/\sqrt{2}$. The experimental results are presented in Fig. \ref{fig:3}\textbf{c} (real part) and \textbf{d} (imaginary part). The process fidelity of braiding operation $\sigma_{1}$ is 0.983 $\pm$ 0.014.      
 
Similarly, for braiding operation $\sigma_{2}^{-1}$, we also characterize its performance according to the same procedure in $\sigma_{1}$. After the braiding operation $L_{\sigma_{2}^{-1}}= \mathbb{1} \otimes (\mathbb{1} \otimes \mathbb{1}-i\sigma_{y}\otimes\sigma_{x})/\sqrt{2}$, the initial state $|\phi_0\rangle=|000\rangle_{l}$ evolves to the final state $|\phi^{\sigma^{-1}_{2}}_f\rangle=(|000\rangle_{l}+|011\rangle_{l})/\sqrt{2}$. The experimental results are presented in Fig. \ref{fig:3}\textbf{e} (real part) and \textbf{f} (imaginary part) with a fidelity of 0.964 $\pm$ 0.028. The average fidelity of the final states according to the Hamiltonians during the braiding evolution is 0.978 $\pm$ 0.015. Moreover, the reconstructed process matrix $(II-iYX)/\sqrt{2}$ (the first chain does not participate) is shown in Fig. \ref{fig:3}\textbf{g} (real part) and \textbf{h} (imaginary part) with a fidelity of 0.980 $\pm$ 0.015. The experimental results of the state tomography and process tomography demonstrate that we have realised both braiding operations, $\sigma_{1}$ and $\sigma_{2}^{-1}$, successfully  with high fidelity for each step of the evolution.
 
In our all-optical platform the imperfections of the experimental results stem from flawed interference, imprecise wave plates and fluctuation in the count of photons. Among these three factors, imperfect interference is the primary reason for the incomplete alignment between the experimental results and theoretical values. During the experimental realisation of $\sigma_{1}$ the main sources of error are imperfections in the five cascaded Sagnac interferometers on Side A, which are experimentally challenging. Meanwhile, the fidelity of both state and process tomography of ${\sigma_{2}^{-1}}$ have decreased slightly compared to $\sigma_{1}$. This is because of the additional multiple beam splitters on side A of the setup required for implementing the ${\sigma_{2}^{-1}}$ evolution. For instance, when we implement the ITE process for evolving the final state corresponding to $H'_{5}$ to the final state corresponding to $H'_{4}$, operation $e^{\sigma_{5}^{y}\sigma_{6}^{z}\sigma_{7}^{x}}$ needs to be applied to the final state of $H'_{5}$. This basis transformation corresponds to experimentally creating eight beams on side A, and then recombining them to four beams which needs interference processes. This process poses significant experimental challenges, especially when these beams have passed through a Sagnac interferometer before recombining. Despite this technical complexity the fidelity of braiding operation ${\sigma_{2}^{-1}}$ still remains exceptionally high, demonstrating the excellent characteristics of our device and its ability to demonstrate Majorana-based Jones polynomial calculations.\\

\begin{figure}[htbp]
	\centering
	\includegraphics[width=1\columnwidth]{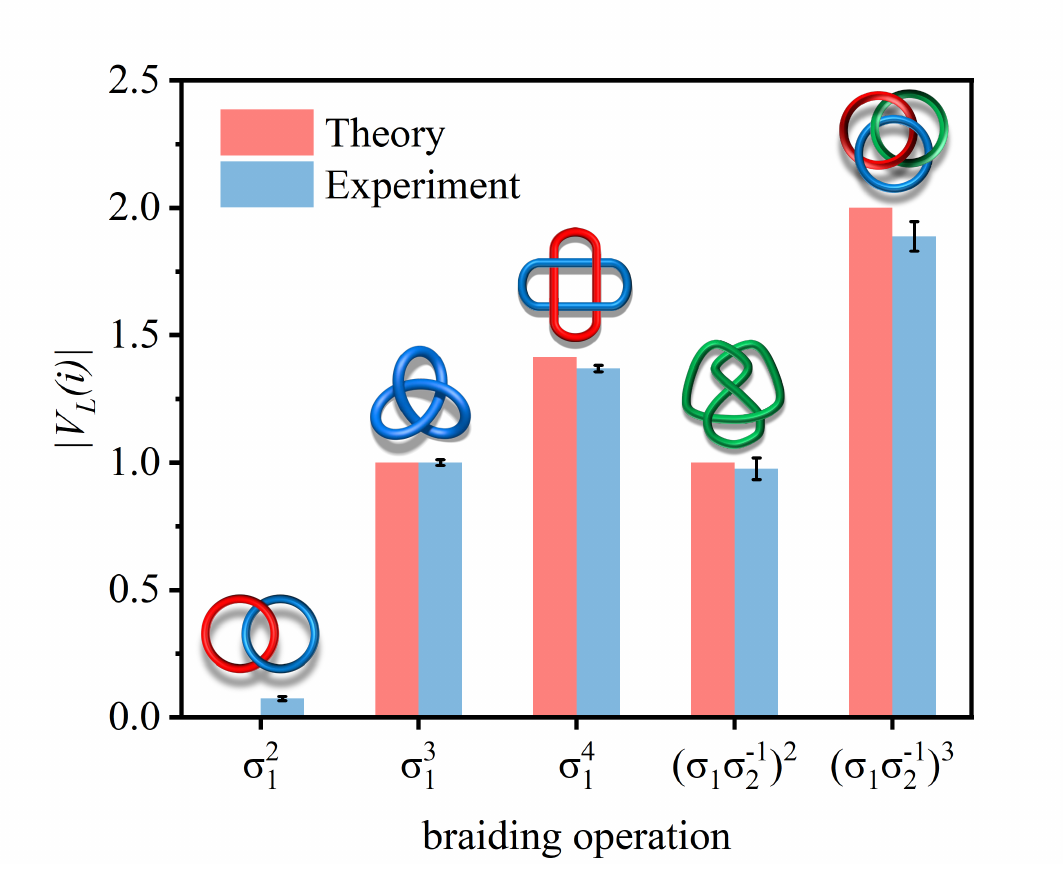}
	\caption{Experimental results. Theoretical (pink) and experimental (blue) values of the Jones polynomial for different braiding operation $\sigma_{1}^{2}$, $\sigma_{1}^{3}$, $\sigma_{1}^{4}$, $(\sigma_{1}\sigma_{2}^{-1})^2$ and $(\sigma_{1}\sigma_{2}^{-1})^3$. The error bars are estimated according to Poissonian counting statistics.}
	\label{fig:4}
\end{figure}

\noindent
\textbf{Calculation of the Majorana-based Jones polynomials.} 
The braiding operations $\sigma_{1}$ and $\sigma_{2}^{-1}$ are the building blocks for generating a wide variety of links. We consider five combinations of these two braiding operations that give rise to the Jones polynomials of five topologically distinct links when their expectation value with respect to $|\phi_0\rangle$ is evaluated. In particular, we realise $\sigma_{1}^{2}$ for the Hopf link, $\sigma_{1}^{3}$ for the Trefoil knot, $\sigma_{1}^{4}$ for the Solomon link,  $(\sigma_{1}\sigma_{2}^{-1})^2$ for the Figure Eight knot and $(\sigma_{1}\sigma_{2}^{-1})^3$ for the Borromean rings, as shown in Figs. \ref{fig:1} and \ref{fig:4}.  The final states after these operations can be directly computed using the experimentally derived process matrix of $\sigma_{1}$ and  $\sigma_{2}^{-1}$.

The resulting values of the Jones polynomial $V_L(i)$ for the realised links $L$ are shown in Fig. \ref{fig:4}. In particular, we obtain that the values of the Jones polynomials for the Hopf link, Trefoil knot, Solomon link, Figure Eight knot and Borromean rings are 0.074 $\pm$ 0.008, 1.000 $\pm$ 0.011, 1.369 $\pm$ 0.013, 0.976 $\pm$ 0.042 and 1.888 $\pm$ 0.058, respectively. All of these experimentally obtained values match well the theoretical predictions, as shown in Fig. \ref{fig:4}. Mathematically, even if the Jones polynomial values of two knots are identical, it does not necessarily mean that they are topologically equivalent \cite{doi:10.1142/4256}. Notably, the Trefoil knot and the Figure Eight knot are topologically distinct despite sharing the same value, $|V_L(i)|=1$. This limitation underscores the difficulty in discerning specific distinct links from their corresponding Jones polynomials when they are evaluated at a specific value $t$ \cite{pachos_2012}. Nevertheless, we can clearly distinguish the links whose values of the Jones polynomials, $V_L(i)$, are different. Such an advance can greatly contribute in the fields of statistical physics \cite{RN142}, molecular synthesis technology \cite{RN140} and integrated DNA replication \cite{RN141}, where intricate topological links and knots emerge frequently. \\

\noindent
\textbf{Discussion}\\
In summary, our work utilizes a photonic quantum platform to simulate the braiding operations of Majorana fermions. Two distinct Majorana braiding evolutions are realised that can generate a wide family of topologically distinct links with their worldlines. By measuring the quantum amplitudes of these evolutions we determined the Jones polynomials of the links. Even though these polynomials are evaluated at a non-universal value of their parameter we can directly use them to distinguish between most of the links. In particular, we have experimentally determined the Jones polynomials of five representative links, achieving high-fidelity results that align closely with theoretical predictions. 

The topological encoding of information with Majorana anyons is equivalent to encoding information in a quantum error correcting code (QECC) \cite{pachos_2012}. The added value of manipulating this logical information by braiding anyons is that the information remains encoded in the QECC at all times during the performance of the logical gates. This required a minimum size of ten fermionic sites to fault-tolerantly encode the manipulations between three pairs of anyons, as shown in Fig. \ref{fig:1}. Hence, the prototype algorithm we performed for evaluating Jones polynomials is encoded in a QECC that protects it against local errors acting on the fermionic sites at all times. We leave the demonstration of fault-tolerance of our encoded quantum algorithm to a future investigation \cite{doi:10.1126/sciadv.aat6533}. The versatility of our platform enables the simulation of a broader range of links, paving the way for the implementation of scalable quantum computation based on fault-tolerant Majorana fermions in various platforms, including superconducting circuits \cite{PhysRevX.10.021054}, integrated silicon chips \cite{PhysRevLett.124.193601}, and Rydberg atoms \cite{PhysRevLett.129.160501}.
\clearpage

\noindent
\textbf{Methods}\\
\textbf{Hamiltonians during braiding opearion $\sigma_1$ and $\sigma_2^{-1}$} 
We use two basic Majorana braiding operations, $\sigma_1$ and $\sigma_2^{-1}$, to generate various knots and links. The first one, $\sigma_1$, corresponds to the clockwise braiding of MZMs B and C, as shown in Fig. \ref{fig:1}{\bf b}. This evolution is achieved with the following sequence of $H_{M_0}$, $H_{M_1}$, $H_{M_2}$, $H_{M_3}$ and back to $H_{M_0}$ at last, where 
\begin{equation}
	\begin{split}
		H_{M_1}=i(&\gamma_{1b} \gamma_{2a}+\gamma_{2b} \gamma_{3a}+\gamma_{4b} \gamma_{5a}+\gamma_{5b} \gamma_{6a}+\gamma_{8b}\gamma_{9a}\\&+\gamma_{9b}\gamma_{10a}+\gamma_{7a}\gamma_{7b}), \\
		H_{M_2}=i(&\gamma_{1b} \gamma_{2a}+\gamma_{2b} \gamma_{3a}+\gamma_{3b} \gamma_{4b}+\gamma_{5b} \gamma_{6a}+\gamma_{8b}\gamma_{9a}\\&+\gamma_{9b}\gamma_{10a}+\gamma_{7a}\gamma_{7b}), \\
		H_{M_3}=i(&\gamma_{1b} \gamma_{2a}+\gamma_{2b} \gamma_{3a}+\gamma_{5b} \gamma_{6a}+\gamma_{8b} \gamma_{9a}+\gamma_{9b}\gamma_{10a}\\&+\gamma_{4a}\gamma_{4b}+\gamma_{7a}\gamma_{7b}).
	\end{split}\tag{1}
	\label{eqn:123}
\end{equation}
The anticlockwise braiding of C and E implements braiding operation $\sigma_{2}^{-1}$, as shown in Fig. \ref{fig:1}{\bf b}. Its corresponding evolution is achieved with the following sequence: $H_{M_0}$, $H'_{M_5}$, $H'_{M_4}$, $H'_{M_3}$, $H'_{M_2}$, $H'_{M_1}$ and $H_{M_0}$, where 
\begin{equation}
	\begin{split}
		H'_{M_5}=i(&\gamma_{1b} \gamma_{2a}+\gamma_{5b} \gamma_{6a}+\gamma_{8b}\gamma_{9a}+\gamma_{9b} \gamma_{10a}+\gamma_{3a} \gamma_{3b}\\&+\gamma_{4a}\gamma_{4b}+\gamma_{7a}\gamma_{7b}),\\
		H'_{M_4}=i(&\gamma_{1b} \gamma_{2a}+\gamma_{5a} \gamma_{7a}+\gamma_{5b} \gamma_{6a}+\gamma_{8b} \gamma_{9a}+\gamma_{9b} \gamma_{10a}\\&+\gamma_{3a} \gamma_{3b}+\gamma_{4a}\gamma_{4b}), \\
		H'_{M_3}=i(&\gamma_{1b} \gamma_{2a}+\gamma_{5a} \gamma_{7a}+\gamma_{5b} \gamma_{6a}+\gamma_{7b} \gamma_{8b}+\gamma_{9b} \gamma_{10a}\\&+\gamma_{3a} \gamma_{3b}+\gamma_{4a}\gamma_{4b}), \\
		H'_{M_2}=i(&\gamma_{1b} \gamma_{2a}+\gamma_{5a} \gamma_{7a}+\gamma_{5b} \gamma_{6a}+\gamma_{9b} \gamma_{10a}+\gamma_{3a}\gamma_{3b}\\&+\gamma_{4a}\gamma_{4b}+\gamma_{8a}\gamma_{8b}), \\
		H'_{M_1}=i(&\gamma_{1b} \gamma_{2a}+\gamma_{5b} \gamma_{6a}+\gamma_{8b}\gamma_{9a} +\gamma_{9b}\gamma_{10a}+\gamma_{3a}\gamma_{3b}\\&+\gamma_{4a}\gamma_{4b}+\gamma_{7a}\gamma_{7b}).
	\end{split}\tag{2}
	\label{eqn:12345}
\end{equation}
After the Jordan-Wigner (JW) transformation,  the  superconducting Hamiltonians employed for realising the braiding operation $\sigma_{1}$ are transformed to spin Hamiltonians 
\begin{equation}
	\begin{split}
		&H_{0} =-\sigma_{1}^{x}\sigma_{2}^{x}-\sigma_{4}^{x}\sigma_{5}^{x}-\sigma_{5}^{x}\sigma_{6}^{x} -\sigma_{8}^{x}\sigma_{9}^{x}-\sigma_{9}^{x}\sigma_{10}^{x}+\sigma_{3}^{z}+\sigma_{7}^{z},\\ &H_{1}=-\sigma_{1}^{x}\sigma_{2}^{x}-\sigma_{2}^{x}\sigma_{3}^{x}-\sigma_{4}^{x}\sigma_{5}^{x} -\sigma_{5}^{x}\sigma_{6}^{x}-\sigma_{8}^{x}\sigma_{9}^{x}-\sigma_{9}^{x}\sigma_{10}^{x}+\sigma_{7}^{z},\\ 
		&H_{2}=-\sigma_{1}^{x}\sigma_{2}^{x}-\sigma_{2}^{x}\sigma_{3}^{x}+\sigma_{3}^{x}\sigma_{4}^{y} -\sigma_{5}^{x}\sigma_{6}^{x}-\sigma_{8}^{x}\sigma_{9}^{x}-\sigma_{9}^{x}\sigma_{10}^{x}+\sigma_{7}^{z},\\
		&H_{3}=-\sigma_{1}^{x}\sigma_{2}^{x}-\sigma_{2}^{x}\sigma_{3}^{x}-\sigma_{5}^{x}\sigma_{6}^{x} -\sigma_{8}^{x}\sigma_{9}^{x}-\sigma_{9}^{x}\sigma_{10}^{x}+\sigma_{4}^{z}+\sigma_{7}^{z}.
	\end{split}\tag{3}
	\label{eq3}
\end{equation}
For braiding operation $\sigma_{2}^{-1}$, the superconducting Hamiltonians in \eqref{eqn:12345} are transformed to the spin Hamiltonians 
\begin{equation}
	\begin{split}
		&H_{0} =-\sigma_{1}^{x}\sigma_{2}^{x}-\sigma_{4}^{x}\sigma_{5}^{x}-\sigma_{5}^{x}\sigma_{6}^{x} -\sigma_{8}^{x}\sigma_{9}^{x}-\sigma_{9}^{x}\sigma_{10}^{x}+\sigma_{3}^{z}+\sigma_{7}^{z},\\ &H'_{5}=-\sigma_{1}^{x}\sigma_{2}^{x}-\sigma_{5}^{x}\sigma_{6}^{x}-\sigma_{8}^{x} \sigma_{9}^{x}-\sigma_{9}^{x} \sigma_{10}^{x}+\sigma_{3}^{z}+\sigma_{4}^{z}+\sigma_{7}^{z},\\ &H'_{4}=-\sigma_{1}^{x}\sigma_{2}^{x}-\sigma_{5}^{x}\sigma_{6}^{x}-\sigma_{5}^{y} \sigma_{6}^{z} \sigma_{7}^{x}-\sigma_{8}^{x} \sigma_{9}^{x}-\sigma_{9}^{x} \sigma_{10}^{x}+\sigma_{3}^{z}+\sigma_{4}^{z},\\
		&H'_{3}=-\sigma_{1}^{x}\sigma_{2}^{x}-\sigma_{5}^{x}\sigma_{6}^{x}-\sigma_{5}^{y} \sigma_{6}^{z} \sigma_{7}^{x}+\sigma_{7}^{x} \sigma_{8}^{y}-\sigma_{9}^{x} \sigma_{10}^{x}+\sigma_{3}^{z}+\sigma_{4}^{z},\\
		&H'_{2}=-\sigma_{1}^{x}\sigma_{2}^{x}-\sigma_{5}^{x}\sigma_{6}^{x}-\sigma_{5}^{y} \sigma_{6}^{z}\sigma_{7}^{x}-\sigma_{9}^{x}\sigma_{10}^{x}+\sigma_{3}^{z}+\sigma_{4}^{z} +\sigma_{8}^{z},\\  &H'_{1}=-\sigma_{1}^{x}\sigma_{2}^{x}-\sigma_{5}^{x}\sigma_{6}^{x}-\sigma_{8}^{x} \sigma_{9}^{x}-\sigma_{9}^{x}\sigma_{10}^{x}+\sigma_{3}^{z}+\sigma_{4}^{z}+\sigma_{7}^{z}.
	\end{split}\tag{4}
	\label{eq4}
\end{equation}
It is worth mentioning that in $\sigma_2^{-1}$ when evolving from the final state of $H'_{5}$ to that of $H'_{4}$, we simultaneously change the basis vectors of three Majorana fermions in one step due to the existence of $-\sigma_{5}^{y} \sigma_{6}^{z} \sigma_{7}^{x}$. Such intricate multi-vector transformation is absent in $\sigma_{1}$. (The specific ground states of the Hamiltonians in Eqs \eqref{eq3} and \eqref{eq4} can be found in Supplemental Material.)\\ 

\begin{table*}[htbp]
	\caption{Preparation of the initial state $\ket{{\phi_{0}'}}$ based on two-photon correlations.} 
	\begin{ruledtabular}
		\begin{tabular}{ccccc}
			&$\ket{a}_{B}$&$\ket{b}_{B}$&$\ket{c}_{B}$&$\ket{d}_{B}$\\
			\colrule
			$\ket{1}_{A}$&$|x_{1}x_{2}\bar{z}_{3}x_{4}x_{5}x_{6}\bar{z}_{7}x_{8}x_{9}x_{10}\rangle$&$-|x_{1}x_{2}\bar{z}_{3}x_{4}x_{5}x_{6}\bar{z}_{7}\bar{x}_{8}\bar{x}_{9}\bar{x}_{10}\rangle$&$|\bar{x}_{1}\bar{x}_{2}\bar{z}_{3}x_{4}x_{5}x_{6}\bar{z}_{7}x_{8}x_{9}x_{10}\rangle$&$-|\bar{x}_{1}\bar{x}_{2}\bar{z}_{3}x_{4}x_{5}x_{6}\bar{z}_{7}\bar{x}_{8}\bar{x}_{9}\bar{x}_{10}\rangle$\\
			\colrule
			$\ket{2}_{A}$&$-|x_{1}x_{2}\bar{z}_{3}\bar{x}_{4}\bar{x}_{5}\bar{x}_{6}\bar{z}_{7}x_{8}x_{9}x_{10}\rangle$&
			$|x_{1}x_{2}\bar{z}_{3}\bar{x}_{4}\bar{x}_{5}\bar{x}_{6}\bar{z}_{7}\bar{x}_{8}\bar{x}_{9}\bar{x}_{10}\rangle$&$-|\bar{x}_{1}\bar{x}_{2}\bar{z}_{3}\bar{x}_{4}\bar{x}_{5}\bar{x}_{6}\bar{z}_{7}x_{8}x_{9}x_{10}\rangle$&$|\bar{x}_{1}\bar{x}_{2}\bar{z}_{3}\bar{x}_{4}\bar{x}_{5}\bar{x}_{6}\bar{z}_{7}\bar{x}_{8}\bar{x}_{9}\bar{x}_{10}\rangle$\\
		\end{tabular}
	\end{ruledtabular}	
\end{table*}

\noindent
\textbf{Two-photon correlation encoding}\\ 
In Fig. \ref{fig:2}, the obtained spatial modes on side A and side B are denoted as $\ket{n}_{A}$ and $\ket{m}_{B}$, respectively, where $n \in \{1,2,3...\}$ and $m \in \{a,b,c...\}$. 
The experiment employs two types of BDs, namely BD3 and BD6, which can spatially separate light beams by 3.0 mm and 6.0 mm, respectively.  The coincidence of the spatial modes between the two photons encodes a specific spin bit. Specifically, the two-photon correlation modes are expressed as $\ket{nm}=\ket{n}_{A} \otimes \ket{m}_{B}$.  For instance, the coincidence between $\ket{1}_{A}$ and $\ket{a}_{B}$ is regarded as the first term of the state $\ket{\phi_{0}'}$, i.e., $\ket{1a}=\ket{1}_{A} \otimes \ket{a}_{B}=|x_{1}x_{2}\bar{z}_{3}x_{4}x_{5}x_{6}\bar{z}_{7}x_{8}x_{9}x_{10}\rangle$, while the rest of the terms of $\ket{\phi_{0}'}$ are encoded similarly. The spatial patterns and corresponding encoding relationships of the initial state $\ket{\phi_{0}'}$ are summarized in Table I (unnormalized). The subsequent encodings of all states during the evolution exhibit similar characteristics.\\  

\noindent
\textbf{Realisation of the CNOT gate}\\ 
In the Sagnac interferometer (as shown in Fig. \ref{fig:2}{\bf a}), the incoming photon's horizontal and vertical polarization components are considered as the control qubit and will be equally rotated by an HWP, resulting in $(\ket{H}+\ket{V})/\sqrt{2}=(\ket{0}+\ket{1})/\sqrt{2}$. Subsequently, the photon will be split into two paths by a PBS, which transmits the horizontal component and reflects vertical component. Thus, the photons with horizontal polarization propagate in a clockwise direction (blue path) are denoted as the ground state $\ket{g}$, and the photons with vertical polarization propagate in a counterclockwise direction (red path) are denoted as the excited state $\ket{e}$, Consequently, the input state can be expressed as $(\ket{H}\ket{g}+\ket{V}\ket{e})/\sqrt{2}=(\ket{0}\ket{g}+\ket{1}\ket{e})/\sqrt{2}$.  Next, the CNOT gate executes the operation: when the control bit is $\ket{0}$, the target bit remains unchanged; when the control bit is $\ket{1}$, the target bit flips, and thus the state becomes $(\ket{H}\ket{g}+\ket{V}\ket{g})/\sqrt{2}=(\ket{0}\ket{g}+\ket{1}\ket{g})/\sqrt{2}$. This process corresponds to the emission of the photon's  horizontal and vertical polarization components exiting from the same port of the PBS.
Finally, after the second Hadamard operation (which is realised through an HWP at 22.5°) on the control qubit, the state becomes $\ket{H}\ket{g}=\ket{0}\ket{g}$. \\
\newpage

\noindent
\textbf{Acknowledgements}\\
This work was supported by the Innovation Program for Quantum Science and Technology (Grants
No. 2021ZD0301200 and No. 2021ZD0301400), the National Natural Science Foundation of China (Grants  No. 61975195, No. 11821404, No. 92365205, No. 12374336 and No. 11874343), the Anhui Initiative in Quantum Information Technologies (Grant No. AHY060300) and USTC Research Funds of
the Double First-Class Initiative (Grant No. YD2030002024). J.K.P would like to thank Iason Sofos and Matthew Yusuf for inspiring conversations. J.K.P acknowledges support by EPSRC, Grant No. EP/R020612/1.\\

\noindent
\textbf{Author contributions}\\
Y.-J.H. proposed this project. K.S. designed the experiment. J.-K.L. carried out the 
experiment and analysed the experimental results assisted by K.S., Z.-Y.H. and J.-H.L.. J.K.P contributed to the theoretical analysis assisted by S.-J.T.. J.-K.L. and J.K.P wrote the manuscript.  J.-S.X., Y.-J.H., C.-F.L. and G.-C.G. supervised the project. 
All the authors discussed the experimental procedures and results.\\

\noindent
\textbf{Competing interests}\\
The authors declare no competing interests.

\bibliography{reference.bib}

\end{document}


\hyphenpenalty=5000
	\tolerance=1000
	
	\title{
		Supplemental Material for:\\
		Photonic simulation of Majorana-based Jones polynomials
	}
	
	\author{Jia-Kun Li}
	\affiliation{CAS Key Laboratory of Quantum Information, University of Science and Technology of China, Hefei 230026, People's Republic of China}
	\affiliation{CAS Center For Excellence in Quantum Information and Quantum Physics, University of Science and Technology of China, Hefei 230026, People's Republic of China}
	
	\author{Kai Sun}
	\email{ksun678@ustc.edu.cn}
	\affiliation{CAS Key Laboratory of Quantum Information, University of Science and Technology of China, Hefei 230026, People's Republic of China}
	\affiliation{CAS Center For Excellence in Quantum Information and Quantum Physics, University of Science and Technology of China, Hefei 230026, People's Republic of China}
	
	\author{Ze-Yan Hao}
	\affiliation{CAS Key Laboratory of Quantum Information, University of Science and Technology of China, Hefei 230026, People's Republic of China}
	\affiliation{CAS Center For Excellence in Quantum Information and Quantum Physics, University of Science and Technology of China, Hefei 230026, People's Republic of China}
	
	\author{Jia-He Liang}
	\affiliation{CAS Key Laboratory of Quantum Information, University of Science and Technology of China, Hefei 230026, People's Republic of China}
	\affiliation{CAS Center For Excellence in Quantum Information and Quantum Physics, University of Science and Technology of China, Hefei 230026, People's Republic of China}
	
	\author{Si-Jing Tao}
	\affiliation{CAS Key Laboratory of Quantum Information, University of Science and Technology of China, Hefei 230026, People's Republic of China}
	\affiliation{CAS Center For Excellence in Quantum Information and Quantum Physics, University of Science and Technology of China, Hefei 230026, People's Republic of China}
	
	\author{Jin-Shi Xu}
	\email{jsxu@ustc.edu.cn}
	\affiliation{CAS Key Laboratory of Quantum Information, University of Science and Technology of China, Hefei 230026, People's Republic of China}
	\affiliation{CAS Center For Excellence in Quantum Information and Quantum Physics, University of Science and Technology of China, Hefei 230026, People's Republic of China}
	
	\author{Yong-Jian Han}
	\email{smhan@ustc.edu.cn}
	\affiliation{CAS Key Laboratory of Quantum Information, University of Science and Technology of China, Hefei 230026, People's Republic of China}
	\affiliation{CAS Center For Excellence in Quantum Information and Quantum Physics, University of Science and Technology of China, Hefei 230026, People's Republic of China}
	
	\author{Chuan-Feng Li}
	\email{cfli@ustc.edu.cn}
	\affiliation{CAS Key Laboratory of Quantum Information, University of Science and Technology of China, Hefei 230026, People's Republic of China}
	\affiliation{CAS Center For Excellence in Quantum Information and Quantum Physics, University of Science and Technology of China, Hefei 230026, People's Republic of China}
	
	\author{Jiannis K. Pachos}
	\affiliation{School of Physics and Astronomy, University of Leeds, Leeds LS2 9JT, United Kingdom}
	
	\author{Guang-Can Guo}
	\affiliation{CAS Key Laboratory of Quantum Information, University of Science and Technology of China, Hefei 230026, People's Republic of China}
	\affiliation{CAS Center For Excellence in Quantum Information and Quantum Physics, University of Science and Technology of China, Hefei 230026, People's Republic of China}
	\maketitle
	\clearpage
	\tableofcontents{}
	\clearpage
	
	\section{Jones polynomials and Majorana fermion amplitudes}
	
	\subsection{Jones polynomials}
	
	Jones polynomials are topological invariants that can be used to distinguish topologically inequivalent links. This property makes Jones polynomials of interest for applications in various sciences, such as biology or statistics \cite{Liu2016,Zhou2021}. Unfortunately, distinguishing if two links are topologically inequivalent is classically an exponentially hard problem. In particular, this complexity increasing with the number of crosses between the different strands of the link. It has been discovered that quantum mechanical amplitude of anyons with world-lines that span the links give an efficient quantum way to determine the Jones polynomials. This inspired efficient quantum algorithms that efficiently approximate Jones polynomials. 
	
	Mathematically, the Jones polynomials $V_L(t)$ of a link $L$ as a function of the parameter $t$ is related to the quantum mechanical amplitude of braided SU(2)$_k$ anyons as
	\begin{equation}
		V_L(t) = (-\sqrt[4]{t})^{-3w(L)} d^{n-1} \langle \phi_0|U(t)|\phi_0\rangle,
		\label{eq:JonesAmp}
	\end{equation}
	where $t = e^{2\pi i \over (k+2)}$, determines the type of employed anyons, $d = -\sqrt{t}-1/\sqrt{t}$ is the quantum dimension of the anyons and $w(L)$ is the writhe of the link $L$ \cite{2009A}, $|\phi_0\rangle$ is the initial state of the system, being taken to correspond to pair creation of $n$ pairs of anyons and $U(t)$ is the unitary that describes the braiding evolution between the anyons that span the link $L$ with their world-lines. For $k\neq 2,4$ the Jones polynomials are known to be exponentially difficult to determine and thus can be used as primitives for quantum computation \cite{2009How}.
	
	\subsection{Jones polynomials for Ising model, SU(2)$_2$}
	
	For the particular case of Ising anyons corresponding to $k=2$ and $d=\sqrt{2}$ the Jones polynomials are given by a simple formula
	\begin{equation}
		V_L (t=i) = \left\{
		\begin{matrix}
			\sqrt{2}^{\#(L)-1} (-1)^{\tt arf}(L),\,\,\, \text{if $L$ is proper}\\
			~~~~~~~~~~0,~~~~~~~~~~~~ \text{if $L$ is not proper}
		\end{matrix}
		\right.
		\label{eqn:JonesArf}
	\end{equation}
	where $\#(L)$ is the number of link components. An oriented link $L$ with components $\{L_k\}$ is proper if each component $L_k$ evenly links the union of all other components, i.e. $\sum_{j\neq k} lk(L_j,L_k) =0\, \rm{mod}\,2$. 
	
	Calculating the Jones polynomials of a link $L$ requires the application of the skein relations at every crossing of $L$. For the special case of a totally proper link, i.e. where $L$ has all pairwise linking numbers even, then an analytic formula can be used to calculate the Jones polynomials ${\tt arf}$ invariant
	\begin{equation}
		{\tt arf}(L) =\!\! \sum_i \! c_1(L_i) +\!\! \sum_{i<j} \! c_2(L_i,L_j) +\!\! \sum_{i<j<k} \!\!\!c_3(L_i,L_j,L_k) \rm{mod}\, 2.
		\label{eq:Arf}
	\end{equation}
	Here $c_1(L_i)$ is the self-linking number of component $L_i$, $c_2(L_i,L_j)$ is given by
	\begin{equation}
		c_2(L_i,L_j) = lk(L_i,L_j)[lk(L_i,L_j)^2-1]/6
	\end{equation}
	and $c_3(L_i,L_j,L_k)$ is the Milnor invariant that counts the number of Borromean rings of the three components $L_i,L_j,L_k$. Hence, the Jones polynomials $V_L (i)$ is always a real number, its magnitude depends on the number of link components $n$ (pairs of anyons) and its sign $(-1)^{\tt arf}(L)$ depends on the self linking of each component, the linking or pairwise components and the number of Borromean rings from each triplet of components. 
	
	\subsection{Ising anyons}
	
	We will now calculate the quantum mechanical amplitude $\langle \phi_0|U_b(A)|\phi_0\rangle$ to determine the Jones polynomials from \eqref{eq:JonesAmp}. The topological Hilbert space is encoded in the fusion outcome of two non-Abelian $\sigma$ anyons of Ising type, $\sigma\times \sigma = 1+ \psi$, namely the vacuum state, $|\sigma\times\sigma\rightarrow 1\rangle$ and the fermion state, $|\sigma\times\sigma\rightarrow \psi\rangle$. In this basis four $\sigma$ anyons, name them 1, 2, 3 and 4, with vacuum total fusion outcome can encode two topological states. Assuming that $1,2$ and $3,4$ pairs have vacuum of fermion fusion outcome. Then the braiding $B_1$ between 1 and 2 and the braiding $B_2$ between 2 and 3 are given by
	\begin{equation}
		B_1 = e^{i\pi/8}
		\left(
		\begin{matrix}
			-1 & 0\\
			0 & i
		\end{matrix}
		\right),\,\,
		B_2 = -{e^{-i\pi/8}\over \sqrt{2}}
		\left(
		\begin{matrix}
			1 & i\\
			i & 1
		\end{matrix}
		\right)
		\label{eq:braids}
	\end{equation}
	Note that $B_2=HB_1H^\dagger$, where H is the fusion matrix, also equal to the Hadamard gate.
	
	In the case of three pairs of anyons the Hilbert space is 4-dim and the braiding operations are
	\begin{eqnarray}
		B_1 &=&  e^{i\pi\over8} \text{diag}(-1,-1,i,i),\,\,B_2 = -{e^{-i\pi\over 8}\over \sqrt{2}}
		\left(
		\begin{matrix}
			1 & i\\
			i & 1
		\end{matrix}
		\right) \otimes 1\!\!1\nonumber\\
		B_3 &=&  e^{i\pi\over 8} \text{diag}(-1,i,i,-1),\,\,B_4 = -{e^{-i\pi\over 8}\over \sqrt{2}}
		1\!\!1\otimes 
		\left(
		\begin{matrix}
			1 & i\\
			i & 1
		\end{matrix}
		\right)
	\end{eqnarray}
	
	For the Borromean link we have
	\begin{equation}
		\langle \phi_0| B_4 (B_2B_3^{-1})^3 B_4^{-1}|\phi_0\rangle = -1.
	\end{equation}
	
	For the Trefoil link we have
	\begin{equation}
		\langle \phi_0| B_2^3|\phi_0\rangle = {e^{-3i\pi\over 8}\over \sqrt{2}}
	\end{equation}
	
	For the Hopf link we have
	\begin{equation}
		\langle \phi_0|B_2^2|\phi_0\rangle = 0
	\end{equation}
	
	Using expression  \eqref{eq:JonesAmp} we find that these amplitudes give the expected Jones polynomials given in the previous section. Indeed, $V_H(i) = (-t^{3/4})^{-w} d^{2-1} \langle \phi_0|B_2^2|\phi_0\rangle =0$, $V_B(i) = (-t^{3/4})^{-w} d^{3-1} \langle \phi_0| B_4 (B_2B_3^{-1})^3 B_4^{-1}|\phi_0\rangle = -2$ for $w=0$ and $d=\sqrt{2}$ and $V_T(i) = (-t^{3/4})^{-w} \langle \phi_0| B_2^3|\phi_0\rangle = -1$ for $w=3$. Note that $n$ in \eqref{eq:JonesAmp} corresponds to the pairs of anyons employed to apply the braiding operations in order to create the desired links and it is not equal necessarily to $\#(L)$.
	\subsection{Four representative links}
	\begin{figure}[!h]
		\centering
		\includegraphics[width=1\columnwidth]{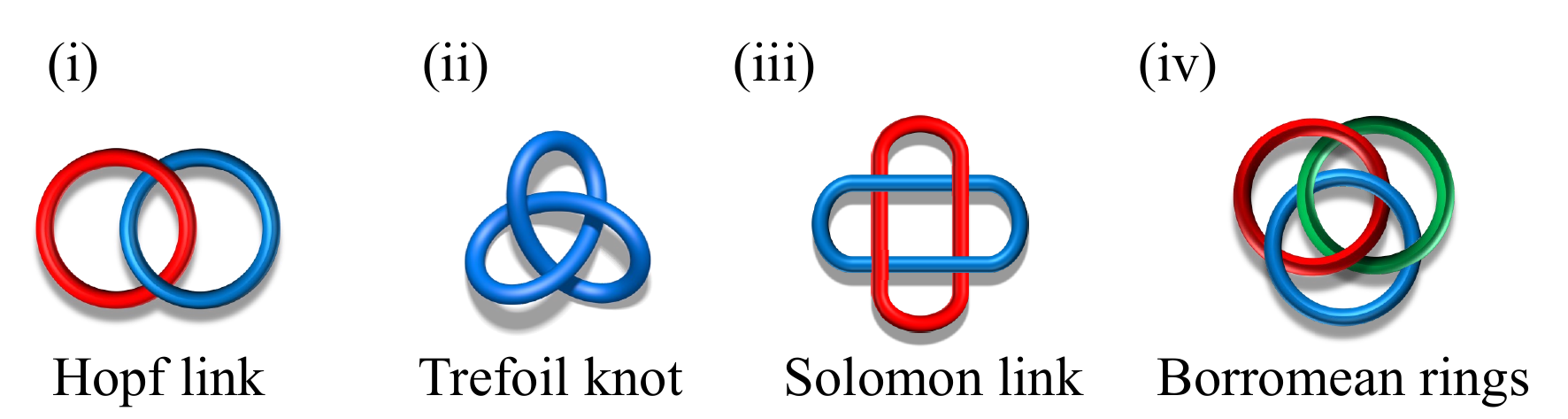}\\
		\normalsize{\begin{flushleft}Fig. S1. (i) The Hopf non-proper link with two components and $lk=1$ ($\#(L)=2$). Proper links: (ii) The trefoil link with a single component and self-linking number $c_1=1$ ($\#(L)=1$). (iii) The Solomon link with $lk=2$ and $c_2=1$ ($\#(L)=2$). (iv) The Borromean link with $c_3=1$ ($\#(L)=3$). \end{flushleft}
		}
		\label{fig:links}
	\end{figure}

	To demonstrate the applicability of \eqref{eqn:JonesArf} in distinguishing between inequivalent links, we can choose representative links that are either not proper, or proper with the topological invariants $c_1$, $c_2$ or $c_3$ being zero apart from one of them. We choose representative links to be (i) the non-proper Hopf link, $L_1$ with two components, and for proper links (ii) the trefoil link $L_2$ with $c_1=1$ and $c_2=c_3=0$, (iii) Solomon link $L_3$ with two components with $lk=2$ giving $c_2=1$ and $c_1=c_3=0$ and (iv) the Borromean link $L_4$ with $c_3=1$ and $c_1=c_2=0$, as shown in Fig. S1.
	
	\subsubsection{Hopf link}
	
	Applying the skein relations we can calculate the Jones polynomials for the Hopf link (see Fig. S1(i))  to be
	\begin{equation}
		V_H(t) = (-t^{3/4})^{-w} (-t-t^{-1}),
	\end{equation}
	where $w=2$. We have that for the Ising model $t=i$ so 
	\begin{equation}
		V_H(i)=0. 
	\end{equation}
	This is in agreement with \eqref{eqn:JonesArf}, as the Hopf link is not proper.
	
	\subsubsection{Trefoil link}
	
	Applying the skein relations to the trefoil link (see Fig. S1(ii)) we obtain
	\begin{equation}
		V_T(t) = t^{-1}+t^{-4} -t^{-4}
	\end{equation}
	as $w=3$. For $t=i$ we have 
	\begin{equation}
		V_T(i) = -1. 
	\end{equation}
	This is in agreement with \eqref{eqn:JonesArf} and \eqref{eq:Arf}, when $\#(L)=1$ and $c_1=1$, $c_2=c_3=0$. This can be contrasted to a single looped link with $c_1=c_2=c_3=0$ that has $V_T(i) = 1$. So these are distinguished by the minus sign.
	
	\subsubsection{Solomon link}
	
	For the link in Fig. S1(iii) we have from \eqref{eqn:JonesArf} and \eqref{eq:Arf} that 
	\begin{equation}
		V_S(i) = -\sqrt{2}, 
	\end{equation}
	where $\#(L)=2$, and as $lk=2$ we have $c_2=1$ while $c_1=c_3=0$. This can be contrasted to two unentangled looped links with $c_1=c_2=c_3=0$ that have $V_S(i) = \sqrt{2}$. So these are distinguished by the minus-sign.
	
	\subsubsection{Borromean link}
	
	For the case of a Borromean link (see Fig. S1(iv)) the Jones polynomials is given by
	\be
	V_B(t) = (-t^{3/4})^{-w} (t^{7/4}+t^{-7/4}-2).
	\ee
	For $t=i$ and $w=0$ we have 
	\be
	V_B(i) =-2.
	\ee
	This is in agreement with \eqref{eqn:JonesArf} and \eqref{eq:Arf}, when $\#(L)=3$ and $c_1=c_2=0$, $c_3=1$. This can be contrasted to three unentangled looped links with $c_1=c_2=c_3=0$ that have $V_B(i) = 2$. So these are distinguished by the minus sign.
	
	\section{Logical basis transformation and final results}
	
	The fermionic system consists of 10 fermions with 6 Majorana zero modes (MZMs).
	According to the initial fermionic Hamiltonian
	\begin{equation}
		H_{M_0}=i(\gamma_{1b} \gamma_{2a}+\gamma_{4b} \gamma_{5a}+\gamma_{5b} \gamma_{6a}+\gamma_{8b} \gamma_{9a}+\gamma_{9b}\gamma_{10a}+\gamma_{3a}\gamma_{3b}+\gamma_{7a}\gamma_{7b}),
	\end{equation}
	it could be transformed to the corresponding spin Hamiltonians $H_{0}$ under the Jordan-Wigner (JW) transformation,
	\begin{equation}
		H_{0} =-\sigma_{1}^{x}\sigma_{2}^{x}-\sigma_{4}^{x}\sigma_{5}^{x}-\sigma_{5}^{x}\sigma_{6}^{x} -\sigma_{8}^{x}\sigma_{9}^{x}-\sigma_{9}^{x}\sigma_{10}^{x}+\sigma_{3}^{z}+\sigma_{7}^{z}.
	\end{equation}
	The ground states of the corresponding Hamiltonian can be expressed in terms of the eigenvectors $\{|x\rangle,|\bar{x}\rangle\}$, $\{|y\rangle,|\bar{y}\rangle\}$ and $\{|z\rangle,|\bar{z}\rangle\}$ of the corresponding Pauli operators $\sigma^x=\left(\begin{smallmatrix}0 &1\\ 1 &0\end{smallmatrix}\right)$, $\sigma^y=\left(\begin{smallmatrix}0 &-i\\ i &0\end{smallmatrix}\right)$ and $\sigma^z=\left(\begin{smallmatrix}1 &0\\ 0 &-1\end{smallmatrix}\right)$, with eigenvalues $\{1,-1\}$, respectively.
	
	After the ITE of the initial Hamiltonian $H_{0}$ which corresponds to creating the MZMs noted as A, B, C, D, E, and F (shown in Fig. 1\textbf{b} in the main text), the state becomes
	\begin{equation}
		\begin{split}
			|\phi_{0}\rangle &=\alpha|x_{1}x_{2}\bar{z}_{3}x_{4}x_{5}x_{6}\bar{z}_{7}x_{8}x_{9}x_{10}\rangle
			+\beta|x_{1}x_{2}\bar{z}_{3}x_{4}x_{5}x_{6}\bar{z}_{7}\bar{x}_{8}\bar{x}_{9}\bar{x}_{10}\rangle \\
			&+\gamma|x_{1}x_{2}\bar{z}_{3}\bar{x}_{4}\bar{x}_{5}\bar{x}_{6}\bar{z}_{7}x_{8}x_{9}x_{10}\rangle
			+\delta|x_{1}x_{2}\bar{z}_{3}\bar{x}_{4}\bar{x}_{5}\bar{x}_{6}\bar{z}_{7}\bar{x}_{8}\bar{x}_{9}\bar{x}_{10}\rangle \\
			&+\eta|\bar{x}_{1}\bar{x}_{2}\bar{z}_{3}x_{4}x_{5}x_{6}\bar{z}_{7}x_{8}x_{9}x_{10}\rangle
			+\kappa|\bar{x}_{1}\bar{x}_{2}\bar{z}_{3}x_{4}x_{5}x_{6}\bar{z}_{7}\bar{x}_{8}\bar{x}_{9}\bar{x}_{10}\rangle \\
			&+\mu|\bar{x}_{1}\bar{x}_{2}\bar{z}_{3}\bar{x}_{4}\bar{x}_{5}\bar{x}_{6}\bar{z}_{7}x_{8}x_{9}x_{10}\rangle
			+\nu|\bar{x}_{1}\bar{x}_{2}\bar{z}_{3}\bar{x}_{4}\bar{x}_{5}\bar{x}_{6}\bar{z}_{7}\bar{x}_{8}\bar{x}_{9}\bar{x}_{10}\rangle,
			\label{initial}
		\end{split}
	\end{equation}
	where $\alpha$, $\beta$, $\gamma$, $\delta$, $\eta$, $\kappa$, $\mu$ and $\nu$ are complex amplitudes satisfying $|\alpha|^{2}+|\beta|^{2}+|\gamma|^{2}+|\delta|^{2}+|\eta|^{2}+|\kappa|^{2}+|\mu|^{2}+|\nu|^{2}=1$.

	We encode the spin qubits to the logical qubits as follows,
	\begin{equation}
		\begin{split}
			&|x_{1}x_{2}\rangle=(|0_{12}\rangle+|1_{12}\rangle)/\sqrt{2},\\
			&|\bar{x}_{1}\bar{x}_{2}\rangle=(|0_{12}\rangle-|1_{12}\rangle)/\sqrt{2},\\
			&|x_{4}x_{5}x_{6}\rangle=(|0_{456}\rangle+|1_{456}\rangle)/\sqrt{2},\\
			&|\bar{x}_{4}\bar{x}_{5}\bar{x}_{6}\rangle=(|1_{456}\rangle-|0_{456}\rangle)/\sqrt{2}),\\
			&|x_{8}x_{9}x_{10}\rangle=(|0_{8910}\rangle+|1_{8910}\rangle)/\sqrt{2},\\
			&|\bar{x}_{8}\bar{x}_{9}\bar{x}_{10}\rangle=(|1_{8910}\rangle-|0_{8910}\rangle)/\sqrt{2}).\\
		\end{split}
	\end{equation}
	Based on this transformation of logical basis, we could obtain that
	the component ($|x_{1}x_{2}\bar{z}_{3}x_{4}x_{5}x_{6}\bar{z}_{7}x_{8}x_{9}x_{10}\rangle$) of the initial state becomes
	\begin{equation}
		\alpha(|000\rangle+|001\rangle+|010\rangle+|011\rangle+|100\rangle+|101\rangle+|110\rangle+|111\rangle).
	\end{equation}
	The component ($|x_{1}x_{2}\bar{z}_{3}x_{4}x_{5}x_{6}\bar{z}_{7}\bar{x}_{8}\bar{x}_{9}\bar{x}_{10}\rangle$) of the initial state becomes
	\begin{equation}
		\beta(|001\rangle-|000\rangle+|011\rangle-|010\rangle+|101\rangle-|100\rangle+|111\rangle-|110\rangle).
	\end{equation}
	The component ($|x_{1}x_{2}\bar{z}_{3}\bar{x}_{4}\bar{x}_{5}\bar{x}_{6}\bar{z}_{7}x_{8}x_{9}x_{10}\rangle$) of the initial state becomes
	\begin{equation}
		\gamma(|010\rangle+|011\rangle-|000\rangle-|001\rangle+|110\rangle+|111\rangle-|100\rangle-|101\rangle).
	\end{equation}
	The component ($|x_{1}x_{2}\bar{z}_{3}\bar{x}_{4}\bar{x}_{5}\bar{x}_{6}\bar{z}_{7}\bar{x}_{8}\bar{x}_{9}\bar{x}_{10}\rangle$) of the initial state becomes
	\begin{equation}
		\delta(|011\rangle-|010\rangle-|001\rangle+|000\rangle+|111\rangle-|110\rangle-|101\rangle+|100\rangle).
	\end{equation}
	The component ($|\bar{x}_{1}\bar{x}_{2}\bar{z}_{3}x_{4}x_{5}x_{6}\bar{z}_{7}x_{8}x_{9}x_{10}\rangle$) of the initial state becomes
	\begin{equation}
		\eta(|000\rangle+|001\rangle+|010\rangle+|011\rangle-|100\rangle-|101\rangle-|110\rangle-|111\rangle).
	\end{equation}
	The component ($|\bar{x}_{1}\bar{x}_{2}\bar{z}_{3}x_{4}x_{5}x_{6}\bar{z}_{7}\bar{x}_{8}\bar{x}_{9}\bar{x}_{10}\rangle$) of the initial state becomes
	\begin{equation}
		\kappa(|001\rangle-|000\rangle+|011\rangle-|010\rangle-|101\rangle+|100\rangle-|111\rangle+|110\rangle).
	\end{equation}
	The component ($|\bar{x}_{1}\bar{x}_{2}\bar{z}_{3}\bar{x}_{4}\bar{x}_{5}\bar{x}_{6}\bar{z}_{7}x_{8}x_{9}x_{10}\rangle$) of the initial state becomes
	\begin{equation}
		\mu(|010\rangle+|011\rangle-|000\rangle-|001\rangle-|110\rangle-|111\rangle+|100\rangle+|101\rangle).
	\end{equation}
	The component ($|\bar{x}_{1}\bar{x}_{2}\bar{z}_{3}\bar{x}_{4}\bar{x}_{5}\bar{x}_{6}\bar{z}_{7}\bar{x}_{8}\bar{x}_{9}\bar{x}_{10}\rangle$) of the initial state becomes
	\begin{equation}
		\nu(|011\rangle-|010\rangle-|001\rangle+|000\rangle-|111\rangle+|110\rangle+|101\rangle-|100\rangle).
	\end{equation}

	To prepare the initial state in the logical ground state ($\ket{\phi_{in}}=\ket{000}$), the initial parameters should be $\alpha=\delta=\eta=\nu=1/2\sqrt{2}$ and $\beta=\gamma=\kappa=\mu=-1/2\sqrt{2}$.
	
	After the Hopf link braiding ($\sigma_{1}^{2}$), the final state becomes 
	\begin{equation}
		\ket{\phi_{in}}\rightarrow\ket{\phi_{f}}=|110\rangle.
	\end{equation}
	The probability is $|\bra{\phi_{in}}\phi_f\rangle|^2=0$.
	
	For the Trefoil knot, we need perform the braiding operations $\sigma_{1}^{3}$, and the final state becomes
	\begin{equation}
		\ket{\phi_{in}}\rightarrow\ket{\phi'_{f}}=\frac{1}{\sqrt{2}}(|000\rangle+i|110\rangle).
	\end{equation}
	The probability is $|\bra{\phi_{in}}\phi'_f\rangle|^2=1/2$.
	
	For the Solomon link, we need perform the braiding operations $\sigma_{1}^{4}$, and the final state becomes
	\begin{equation}
		\ket{\phi_{in}}\rightarrow\ket{\phi'_{f}}=|000\rangle.
	\end{equation}
	The probability is $|\bra{\phi_{in}}\phi'_f\rangle|^2=1$.
	
	For the Figure Eight knot, we need perform the braiding operations $(\sigma_{1}\sigma_{2}^{-1})^{2}$, and the final state becomes
	\begin{equation}
		\ket{\phi_{in}}\rightarrow\ket{\phi'_{f}}=\frac{1}{2}(|000\rangle-|011\rangle-i|101\rangle-i|110\rangle).
	\end{equation}
	The probability is $|\bra{\phi_{in}}\phi'_f\rangle|^2=1/4$.
	
	For the Borromean rings, we need perform the braiding operations $(\sigma_{1}\sigma_{2}^{-1})^{3}$, and the final state becomes
	\begin{equation}
		\ket{\phi_{in}}\rightarrow\ket{\phi'_{f}}=|000\rangle.
	\end{equation}
	The probability is $|\bra{\phi_{in}}\phi'_f\rangle|^2=1$.

	\section{The process of imaginary-time evolution}
	
	To experimentally simulate the braiding operation of MZMs, the non-dissipative imaginary-time evolution (ITE) evolution is proposed to simulate the correspinding Hamiltonian. For a given Hamiltonian $H$ with a complete set of eigenstates $|e_{k}\rangle$ and the corresponding eigenvalues $E_{k}$, any arbitrary pure state $|\phi\rangle$ can be expressed as
	\begin{equation}
		|\phi\rangle=\sum_{k}q_{k}|e_{k}\rangle,
	\end{equation}
	with $q_{k}$ representing the corresponding complex amplitude. Here, we focus on pure states, but the argument is also valid for mixed states. The corresponding imaginary-time evolution (ITE) operator ($U$) on the state becomes
	\begin{equation}
		U|\phi\rangle=\exp(-H*t)\sum_{k}q_{k}|e_{k}\rangle=\sum_{k}q_{k}\exp(-t*E_{k})|e_{k}\rangle.\label{imaginaryevolution}
	\end{equation}
	After the ITE, the amplitude $q_{k}$ is changed to be $q_{k}\exp(-t*E_{k})$.
	We can see that the decay of the amplitude is strongly (exponentially) dependent on the energy: the higher energy, the faster decay of the amplitude.
	In the previous works \cite{jsxu2016,jsxu2018}, we choose a long evolution time $t$ leading that only the ground states (with lowest energy), represented as $\ket{g}$, can be survived during the evolution and the excited states (with high energy), represented as $\ket{e}$, are dissipative in the evolution.
	
	Here, we design a new circuit to avoid the dissipation of the excited states with transferring the excited states to ground states, as shown in Fig. 2\textbf{a} in the main text. The circuit is composed of several gates: (a) two Hadamard gates
	\begin{equation}
		H=\left(
		\begin{array}{cc}
			1 & 1\\
			1 & -1
		\end{array}
		\right)/\sqrt{2},
	\end{equation} 
	applying to the two-level ancilla qubit $\ket{0}$ and $\ket{1}$ at the initial step and the final step of the circuit; (b) a local phase gate,  
	\begin{equation}
		R(\alpha)=\left(
		\begin{array}{cc}
			1 & 0\\
			0 & -ie^{i\alpha}
		\end{array}
		\right),
	\end{equation}
	where the parameter $\alpha$ is chosen to optimize the efficiency of the algorithm; (c) a two-qubit control unitary operation,
	\begin{equation}
		I\otimes\ket{0}\bra{0}+U\otimes\ket{1}\bra{1},
	\end{equation}
	where $I$ represent identity operator, and 
	\begin{equation}
		U(t)=e^{-iH_{s}t},
	\end{equation} 
	is the real time evolution operator for the system. $H_{s}$
	is the Hamiltonian of the considered
	system ($L$-site KCM here), and $t$ is the time of evolution which is another parameter we can
	use to optimize the efficiency of the algorithm. For a many-body Hamiltonian, we can well
	approximate the unitary evolution operator $U(t)$ by the product of a set of local unitary
	operators through the Trotter-Suzuki expansion \cite{1976Generalized}.
	
	After the unitary operation $U$,  the system state could be written as (omit the normalization)
	\begin{equation}
		a\ket{0}\otimes\exp(-t*E_{0})\ket{g}+b\ket{1}\otimes\exp(-t*E_{1})\ket{e},
	\end{equation}
	The following newly introduced control-not gate reserve the ground state and cool the excited state back to the ground state without dissipation (where $t$ is taken to be short enough and approaching zero).
	\begin{equation}
		(a\ket{0}+b\ket{1})*\ket{g}.
	\end{equation}

	\section{The case of Jones polynomials with topological protection for $\sigma_{1}$}
	
	\begin{figure}[htbp]
		\centering
		\includegraphics[width=1\columnwidth]{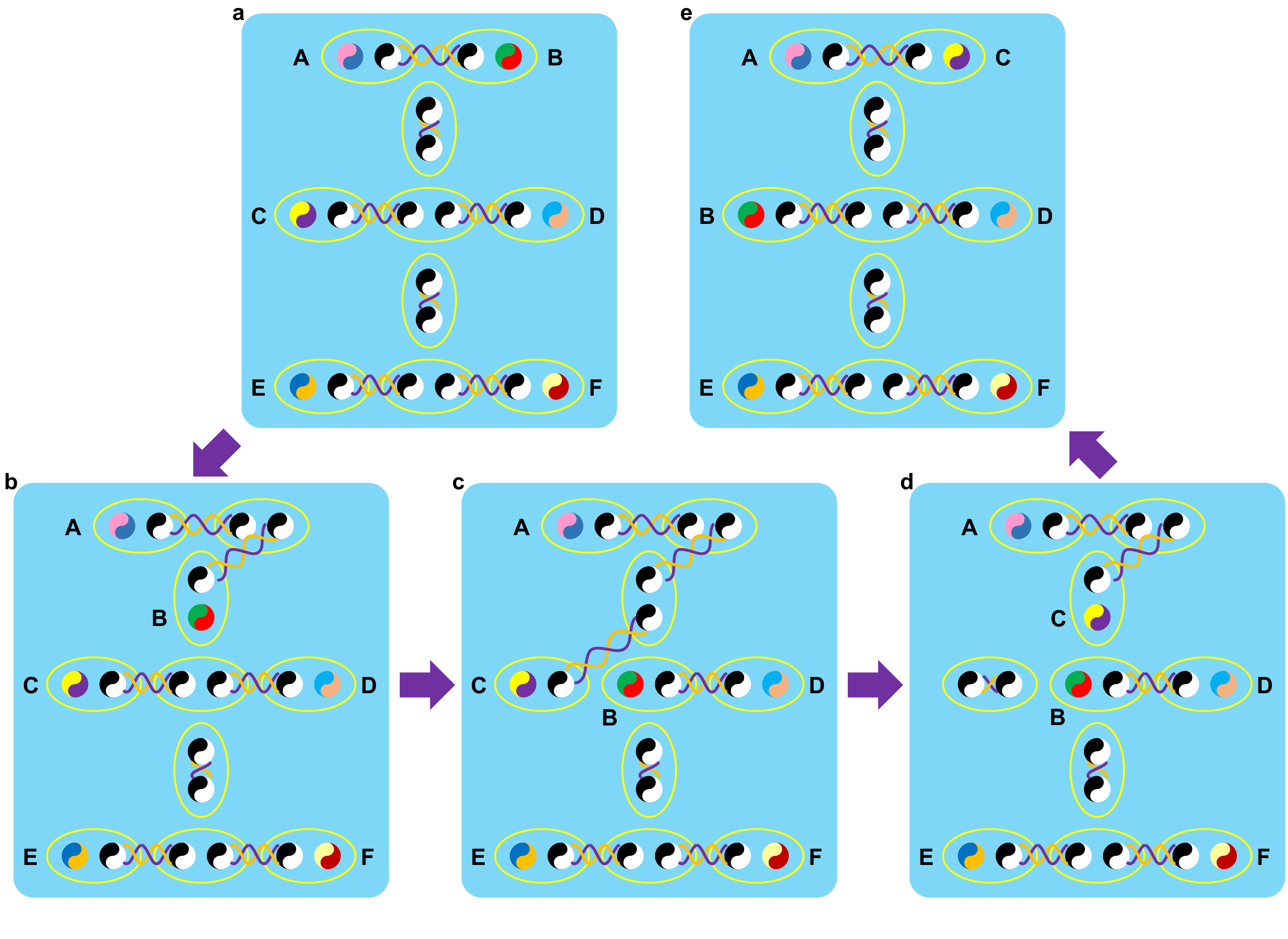}
		\small{\begin{flushleft}Fig. S2. The process of clockwise Majorana zero modes (MZMs) B and C. The Kitaev chain consists of ten fermions with six MZMs from A to F. Each dual-color circle represents a Majorana fermion, and a pair of Majorana fermions in the yellow circles constitutes a normal fermion. The helix lines between different Majorana fermions represent the interactions $i\gamma_{k}\gamma_{l}$ between them ($k$ and $l$= 1a, 1b, ..., 10b). $\textbf{a}, \textbf{b}, \textbf{c}, \textbf{d}$ and $ \textbf{e}$ corresponds  to the Hamiltonians $H_{0}$, $H_{1}$, $H_{2}$, $H_{3}$ and $H_{0}$, respectively.\end{flushleft}}	
		
	\end{figure}
	
	To implement the $\sigma_{1}$ of the braid group, we clockwise exchange Majorana zero modes (MZMs) B and C, as shown in Fig. S2.
	The exchange process is controlled by ten Hamiltonians
	\begin{equation}
		\begin{split}
			&H_{M_0}=i(\gamma_{1b} \gamma_{2a}+\gamma_{4b} \gamma_{5a}+\gamma_{5b} \gamma_{6a}+\gamma_{8b} \gamma_{9a}+\gamma_{9b}\gamma_{10a}+\gamma_{3a}\gamma_{3b}+\gamma_{7a}\gamma_{7b}), \\
			&H_{M_1}=i(\gamma_{1b} \gamma_{2a}+\gamma_{2b} \gamma_{3a}+\gamma_{4b} \gamma_{5a}+\gamma_{5b} \gamma_{6a}+\gamma_{8b}\gamma_{9a}+\gamma_{9b}\gamma_{10a}+\gamma_{7a}\gamma_{7b}), \\
			&H_{M_2}=i(\gamma_{1b} \gamma_{2a}+\gamma_{2b} \gamma_{3a}+\gamma_{3b} \gamma_{4b}+\gamma_{5b} \gamma_{6a}+\gamma_{8b}\gamma_{9a}+\gamma_{9b}\gamma_{10a}+\gamma_{7a}\gamma_{7b}), \\
			&H_{M_3}=i(\gamma_{1b} \gamma_{2a}+\gamma_{2b} \gamma_{3a}+\gamma_{5b} \gamma_{6a}+\gamma_{8b} \gamma_{9a}+\gamma_{9b}\gamma_{10a}+\gamma_{4a}\gamma_{4b}+\gamma_{7a}\gamma_{7b}).
		\end{split}
	\end{equation}
	The consecutive Hamiltonians are adiabatically connected and $H_{M_3}$ is finally adiabatically evolved to the initial Hamiltonian $H_{M_0}$. The adiabatic transport of the MZMs is implemented with our photonic simulator by imaginary-time evolution operators~\cite{jsxu2016}.
	
	Under the Jordan-Wigner (JW) transformation, the fermionic Hamiltonians $H_{M_0}$, $H_{M_1}$, $H_{M_2}$ and $H_{M_3}$ can be transformed to the corresponding spin Hamiltonians $H_{0}$, $H_{1}$, $H_{2}$ and $H_{3}$, respectively, where
	\begin{equation}
		\begin{split}
			&H_{0} =-\sigma_{1}^{x}\sigma_{2}^{x}-\sigma_{4}^{x}\sigma_{5}^{x}-\sigma_{5}^{x}\sigma_{6}^{x} -\sigma_{8}^{x}\sigma_{9}^{x}-\sigma_{9}^{x}\sigma_{10}^{x}+\sigma_{3}^{z}+\sigma_{7}^{z}, \\
			&H_{1}=-\sigma_{1}^{x}\sigma_{2}^{x}-\sigma_{2}^{x}\sigma_{3}^{x}-\sigma_{4}^{x}\sigma_{5}^{x} -\sigma_{5}^{x}\sigma_{6}^{x}-\sigma_{8}^{x}\sigma_{9}^{x}-\sigma_{9}^{x}\sigma_{10}^{x}+\sigma_{7}^{z}, \\
			&H_{2}=-\sigma_{1}^{x}\sigma_{2}^{x}-\sigma_{2}^{x}\sigma_{3}^{x}+\sigma_{3}^{x}\sigma_{4}^{y} -\sigma_{5}^{x}\sigma_{6}^{x}-\sigma_{8}^{x}\sigma_{9}^{x}-\sigma_{9}^{x}\sigma_{10}^{x}+\sigma_{7}^{z}, \\
			&H_{3}=-\sigma_{1}^{x}\sigma_{2}^{x}-\sigma_{2}^{x}\sigma_{3}^{x}-\sigma_{5}^{x}\sigma_{6}^{x} -\sigma_{8}^{x}\sigma_{9}^{x}-\sigma_{9}^{x}\sigma_{10}^{x}+\sigma_{4}^{z}+\sigma_{7}^{z}.
		\end{split}
	\end{equation}
	The JW transformation between the fermionic and spin Hamiltonians preserves their spectrum.
	So the adiabatic transport of the fermions system can be studied in the corresponding spin description.
	We can further simplify the process of ITE as many of the terms in $H_{0}$, $H_{1}$, $H_{2}$ and $H_{3}$ commute with each other.
	The ground states of the corresponding Hamiltonians can be expressed in terms of the eigenvectors $\{|x\rangle,|\bar{x}\rangle\}$, $\{|y\rangle,|\bar{y}\rangle\}$ and $\{|z\rangle,|\bar{z}\rangle\}$ of the corresponding Pauli operators $\sigma^x$, $\sigma^y$ and $\sigma^z$, with eigenvalues $\{1,-1\}$, respectively.
	The detailed evolution in the spin system during the exchange is calculated as follows.
	
	After the ITE of the initial Hamiltonian
	\begin{equation}
		H_{0} =-\sigma_{1}^{x}\sigma_{2}^{x}-\sigma_{4}^{x}\sigma_{5}^{x}-\sigma_{5}^{x}\sigma_{6}^{x} -\sigma_{8}^{x}\sigma_{9}^{x}-\sigma_{9}^{x}\sigma_{10}^{x}+\sigma_{3}^{z}+\sigma_{7}^{z},
	\end{equation}
	which corresponds to creating the MZMs A, B, C, D, E and F, the state becomes
	\begin{equation}
		\begin{split}
			|\phi_{0}\rangle &=\alpha|x_{1}x_{2}\bar{z}_{3}x_{4}x_{5}x_{6}\bar{z}_{7}x_{8}x_{9}x_{10}\rangle
			+\beta|x_{1}x_{2}\bar{z}_{3}x_{4}x_{5}x_{6}\bar{z}_{7}\bar{x}_{8}\bar{x}_{9}\bar{x}_{10}\rangle \\
			&+\gamma|x_{1}x_{2}\bar{z}_{3}\bar{x}_{4}\bar{x}_{5}\bar{x}_{6}\bar{z}_{7}x_{8}x_{9}x_{10}\rangle
			+\delta|x_{1}x_{2}\bar{z}_{3}\bar{x}_{4}\bar{x}_{5}\bar{x}_{6}\bar{z}_{7}\bar{x}_{8}\bar{x}_{9}\bar{x}_{10}\rangle \\
			&+\eta|\bar{x}_{1}\bar{x}_{2}\bar{z}_{3}x_{4}x_{5}x_{6}\bar{z}_{7}x_{8}x_{9}x_{10}\rangle
			+\kappa|\bar{x}_{1}\bar{x}_{2}\bar{z}_{3}x_{4}x_{5}x_{6}\bar{z}_{7}\bar{x}_{8}\bar{x}_{9}\bar{x}_{10}\rangle \\
			&+\mu|\bar{x}_{1}\bar{x}_{2}\bar{z}_{3}\bar{x}_{4}\bar{x}_{5}\bar{x}_{6}\bar{z}_{7}x_{8}x_{9}x_{10}\rangle
			+\nu|\bar{x}_{1}\bar{x}_{2}\bar{z}_{3}\bar{x}_{4}\bar{x}_{5}\bar{x}_{6}\bar{z}_{7}\bar{x}_{8}\bar{x}_{9}\bar{x}_{10}\rangle,
			\label{initial}
		\end{split}
	\end{equation}
	where $\alpha$, $\beta$, $\gamma$, $\delta$, $\eta$, $\kappa$, $\mu$ and $\nu$ are complex amplitudes satisfying $|\alpha|^{2}+|\beta|^{2}+|\gamma|^{2}+|\delta|^{2}+|\eta|^{2}+|\kappa|^{2}+|\mu|^{2}+|\nu|^{2}=1$.

	For the ITE of $H_{1}$, we first transfer the eigenstates of $\sigma_{3}^{z}$ to that of $\sigma_{3}^{x}$, and the initial state becomes
	\begin{equation}
		\begin{split}
			|\phi_{0\rightarrow 1}\rangle &=\alpha|x_{1}x_{2}(x_{3}-\bar{x}_{3})x_{4}x_{5}x_{6}\bar{z}_{7}x_{8}x_{9}x_{10}\rangle
			+\beta|x_{1}x_{2}(x_{3}-\bar{x}_{3})x_{4}x_{5}x_{6}\bar{z}_{7}\bar{x}_{8}\bar{x}_{9}\bar{x}_{10}\rangle \\
			&+\gamma|x_{1}x_{2}(x_{3}-\bar{x}_{3})\bar{x}_{4}\bar{x}_{5}\bar{x}_{6}\bar{z}_{7}x_{8}x_{9}x_{10}\rangle
			+\delta|x_{1}x_{2}(x_{3}-\bar{x}_{3})\bar{x}_{4}\bar{x}_{5}\bar{x}_{6}\bar{z}_{7}\bar{x}_{8}\bar{x}_{9}\bar{x}_{10}\rangle \\
			&+\eta|\bar{x}_{1}\bar{x}_{2}(x_{3}-\bar{x}_{3})x_{4}x_{5}x_{6}\bar{z}_{7}x_{8}x_{9}x_{10}\rangle
			+\kappa|\bar{x}_{1}\bar{x}_{2}(x_{3}-\bar{x}_{3})x_{4}x_{5}x_{6}\bar{z}_{7}\bar{x}_{8}\bar{x}_{9}\bar{x}_{10}\rangle \\
			&+\mu|\bar{x}_{1}\bar{x}_{2}(x_{3}-\bar{x}_{3})\bar{x}_{4}\bar{x}_{5}\bar{x}_{6}\bar{z}_{7}x_{8}x_{9}x_{10}\rangle
			+\nu|\bar{x}_{1}\bar{x}_{2}(x_{3}-\bar{x}_{3})\bar{x}_{4}\bar{x}_{5}\bar{x}_{6}\bar{z}_{7}\bar{x}_{8}\bar{x}_{9}\bar{x}_{10}\rangle,
		\end{split}
	\end{equation}
	we then implement the ITE of $-\sigma_{2}^{x}\sigma_{3}^{x}$.
	The state becomes
	\begin{equation}
		\begin{split}
			|\phi_{1}\rangle
			&=\alpha|x_{1}x_{2}x_{3}x_{4}x_{5}x_{6}\bar{z}_{7}x_{8}x_{9}x_{10}\rangle
			+\beta|x_{1}x_{2}x_{3}x_{4}x_{5}x_{6}\bar{z}_{7}\bar{x}_{8}\bar{x}_{9}\bar{x}_{10}\rangle \\
			&+\gamma|x_{1}x_{2}x_{3}\bar{x}_{4}\bar{x}_{5}\bar{x}_{6}\bar{z}_{7}x_{8}x_{9}x_{10}\rangle
			+\delta|x_{1}x_{2}x_{3}\bar{x}_{4}\bar{x}_{5}\bar{x}_{6}\bar{z}_{7}\bar{x}_{8}\bar{x}_{9}\bar{x}_{10}\rangle \\
			&-\eta|\bar{x}_{1}\bar{x}_{2}\bar{x}_{3}x_{4}x_{5}x_{6}\bar{z}_{7}x_{8}x_{9}x_{10}\rangle
			-\kappa|\bar{x}_{1}\bar{x}_{2}\bar{x}_{3}x_{4}x_{5}x_{6}\bar{z}_{7}\bar{x}_{8}\bar{x}_{9}\bar{x}_{10}\rangle \\
			&-\mu|\bar{x}_{1}\bar{x}_{2}\bar{x}_{3}\bar{x}_{4}\bar{x}_{5}\bar{x}_{6}\bar{z}_{7}x_{8}x_{9}x_{10}\rangle
			-\nu|\bar{x}_{1}\bar{x}_{2}\bar{x}_{3}\bar{x}_{4}\bar{x}_{5}\bar{x}_{6}\bar{z}_{7}\bar{x}_{8}\bar{x}_{9}\bar{x}_{10}\rangle.
		\end{split}
	\end{equation}

	For the evolution of $H_{2}$, we need to transfer the eigenstates of $\sigma_{4}^{x}$ to that of $\sigma_{4}^{y}$, and the spatial states becomes 16 points in spatial modes which could be written as
	\begin{equation}
		\begin{split}
			|\phi_{1\rightarrow 2}\rangle
			&=\alpha|x_{1}x_{2}x_{3}(y_{4}+i\bar{y}_{4})x_{5}x_{6}\bar{z}_{7}x_{8}x_{9}x_{10}\rangle
			+\beta|x_{1}x_{2}x_{3}(y_{4}+i\bar{y}_{4})x_{5}x_{6}\bar{z}_{7}\bar{x}_{8}\bar{x}_{9}\bar{x}_{10}\rangle \\
			&+\gamma|x_{1}x_{2}x_{3}(iy_{4}+\bar{y}_{4})\bar{x}_{5}\bar{x}_{6}\bar{z}_{7}x_{8}x_{9}x_{10}\rangle
			+\delta|x_{1}x_{2}x_{3}(iy_{4}+\bar{y}_{4})\bar{x}_{5}\bar{x}_{6}\bar{z}_{7}\bar{x}_{8}\bar{x}_{9}\bar{x}_{10}\rangle \\
			&-\eta|\bar{x}_{1}\bar{x}_{2}\bar{x}_{3}(y_{4}+i\bar{y}_{4})x_{5}x_{6}\bar{z}_{7}x_{8}x_{9}x_{10}\rangle
			-\kappa|\bar{x}_{1}\bar{x}_{2}\bar{x}_{3}(y_{4}+i\bar{y}_{4})x_{5}x_{6}\bar{z}_{7}\bar{x}_{8}\bar{x}_{9}\bar{x}_{10}\rangle \\
			&-\mu|\bar{x}_{1}\bar{x}_{2}\bar{x}_{3}(iy_{4}+\bar{y}_{4})\bar{x}_{5}\bar{x}_{6}\bar{z}_{7}x_{8}x_{9}x_{10}\rangle
			-\nu|\bar{x}_{1}\bar{x}_{2}\bar{x}_{3}(iy_{4}+\bar{y}_{4})\bar{x}_{5}\bar{x}_{6}\bar{z}_{7}\bar{x}_{8}\bar{x}_{9}\bar{x}_{10}\rangle,
		\end{split}
	\end{equation}
	then we implement ITE of $\sigma_{3}^{x}\sigma_{4}^{y}$ leading to the state presented by 8 points in spatial modes,
	\begin{equation}
		\begin{split}
			|\phi_{2}\rangle
			&=i\alpha|x_{1}x_{2}x_{3}\bar{y}_{4}x_{5}x_{6}\bar{z}_{7}x_{8}x_{9}x_{10}\rangle
			+i\beta|x_{1}x_{2}x_{3}\bar{y}_{4}x_{5}x_{6}\bar{z}_{7}\bar{x}_{8}\bar{x}_{9}\bar{x}_{10}\rangle \\
			&+\gamma|x_{1}x_{2}x_{3}\bar{y}_{4}\bar{x}_{5}\bar{x}_{6}\bar{z}_{7}x_{8}x_{9}x_{10}\rangle
			+\delta|x_{1}x_{2}x_{3}\bar{y}_{4}\bar{x}_{5}\bar{x}_{6}\bar{z}_{7}\bar{x}_{8}\bar{x}_{9}\bar{x}_{10}\rangle \\
			&-\eta|\bar{x}_{1}\bar{x}_{2}\bar{x}_{3}y_{4}x_{5}x_{6}\bar{z}_{7}x_{8}x_{9}x_{10}\rangle
			-\kappa|\bar{x}_{1}\bar{x}_{2}\bar{x}_{3}y_{4}x_{5}x_{6}\bar{z}_{7}\bar{x}_{8}\bar{x}_{9}\bar{x}_{10}\rangle \\
			&-i\mu|\bar{x}_{1}\bar{x}_{2}\bar{x}_{3}y_{4}\bar{x}_{5}\bar{x}_{6}\bar{z}_{7}x_{8}x_{9}x_{10}\rangle
			-i\nu|\bar{x}_{1}\bar{x}_{2}\bar{x}_{3}y_{4}\bar{x}_{5}\bar{x}_{6}\bar{z}_{7}\bar{x}_{8}\bar{x}_{9}\bar{x}_{10}\rangle.
		\end{split}
	\end{equation}

	For the evolution of $H_{3}$, we need to transfer the eigenstates of $\sigma_{4}^{y}$ to that of $\sigma_{4}^{z}$ which are written as 16 points in spatial modes,
	\begin{equation}
		\begin{split}
			|\phi_{2\rightarrow 3}\rangle
			&=i\alpha|x_{1}x_{2}x_{3}(z_4-i\bar{z}_{4})x_{5}x_{6}\bar{z}_{7}x_{8}x_{9}x_{10}\rangle
			+i\beta|x_{1}x_{2}x_{3}(z_4-i\bar{z}_{4})x_{5}x_{6}\bar{z}_{7}\bar{x}_{8}\bar{x}_{9}\bar{x}_{10}\rangle \\
			&+\gamma|x_{1}x_{2}x_{3}(z_4-i\bar{z}_{4})\bar{x}_{5}\bar{x}_{6}\bar{z}_{7}x_{8}x_{9}x_{10}\rangle
			+\delta|x_{1}x_{2}x_{3}(z_4-i\bar{z}_{4})\bar{x}_{5}\bar{x}_{6}\bar{z}_{7}\bar{x}_{8}\bar{x}_{9}\bar{x}_{10}\rangle \\
			&-\eta|\bar{x}_{1}\bar{x}_{2}\bar{x}_{3}(z_4+i\bar{z}_{4})x_{5}x_{6}\bar{z}_{7}x_{8}x_{9}x_{10}\rangle
			-\kappa|\bar{x}_{1}\bar{x}_{2}\bar{x}_{3}(z_4+i\bar{z}_{4})x_{5}x_{6}\bar{z}_{7}\bar{x}_{8}\bar{x}_{9}\bar{x}_{10}\rangle \\
			&-i\mu|\bar{x}_{1}\bar{x}_{2}\bar{x}_{3}(z_4+i\bar{z}_{4})\bar{x}_{5}\bar{x}_{6}\bar{z}_{7}x_{8}x_{9}x_{10}\rangle
			-i\nu|\bar{x}_{1}\bar{x}_{2}\bar{x}_{3}(z_4+i\bar{z}_{4})\bar{x}_{5}\bar{x}_{6}\bar{z}_{7}\bar{x}_{8}\bar{x}_{9}\bar{x}_{10}\rangle,
		\end{split}
	\end{equation}
	Then we implement ITE of $\sigma_{4}^{z}$. The state becomes
	\begin{equation}
		\begin{split}
			|\phi_{3}\rangle &=\alpha|x_{1}x_{2}x_{3}\bar{z}_{4}x_{5}x_{6}\bar{z}_{7}x_{8}x_{9}x_{10}\rangle
			+\beta|x_{1}x_{2}x_{3}\bar{z}_{4}x_{5}x_{6}\bar{z}_{7}\bar{x}_{8}\bar{x}_{9}\bar{x}_{10}\rangle \\
			&-i\gamma|x_{1}x_{2}x_{3}\bar{z}_{4}\bar{x}_{5}\bar{x}_{6}\bar{z}_{7}x_{8}x_{9}x_{10}\rangle
			-i\delta|x_{1}x_{2}x_{3}\bar{z}_{4}\bar{x}_{5}\bar{x}_{6}\bar{z}_{7}\bar{x}_{8}\bar{x}_{9}\bar{x}_{10}\rangle \\
			&-i\eta|\bar{x}_{1}\bar{x}_{2}\bar{x}_{3}\bar{z}_{4}x_{5}x_{6}\bar{z}_{7}x_{8}x_{9}x_{10}\rangle
			-i\kappa|\bar{x}_{1}\bar{x}_{2}\bar{x}_{3}\bar{z}_{4}x_{5}x_{6}\bar{z}_{7}\bar{x}_{8}\bar{x}_{9}\bar{x}_{10}\rangle \\
			&+\mu|\bar{x}_{1}\bar{x}_{2}\bar{x}_{3}\bar{z}_{4}\bar{x}_{5}\bar{x}_{6}\bar{z}_{7}x_{8}x_{9}x_{10}\rangle
			+\nu|\bar{x}_{1}\bar{x}_{2}\bar{x}_{3}\bar{z}_{4}\bar{x}_{5}\bar{x}_{6}\bar{z}_{7}\bar{x}_{8}\bar{x}_{9}\bar{x}_{10}\rangle.
		\end{split}
	\end{equation}

	We now back to the initial Hamiltonian. We need to implement ITE of $\sigma_{3}^{z}$, and then $-\sigma_{4}^{x}\sigma_{5}^{x}$. For the $\sigma_{3}^{z}$, its eigenstates is transferred from that of $\sigma_{3}^{x}$, which could be written as
	\begin{equation}
		\begin{split}
			|\phi_{3\rightarrow 4}'\rangle &=\alpha|x_{1}x_{2}(z_{3}+\bar{z}_{3})\bar{z}_{4}x_{5}x_{6}\bar{z}_{7}x_{8}x_{9}x_{10}\rangle
			+\beta|x_{1}x_{2}(z_{3}+\bar{z}_{3})\bar{z}_{4}x_{5}x_{6}\bar{z}_{7}\bar{x}_{8}\bar{x}_{9}\bar{x}_{10}\rangle \\
			&-i\gamma|x_{1}x_{2}(z_{3}+\bar{z}_{3})\bar{z}_{4}\bar{x}_{5}\bar{x}_{6}\bar{z}_{7}x_{8}x_{9}x_{10}\rangle
			-i\delta|x_{1}x_{2}(z_{3}+\bar{z}_{3})\bar{z}_{4}\bar{x}_{5}\bar{x}_{6}\bar{z}_{7}\bar{x}_{8}\bar{x}_{9}\bar{x}_{10}\rangle \\
			&-i\eta|\bar{x}_{1}\bar{x}_{2}(z_{3}-\bar{z}_{3})\bar{z}_{4}x_{5}x_{6}\bar{z}_{7}x_{8}x_{9}x_{10}\rangle
			-i\kappa|\bar{x}_{1}\bar{x}_{2}(z_{3}-\bar{z}_{3})\bar{z}_{4}x_{5}x_{6}\bar{z}_{7}\bar{x}_{8}\bar{x}_{9}\bar{x}_{10}\rangle \\
			&+\mu|\bar{x}_{1}\bar{x}_{2}(z_{3}-\bar{z}_{3})\bar{z}_{4}\bar{x}_{5}\bar{x}_{6}\bar{z}_{7}x_{8}x_{9}x_{10}\rangle
			+\nu|\bar{x}_{1}\bar{x}_{2}(z_{3}-\bar{z}_{3})\bar{z}_{4}\bar{x}_{5}\bar{x}_{6}\bar{z}_{7}\bar{x}_{8}\bar{x}_{9}\bar{x}_{10}\rangle,
		\end{split}
	\end{equation}
	After the ITE of $\sigma_{3}^{z}$, the state becomes
	\begin{equation}
		\begin{split}
			|\phi_{4}'\rangle &=\alpha|x_{1}x_{2}\bar{z}_{3}\bar{z}_{4}x_{5}x_{6}\bar{z}_{7}x_{8}x_{9}x_{10}\rangle
			+\beta|x_{1}x_{2}\bar{z}_{3}\bar{z}_{4}x_{5}x_{6}\bar{z}_{7}\bar{x}_{8}\bar{x}_{9}\bar{x}_{10}\rangle \\
			&-i\gamma|x_{1}x_{2}\bar{z}_{3}\bar{z}_{4}\bar{x}_{5}\bar{x}_{6}\bar{z}_{7}x_{8}x_{9}x_{10}\rangle
			-i\delta|x_{1}x_{2}\bar{z}_{3}\bar{z}_{4}\bar{x}_{5}\bar{x}_{6}\bar{z}_{7}\bar{x}_{8}\bar{x}_{9}\bar{x}_{10}\rangle \\
			&+i\eta|\bar{x}_{1}\bar{x}_{2}\bar{z}_{3}\bar{z}_{4}x_{5}x_{6}\bar{z}_{7}x_{8}x_{9}x_{10}\rangle
			+i\kappa|\bar{x}_{1}\bar{x}_{2}\bar{z}_{3}\bar{z}_{4}x_{5}x_{6}\bar{z}_{7}\bar{x}_{8}\bar{x}_{9}\bar{x}_{10}\rangle \\
			&-\mu|\bar{x}_{1}\bar{x}_{2}\bar{z}_{3}\bar{z}_{4}\bar{x}_{5}\bar{x}_{6}\bar{z}_{7}x_{8}x_{9}x_{10}\rangle
			-\nu|\bar{x}_{1}\bar{x}_{2}\bar{z}_{3}\bar{z}_{4}\bar{x}_{5}\bar{x}_{6}\bar{z}_{7}\bar{x}_{8}\bar{x}_{9}\bar{x}_{10}\rangle.
		\end{split}
	\end{equation}
	For the operation $-\sigma_{4}^{x}\sigma_{5}^{x}$, we need to transfer the eigenstates of $\sigma_{4}^{z}$ to that of $\sigma_{4}^{x}$ as follows,
	\begin{equation}
		\begin{split}
			|\phi_{3\rightarrow4}''\rangle &=\alpha|x_{1}x_{2}\bar{z}_{3}(x_{4}-\bar{x}_{4})x_{5}x_{6}\bar{z}_{7}x_{8}x_{9}x_{10}\rangle
			+\beta|x_{1}x_{2}\bar{z}_{3}(x_{4}-\bar{x}_{4})x_{5}x_{6}\bar{z}_{7}\bar{x}_{8}\bar{x}_{9}\bar{x}_{10}\rangle \\
			&-i\gamma|x_{1}x_{2}\bar{z}_{3}(x_{4}-\bar{x}_{4})\bar{x}_{5}\bar{x}_{6}\bar{z}_{7}x_{8}x_{9}x_{10}\rangle
			-i\delta|x_{1}x_{2}\bar{z}_{3}(x_{4}-\bar{x}_{4})\bar{x}_{5}\bar{x}_{6}\bar{z}_{7}\bar{x}_{8}\bar{x}_{9}\bar{x}_{10}\rangle \\
			&+i\eta|\bar{x}_{1}\bar{x}_{2}\bar{z}_{3}(x_{4}-\bar{x}_{4})x_{5}x_{6}\bar{z}_{7}x_{8}x_{9}x_{10}\rangle
			+i\kappa|\bar{x}_{1}\bar{x}_{2}\bar{z}_{3}(x_{4}-\bar{x}_{4})x_{5}x_{6}\bar{z}_{7}\bar{x}_{8}\bar{x}_{9}\bar{x}_{10}\rangle \\
			&-\mu|\bar{x}_{1}\bar{x}_{2}\bar{z}_{3}(x_{4}-\bar{x}_{4})\bar{x}_{5}\bar{x}_{6}\bar{z}_{7}x_{8}x_{9}x_{10}\rangle
			-\nu|\bar{x}_{1}\bar{x}_{2}\bar{z}_{3}(x_{4}-\bar{x}_{4})\bar{x}_{5}\bar{x}_{6}\bar{z}_{7}\bar{x}_{8}\bar{x}_{9}\bar{x}_{10}\rangle,
		\end{split}
	\end{equation}
	After the ITE of $-\sigma_{4}^{x}\sigma_{5}^{x}$, the state becomes
	\begin{equation}
		\begin{split}
			|\phi_{4}\rangle &=\alpha|x_{1}x_{2}\bar{z}_{3}x_{4}x_{5}x_{6}\bar{z}_{7}x_{8}x_{9}x_{10}\rangle
			+\beta|x_{1}x_{2}\bar{z}_{3}x_{4}x_{5}x_{6}\bar{z}_{7}\bar{x}_{8}\bar{x}_{9}\bar{x}_{10}\rangle \\
			&+i\gamma|x_{1}x_{2}\bar{z}_{3}\bar{x}_{4}\bar{x}_{5}\bar{x}_{6}\bar{z}_{7}x_{8}x_{9}x_{10}\rangle
			+i\delta|x_{1}x_{2}\bar{z}_{3}\bar{x}_{4}\bar{x}_{5}\bar{x}_{6}\bar{z}_{7}\bar{x}_{8}\bar{x}_{9}\bar{x}_{10}\rangle \\
			&+i\eta|\bar{x}_{1}\bar{z}_{2}\bar{z}_{3}x_{4}x_{5}x_{6}\bar{z}_{7}x_{8}x_{9}x_{10}\rangle
			+i\kappa|\bar{x}_{1}\bar{x}_{2}\bar{z}_{3}x_{4}x_{5}x_{6}\bar{z}_{7}\bar{x}_{8}\bar{x}_{9}\bar{x}_{10}\rangle \\
			&+\mu|\bar{x}_{1}\bar{x}_{2}\bar{z}_{3}\bar{x}_{4}\bar{x}_{5}\bar{x}_{6}\bar{z}_{7}x_{8}x_{9}x_{10}\rangle
			+\nu|\bar{x}_{1}\bar{x}_{2}\bar{z}_{3}\bar{x}_{4}\bar{x}_{5}\bar{x}_{6}\bar{z}_{7}\bar{x}_{8}\bar{x}_{9}\bar{x}_{10}\rangle.
		\end{split}
	\end{equation}
	
	For the operation $\sigma_1$ of the braid group, the final output state is $\ket{\phi_f} = \ket{\phi_4}$ and the whole transformations are implemented with five Sagnac interferometers. Regardless of transformations of the basis of every Hamiltonian, the bases in the initial state and final state are the same, and the corresponding transformation of this clockwise braiding could be written as
	\begin{equation}
		U_{\sigma_1}=\left(
		\begin{array}{cccccccc}
			1\ & 0\ & 0\ & 0\ & 0\ & 0\ & 0\ & 0\\
			0 & 1 & 0 & 0 & 0 & 0 & 0 & 0\\
			0 & 0 & i & 0 & 0 & 0 & 0 & 0\\
			0 & 0 & 0 & i & 0 & 0 & 0 & 0\\
			0 & 0 & 0 & 0 & i & 0 & 0 & 0\\
			0 & 0 & 0 & 0 & 0 & i & 0 & 0\\
			0 & 0 & 0 & 0 & 0 & 0 & 1 & 0\\
			0 & 0 & 0 & 0 & 0 & 0 & 0 & 1\\
		\end{array}
		\right)/\sqrt{2},
	\end{equation}
	With its transformation matrix in the logical basis
	\begin{equation}
		L_{\sigma_1}=\left(
		\begin{array}{cccccccc}
			1\ & 0\ & 0\ & 0\ & 0\ & 0\ & i\ & 0\\
			0 & 1 & 0 & 0 & 0 & 0 & 0 & i\\
			0 & 0 & 1 & 0 & i & 0 & 0 & 0\\
			0 & 0 & 0 & 1 & 0 & i & 0 & 0\\
			0 & 0 & i & 0 & 1 & 0 & 0 & 0\\
			0 & 0 & 0 & i & 0 & 1 & 0 & 0\\
			i & 0 & 0 & 0 & 0 & 0 & 1 & 0\\
			0 & i & 0 & 0 & 0 & 0 & 0 & 1\\
		\end{array}
		\right)/\sqrt{2},
	\end{equation}
	Projected to the initial state
	\begin{equation}
		P=|\langle\phi_{0}|\phi_{4}\rangle|^{2} =||\alpha|^2+|\beta|^2+i|\gamma|^2+i|\delta|^2+i|\eta|^2+i|\kappa|^2+|\mu|^2+|\nu|^2|^2.
	\end{equation}
	
	After the second $\sigma_{1}$ operation ($\sigma_{1}^{2}$), the state would become
	\begin{equation}
		\begin{split}
			|\phi_{4}\rangle &=\alpha|x_{1}x_{2}\bar{z}_{3}x_{4}x_{5}x_{6}\bar{z}_{7}x_{8}x_{9}x_{10}\rangle
			+\beta|x_{1}x_{2}\bar{z}_{3}x_{4}x_{5}x_{6}\bar{z}_{7}\bar{x}_{8}\bar{x}_{9}\bar{x}_{10}\rangle \\
			&-\gamma|x_{1}x_{2}\bar{z}_{3}\bar{x}_{4}\bar{x}_{5}\bar{x}_{6}\bar{z}_{7}x_{8}x_{9}x_{10}\rangle
			-\delta|x_{1}x_{2}\bar{z}_{3}\bar{x}_{4}\bar{x}_{5}\bar{x}_{6}\bar{z}_{7}\bar{x}_{8}\bar{x}_{9}\bar{x}_{10}\rangle \\
			&-\eta|\bar{x}_{1}\bar{z}_{2}\bar{z}_{3}x_{4}x_{5}x_{6}\bar{z}_{7}x_{8}x_{9}x_{10}\rangle
			-\kappa|\bar{x}_{1}\bar{x}_{2}\bar{z}_{3}x_{4}x_{5}x_{6}\bar{z}_{7}\bar{x}_{8}\bar{x}_{9}\bar{x}_{10}\rangle \\
			&+\mu|\bar{x}_{1}\bar{x}_{2}\bar{z}_{3}\bar{x}_{4}\bar{x}_{5}\bar{x}_{6}\bar{z}_{7}x_{8}x_{9}x_{10}\rangle
			+\nu|\bar{x}_{1}\bar{x}_{2}\bar{z}_{3}\bar{x}_{4}\bar{x}_{5}\bar{x}_{6}\bar{z}_{7}\bar{x}_{8}\bar{x}_{9}\bar{x}_{10}\rangle.
		\end{split}
	\end{equation}
	
	Projected to the initial state
	\begin{equation}
		P=|\langle\phi_{0}|\phi_{4}\rangle|^{2} =||\alpha|^2+|\beta|^2-|\gamma|^2-|\delta|^2-|\eta|^2-|\kappa|^2+|\mu|^2+|\nu|^2|^2.
	\end{equation}
	
	After the third $\sigma_{1}$ operation ($\sigma_{1}^{3}$), the state would become
	\begin{equation}
		\begin{split}
			|\phi_{4}\rangle &=\alpha|x_{1}x_{2}\bar{z}_{3}x_{4}x_{5}x_{6}\bar{z}_{7}x_{8}x_{9}x_{10}\rangle
			+\beta|x_{1}x_{2}\bar{z}_{3}x_{4}x_{5}x_{6}\bar{z}_{7}\bar{x}_{8}\bar{x}_{9}\bar{x}_{10}\rangle \\
			&-i\gamma|x_{1}x_{2}\bar{z}_{3}\bar{x}_{4}\bar{x}_{5}\bar{x}_{6}\bar{z}_{7}x_{8}x_{9}x_{10}\rangle
			-i\delta|x_{1}x_{2}\bar{z}_{3}\bar{x}_{4}\bar{x}_{5}\bar{x}_{6}\bar{z}_{7}\bar{x}_{8}\bar{x}_{9}\bar{x}_{10}\rangle \\
			&-i\eta|\bar{x}_{1}\bar{z}_{2}\bar{z}_{3}x_{4}x_{5}x_{6}\bar{z}_{7}x_{8}x_{9}x_{10}\rangle
			-i\kappa|\bar{x}_{1}\bar{x}_{2}\bar{z}_{3}x_{4}x_{5}x_{6}\bar{z}_{7}\bar{x}_{8}\bar{x}_{9}\bar{x}_{10}\rangle \\
			&+\mu|\bar{x}_{1}\bar{x}_{2}\bar{z}_{3}\bar{x}_{4}\bar{x}_{5}\bar{x}_{6}\bar{z}_{7}x_{8}x_{9}x_{10}\rangle
			+\nu|\bar{x}_{1}\bar{x}_{2}\bar{z}_{3}\bar{x}_{4}\bar{x}_{5}\bar{x}_{6}\bar{z}_{7}\bar{x}_{8}\bar{x}_{9}\bar{x}_{10}\rangle.
		\end{split}
	\end{equation}
	
	Projected to the initial state
	\begin{equation}
		P=|\langle\phi_{0}|\phi_{4}\rangle|^{2} =||\alpha|^2+|\beta|^2-i|\gamma|^2-i|\delta|^2-i|\eta|^2-i|\kappa|^2+|\mu|^2+|\nu|^2|^2.
	\end{equation}
	
	\begin{figure}[htbp]
		\centering
		\includegraphics[width=1\columnwidth]{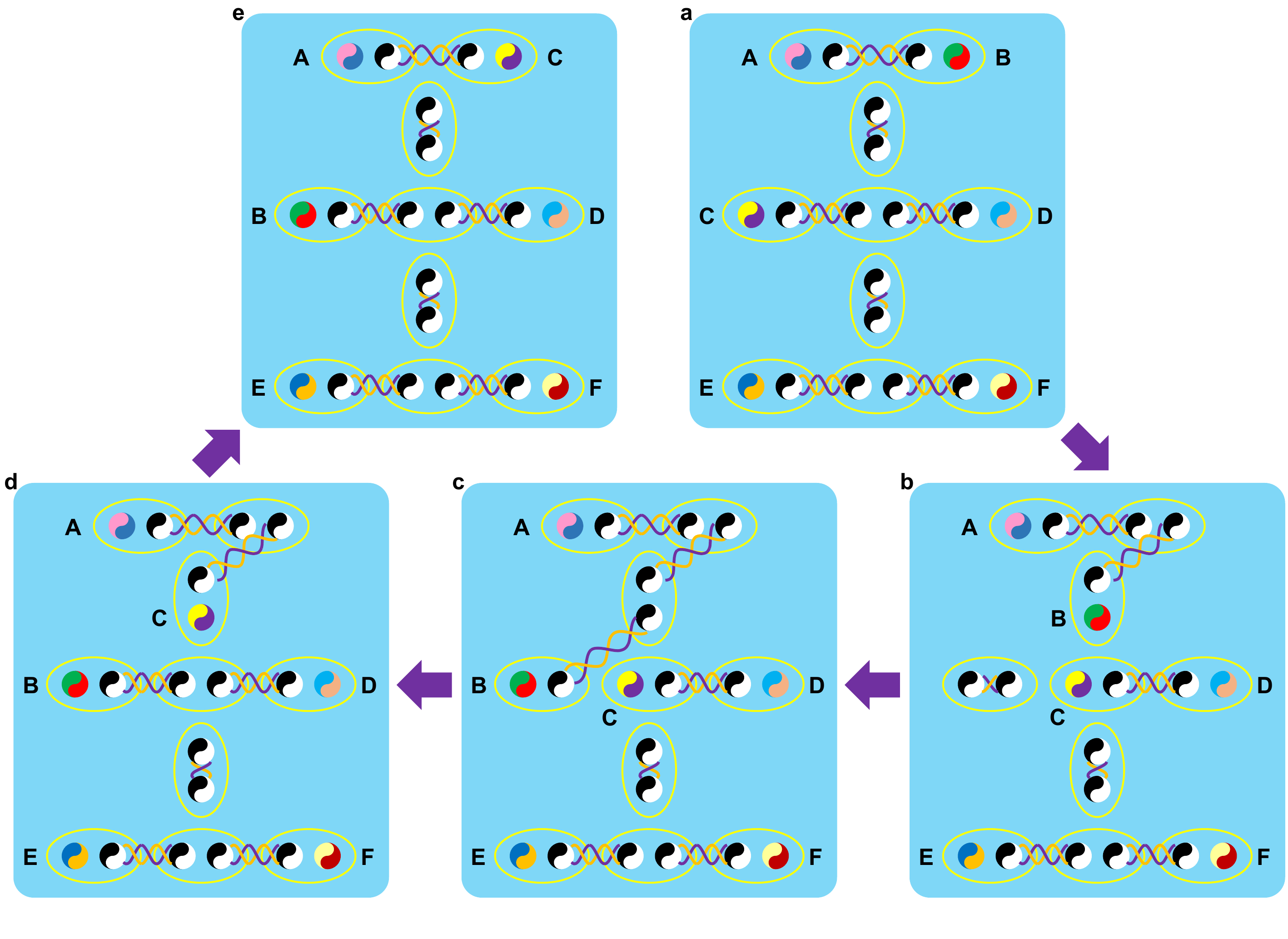}
		\small{\begin{flushleft}Fig. S3. The process of anticlockwise Majorana zero modes (MZMs) B and C. The Kitaev chain consists of ten fermions with six MZMs from A to F. Each dual-color circle represents a Majorana fermion, and a pair of Majorana fermions in the yellow circles constitutes a normal fermion. The helix lines between different Majorana fermions represent the interactions $i\gamma_{k}\gamma_{l}$ between them ($k$ and $l$= 1a, 1b, ..., 10b). $\textbf{a}, \textbf{b}, \textbf{c}, \textbf{d}$ and $ \textbf{e}$ corresponds  to the Hamiltonians $H_{0}$, $H_{3}$, $H_{2}$, $H_{1}$ and $H_{0}$, respectively.\end{flushleft}}	
		
	\end{figure}	
	
	While to implement the $\sigma_{1}^{-1}$ of the braid group, we anticlockwise exchange Majorana B and C, which means the exchange process follows $H_{0}\rightarrow H_{3}\rightarrow H_{2}\rightarrow H_{1}\rightarrow H_{0}$ as shown in Fig. S3. Still beginning from the initial state in (19), the corresponding final states obtained via ITE of every Hamiltonian are listed below.
	
	The initial state after the ITE of $H_{0}$ is
	\begin{equation}
		\begin{split}
			|\psi_{0}\rangle=|\phi_{0}\rangle &= \alpha|x_{1}x_{2}\bar{z}_{3}x_{4}x_{5}x_{6}\bar{z}_{7}x_{8}x_{9}x_{10}\rangle
			+\beta|x_{1}x_{2}\bar{z}_{3}x_{4}x_{5}x_{6}\bar{z}_{7}\bar{x}_{8}\bar{x}_{9}\bar{x}_{10}\rangle \\
			&+\gamma|x_{1}x_{2}\bar{z}_{3}\bar{x}_{4}\bar{x}_{5}\bar{x}_{6}\bar{z}_{7}x_{8}x_{9}x_{10}\rangle
			+\delta|x_{1}x_{2}\bar{z}_{3}\bar{x}_{4}\bar{x}_{5}\bar{x}_{6}\bar{z}_{7}\bar{x}_{8}\bar{x}_{9}\bar{x}_{10}\rangle \\
			&+\eta|\bar{x}_{1}\bar{x}_{2}\bar{z}_{3}x_{4}x_{5}x_{6}\bar{z}_{7}x_{8}x_{9}x_{10}\rangle
			+\kappa|\bar{x}_{1}\bar{x}_{2}\bar{z}_{3}x_{4}x_{5}x_{6}\bar{z}_{7}\bar{x}_{8}\bar{x}_{9}\bar{x}_{10}\rangle \\
			&+\mu|\bar{x}_{1}\bar{x}_{2}\bar{z}_{3}\bar{x}_{4}\bar{x}_{5}\bar{x}_{6}\bar{z}_{7}x_{8}x_{9}x_{10}\rangle
			+\nu|\bar{x}_{1}\bar{x}_{2}\bar{z}_{3}\bar{x}_{4}\bar{x}_{5}\bar{x}_{6}\bar{z}_{7}\bar{x}_{8}\bar{x}_{9}\bar{x}_{10}\rangle.
		\end{split}
	\end{equation}
	
	After ITE of Hamiltonian $H_{3}$, the state is written as
	\begin{equation}
		\begin{split}
			|\psi_{1}\rangle &=\alpha|x_{1}x_{2}x_{3}\bar{z}_{4}x_{5}x_{6}\bar{z}_{7}x_{8}x_{9}x_{10}\rangle
			+\beta|x_{1}x_{2}x_{3}\bar{z}_{4}x_{5}x_{6}\bar{z}_{7}\bar{x}_{8}\bar{x}_{9}\bar{x}_{10}\rangle \\
			&-\gamma|x_{1}x_{2}x_{3}\bar{z}_{4}\bar{x}_{5}\bar{x}_{6}\bar{z}_{7}x_{8}x_{9}x_{10}\rangle
			-\delta|x_{1}x_{2}x_{3}\bar{z}_{4}\bar{x}_{5}\bar{x}_{6}\bar{z}_{7}\bar{x}_{8}\bar{x}_{9}\bar{x}_{10}\rangle \\
			&-\eta|\bar{x}_{1}\bar{x}_{2}\bar{x}_{3}\bar{z}_{4}x_{5}x_{6}\bar{z}_{7}x_{8}x_{9}x_{10}\rangle
			-\kappa|\bar{x}_{1}\bar{x}_{2}\bar{x}_{3}\bar{z}_{4}x_{5}x_{6}\bar{z}_{7}\bar{x}_{8}\bar{x}_{9}\bar{x}_{10}\rangle \\
			&+\mu|\bar{x}_{1}\bar{x}_{2}\bar{x}_{3}\bar{z}_{4}\bar{x}_{5}\bar{x}_{6}\bar{z}_{7}x_{8}x_{9}x_{10}\rangle
			+\nu|\bar{x}_{1}\bar{x}_{2}\bar{x}_{3}\bar{z}_{4}\bar{x}_{5}\bar{x}_{6}\bar{z}_{7}\bar{x}_{8}\bar{x}_{9}\bar{x}_{10}\rangle.
		\end{split}
	\end{equation}
	
	After ITE of Hamiltonian $H_{2}$, the state is written as
	\begin{equation}
		\begin{split}
			|\psi_{2}\rangle
			&=\alpha|x_{1}x_{2}x_{3}\bar{y}_{4}x_{5}x_{6}\bar{z}_{7}x_{8}x_{9}x_{10}\rangle
			+\beta|x_{1}x_{2}x_{3}\bar{y}_{4}x_{5}x_{6}\bar{z}_{7}\bar{x}_{8}\bar{x}_{9}\bar{x}_{10}\rangle \\
			&-\gamma|x_{1}x_{2}x_{3}\bar{y}_{4}\bar{x}_{5}\bar{x}_{6}\bar{z}_{7}x_{8}x_{9}x_{10}\rangle
			-\delta|x_{1}x_{2}x_{3}\bar{y}_{4}\bar{x}_{5}\bar{x}_{6}\bar{z}_{7}\bar{x}_{8}\bar{x}_{9}\bar{x}_{10}\rangle \\
			&+\eta|\bar{x}_{1}\bar{x}_{2}\bar{x}_{3}y_{4}x_{5}x_{6}\bar{z}_{7}x_{8}x_{9}x_{10}\rangle
			+\kappa|\bar{x}_{1}\bar{x}_{2}\bar{x}_{3}y_{4}x_{5}x_{6}\bar{z}_{7}\bar{x}_{8}\bar{x}_{9}\bar{x}_{10}\rangle \\
			&-\mu|\bar{x}_{1}\bar{x}_{2}\bar{x}_{3}y_{4}\bar{x}_{5}\bar{x}_{6}\bar{z}_{7}x_{8}x_{9}x_{10}\rangle
			-\nu|\bar{x}_{1}\bar{x}_{2}\bar{x}_{3}y_{4}\bar{x}_{5}\bar{x}_{6}\bar{z}_{7}\bar{x}_{8}\bar{x}_{9}\bar{x}_{10}\rangle.
		\end{split}
	\end{equation}
	
	After ITE of Hamiltonian $H_{1}$, the state is written as
	\begin{equation}
		\begin{split}
			|\psi_{3}\rangle
			&=\alpha|x_{1}x_{2}x_{3}x_{4}x_{5}x_{6}\bar{z}_{7}x_{8}x_{9}x_{10}\rangle
			+\beta|x_{1}x_{2}x_{3}x_{4}x_{5}x_{6}\bar{z}_{7}\bar{x}_{8}\bar{x}_{9}\bar{x}_{10}\rangle \\
			&-i\gamma|x_{1}x_{2}x_{3}\bar{x}_{4}\bar{x}_{5}\bar{x}_{6}\bar{z}_{7}x_{8}x_{9}x_{10}\rangle
			-i\delta|x_{1}x_{2}x_{3}\bar{x}_{4}\bar{x}_{5}\bar{x}_{6}\bar{z}_{7}\bar{x}_{8}\bar{x}_{9}\bar{x}_{10}\rangle \\
			&+i\eta|\bar{x}_{1}\bar{x}_{2}\bar{x}_{3}x_{4}x_{5}x_{6}\bar{z}_{7}x_{8}x_{9}x_{10}\rangle
			+i\kappa|\bar{x}_{1}\bar{x}_{2}\bar{x}_{3}x_{4}x_{5}x_{6}\bar{z}_{7}\bar{x}_{8}\bar{x}_{9}\bar{x}_{10}\rangle \\
			&-\mu|\bar{x}_{1}\bar{x}_{2}\bar{x}_{3}\bar{x}_{4}\bar{x}_{5}\bar{x}_{6}\bar{z}_{7}x_{8}x_{9}x_{10}\rangle
			-\nu|\bar{x}_{1}\bar{x}_{2}\bar{x}_{3}\bar{x}_{4}\bar{x}_{5}\bar{x}_{6}\bar{z}_{7}\bar{x}_{8}\bar{x}_{9}\bar{x}_{10}\rangle.
		\end{split}
	\end{equation}
	
	After ITE of Hamiltonian $H_{0}$, the state is written as
	\begin{equation}
		\begin{split}
			|\psi_{4}\rangle &=\alpha|x_{1}x_{2}\bar{z}_{3}x_{4}x_{5}x_{6}\bar{z}_{7}x_{8}x_{9}x_{10}\rangle
			+\beta|x_{1}x_{2}\bar{z}_{3}x_{4}x_{5}x_{6}\bar{z}_{7}\bar{x}_{8}\bar{x}_{9}\bar{x}_{10}\rangle \\
			&-i\gamma|x_{1}x_{2}\bar{z}_{3}\bar{x}_{4}\bar{x}_{5}\bar{x}_{6}\bar{z}_{7}x_{8}x_{9}x_{10}\rangle
			-i\delta|x_{1}x_{2}\bar{z}_{3}\bar{x}_{4}\bar{x}_{5}\bar{x}_{6}\bar{z}_{7}\bar{x}_{8}\bar{x}_{9}\bar{x}_{10}\rangle \\
			&-i\eta|\bar{x}_{1}\bar{x}_{2}\bar{z}_{3}x_{4}x_{5}x_{6}\bar{z}_{7}x_{8}x_{9}x_{10}\rangle
			-i\kappa|\bar{x}_{1}\bar{x}_{2}\bar{z}_{3}x_{4}x_{5}x_{6}\bar{z}_{7}\bar{x}_{8}\bar{x}_{9}\bar{x}_{10}\rangle \\
			&+\mu|\bar{x}_{1}\bar{x}_{2}\bar{z}_{3}\bar{x}_{4}\bar{x}_{5}\bar{x}_{6}\bar{z}_{7}x_{8}x_{9}x_{10}\rangle
			+\nu|\bar{x}_{1}\bar{x}_{2}\bar{z}_{3}\bar{x}_{4}\bar{x}_{5}\bar{x}_{6}\bar{z}_{7}\bar{x}_{8}\bar{x}_{9}\bar{x}_{10}\rangle.
		\end{split}
	\end{equation}
	
	the corresponding transformation of this clockwise braiding could be written as
	\begin{equation}
		U_{\sigma_1^{-1}}=\left(
		\begin{array}{cccccccc}
			1\ & 0\ & 0\ & 0\ & 0\ & 0\ & 0\ & 0\\
			0 & 1 & 0 & 0 & 0 & 0 & 0 & 0\\
			0 & 0 & -i & 0 & 0 & 0 & 0 & 0\\
			0 & 0 & 0 & -i & 0 & 0 & 0 & 0\\
			0 & 0 & 0 & 0 & -i & 0 & 0 & 0\\
			0 & 0 & 0 & 0 & 0 & -i & 0 & 0\\
			0 & 0 & 0 & 0 & 0 & 0 & 1 & 0\\
			0 & 0 & 0 & 0 & 0 & 0 & 0 & 1\\
		\end{array}
		\right)/\sqrt{2},
	\end{equation}
	
	With its transformation matrix in the logical basis
	\begin{equation}
		L_{\sigma_1^{-1}}=\left(
		\begin{array}{cccccccc}
			1\ & 0\ & 0\ & 0\ & 0\ & 0\ & -i\ & 0\\
			0 & 1 & 0 & 0 & 0 & 0 & 0 & -i\\
			0 & 0 & 1 & 0 & -i & 0 & 0 & 0\\
			0 & 0 & 0 & 1 & 0 & -i & 0 & 0\\
			0 & 0 & -i & 0 & 1 & 0 & 0 & 0\\
			0 & 0 & 0 & -i & 0 & 1 & 0 & 0\\
			-i & 0 & 0 & 0 & 0 & 0 & 1 & 0\\
			0 & -i & 0 & 0 & 0 & 0 & 0 & 1\\
		\end{array}
		\right)/\sqrt{2},
	\end{equation}
	\clearpage
	
	\section{The case of Jones polynomials with topological protection for $\sigma_{2}$}

	To implement the $\sigma_{2}$ of the braid group, we clockwise exchange Majorana C and E, as shown in Fig. S4. The exchange process is controlled by ten Hamiltonians
	\begin{equation}
		\begin{split}
			&H_{M_0}=i(\gamma_{1b} \gamma_{2a}+\gamma_{4b} \gamma_{5a}+\gamma_{5b} \gamma_{6a}+\gamma_{8b} \gamma_{9a}+\gamma_{9b}\gamma_{10a}+\gamma_{3a}\gamma_{3b}+\gamma_{7a}\gamma_{7b}), \\
			&H_{M_1}=i(\gamma_{1b} \gamma_{2a}+\gamma_{5b} \gamma_{6a}+\gamma_{8b}\gamma_{9a} +\gamma_{9b}\gamma_{10a}+\gamma_{3a}\gamma_{3b}+\gamma_{4a}\gamma_{4b}+\gamma_{7a}\gamma_{7b}), \\
			&H_{M_2}=i(\gamma_{1b} \gamma_{2a}+\gamma_{5a} \gamma_{7a}+\gamma_{5b} \gamma_{6a}+\gamma_{9b} \gamma_{10a}+\gamma_{3a}\gamma_{3b}+\gamma_{4a}\gamma_{4b}+\gamma_{8a}\gamma_{8b}), \\
			&H_{M_3}=i(\gamma_{1b} \gamma_{2a}+\gamma_{5a} \gamma_{7a}+\gamma_{5b} \gamma_{6a}+\gamma_{7b} \gamma_{8b}+\gamma_{9b} \gamma_{10a}+\gamma_{3a} \gamma_{3b}+\gamma_{4a}\gamma_{4b}), \\
			&H_{M_4}=i(\gamma_{1b} \gamma_{2a}+\gamma_{5a} \gamma_{7a}+\gamma_{5b} \gamma_{6a}+\gamma_{8b} \gamma_{9a}+\gamma_{9b} \gamma_{10a}+\gamma_{3a} \gamma_{3b}+\gamma_{4a}\gamma_{4b}), \\
			&H_{M_5}=i(\gamma_{1b} \gamma_{2a}+\gamma_{5b} \gamma_{6a}+\gamma_{8b} \gamma_{9a}+\gamma_{9b} \gamma_{10a}+\gamma_{3a} \gamma_{3b}+\gamma_{4a}\gamma_{4b}+\gamma_{7a}\gamma_{7b}).
		\end{split}
	\end{equation}
	
	The consecutive Hamiltonians are adiabatically connected and $H_{M_5}$ is finally adiabatically evolved to the initial Hamiltonian $H_{M_0}$. The adiabatic transport of the MZMs is implemented with our photonic simulator by ITE operators~\cite{jsxu2016}.
	
	\begin{figure}[htbp]
		\centering
		\includegraphics[width=1\columnwidth]{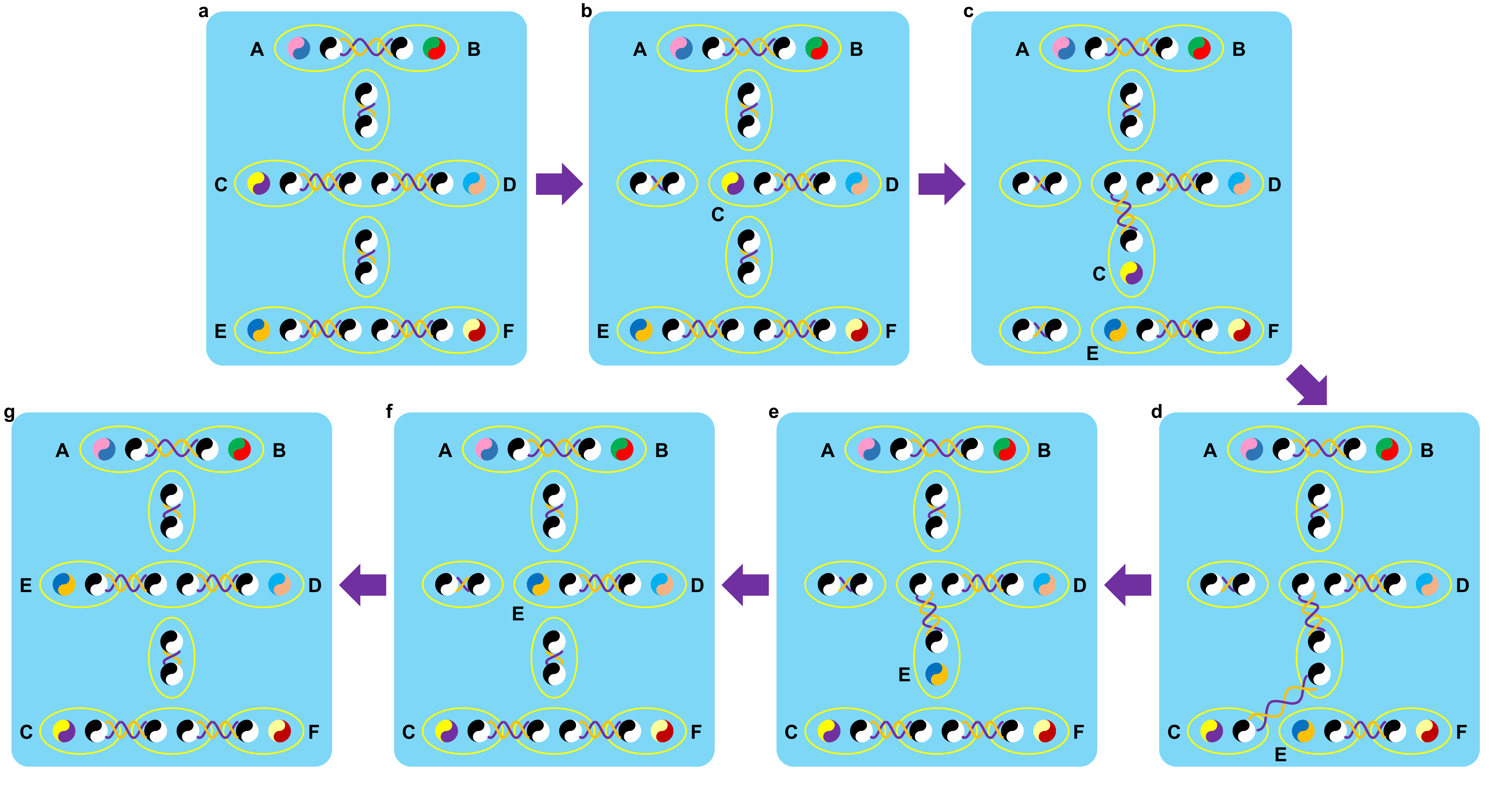}
		\small{\begin{flushleft}Fig. S4. The process of clockwise Majorana zero modes (MZMs) C and E. The Kitaev chain consists of ten fermions with six MZMs from A to F. Each dual-color circle represents a Majorana fermion, and a pair of Majorana fermions in the yellow circles constitutes a normal fermion. The helix lines between different Majorana fermions represent the interactions $i\gamma_{k}\gamma_{l}$ between them ($k$ and $l$= 1a, 1b, ..., 10b). $\textbf{a}, \textbf{b}, \textbf{c}, \textbf{d},  \textbf{e}, \textbf{f}$ and $\textbf{g}$ corresponds  to the Hamiltonians $H_{0}$, $H_{1}$, $H_{2}$, $H_{3}$, $H_{4}$, $H_{5}$ and $H_{0}$, respectively.\end{flushleft}}	
		
	\end{figure}
	
	Under the JW transformation, the fermionic Hamiltonians $H_{M_0}$, $H_{M_1}$, $H_{M_2}$, $H_{M_3}$, $H_{M_4}$, and $H_{M_5}$ can be transformed to the corresponding spin Hamiltonians $H_{0}$, $H_{1}$, $H_{2}$, $H_{3}$, $H_{4}$, and $H_{5}$, respectively, where
	\begin{equation}
		\begin{split}
			&H_{0} =-\sigma_{1}^{x}\sigma_{2}^{x}-\sigma_{4}^{x}\sigma_{5}^{x}-\sigma_{5}^{x}\sigma_{6}^{x} -\sigma_{8}^{x}\sigma_{9}^{x}-\sigma_{9}^{x}\sigma_{10}^{x}+\sigma_{3}^{z}+\sigma_{7}^{z}, \\
			&H_{1}=-\sigma_{1}^{x}\sigma_{2}^{x}-\sigma_{5}^{x}\sigma_{6}^{x}-\sigma_{8}^{x} \sigma_{9}^{x}-\sigma_{9}^{x}\sigma_{10}^{x}+\sigma_{3}^{z}+\sigma_{4}^{z}+\sigma_{7}^{z}, \\
			&H_{2}=-\sigma_{1}^{x}\sigma_{2}^{x}-\sigma_{5}^{x}\sigma_{6}^{x}-\sigma_{5}^{y} \sigma_{6}^{z}\sigma_{7}^{x}-\sigma_{9}^{x}\sigma_{10}^{x}+\sigma_{3}^{z}+\sigma_{4}^{z} +\sigma_{8}^{z}, \\
			&H_{3}=-\sigma_{1}^{x}\sigma_{2}^{x}-\sigma_{5}^{x}\sigma_{6}^{x}-\sigma_{5}^{y} \sigma_{6}^{z} \sigma_{7}^{x}+\sigma_{7}^{x} \sigma_{8}^{y}-\sigma_{9}^{x} \sigma_{10}^{x}+\sigma_{3}^{z}+\sigma_{4}^{z}, \\
			&H_{4}=-\sigma_{1}^{x}\sigma_{2}^{x}-\sigma_{5}^{x}\sigma_{6}^{x}-\sigma_{5}^{y} \sigma_{6}^{z} \sigma_{7}^{x}-\sigma_{8}^{x} \sigma_{9}^{x}-\sigma_{9}^{x} \sigma_{10}^{x}+\sigma_{3}^{z}+\sigma_{4}^{z}, \\
			&H_{5}=H_1=-\sigma_{1}^{x}\sigma_{2}^{x}-\sigma_{5}^{x}\sigma_{6}^{x}-\sigma_{8}^{x} \sigma_{9}^{x}-\sigma_{9}^{x} \sigma_{10}^{x}+\sigma_{3}^{z}+\sigma_{4}^{z}+\sigma_{7}^{z}.
		\end{split}
	\end{equation}
	The JW transformation between the fermionic and spin Hamiltonians preserves their spectrum. So the adiabatic transport of the fermions system can be studied in the corresponding spin description. We can further simplify the process of ITE as many of the terms in $H_{0}$, $H_{1}$, $H_{2}$, $H_{3}$, $H_{4}$, and $H_{5}$ commute with each other. The ground states of the corresponding Hamiltonians can be expressed in terms of the eigenvectors $\{|x\rangle,|\bar{x}\rangle\}$, $\{|y\rangle,|\bar{y}\rangle\}$ and $\{|z\rangle,|\bar{z}\rangle\}$ of the corresponding Pauli operators $\sigma^x$, $\sigma^y$ and $\sigma^z$, with eigenvalues $\{1,-1\}$, respectively. The detailed evolution in the spin system during the exchange is calculated as follows.
	
	After the ITE of the initial Hamiltonian
	\begin{equation}
		H_{0} =-\sigma_{1}^{x}\sigma_{2}^{x}-\sigma_{4}^{x}\sigma_{5}^{x}-\sigma_{5}^{x}\sigma_{6}^{x} -\sigma_{8}^{x}\sigma_{9}^{x}-\sigma_{9}^{x}\sigma_{10}^{x}+\sigma_{3}^{z}+\sigma_{7}^{z},
	\end{equation}
	which corresponds to creating the MZMs A, B, C, D, E and F, the state becomes
	\begin{equation}
		\begin{split}
			|\phi_{0}\rangle &=\alpha|x_{1}x_{2}\bar{z}_{3}x_{4}x_{5}x_{6}\bar{z}_{7}x_{8}x_{9}x_{10}\rangle
			+\beta|x_{1}x_{2}\bar{z}_{3}x_{4}x_{5}x_{6}\bar{z}_{7}\bar{x}_{8}\bar{x}_{9}\bar{x}_{10}\rangle \\
			&+\gamma|x_{1}x_{2}\bar{z}_{3}\bar{x}_{4}\bar{x}_{5}\bar{x}_{6}\bar{z}_{7}x_{8}x_{9}x_{10}\rangle
			+\delta|x_{1}x_{2}\bar{z}_{3}\bar{x}_{4}\bar{x}_{5}\bar{x}_{6}\bar{z}_{7}\bar{x}_{8}\bar{x}_{9}\bar{x}_{10}\rangle \\
			&+\eta|\bar{x}_{1}\bar{x}_{2}\bar{z}_{3}x_{4}x_{5}x_{6}\bar{z}_{7}x_{8}x_{9}x_{10}\rangle
			+\kappa|\bar{x}_{1}\bar{x}_{2}\bar{z}_{3}x_{4}x_{5}x_{6}\bar{z}_{7}\bar{x}_{8}\bar{x}_{9}\bar{x}_{10}\rangle \\
			&+\mu|\bar{x}_{1}\bar{x}_{2}\bar{z}_{3}\bar{x}_{4}\bar{x}_{5}\bar{x}_{6}\bar{z}_{7}x_{8}x_{9}x_{10}\rangle
			+\nu|\bar{x}_{1}\bar{x}_{2}\bar{z}_{3}\bar{x}_{4}\bar{x}_{5}\bar{x}_{6}\bar{z}_{7}\bar{x}_{8}\bar{x}_{9}\bar{x}_{10}\rangle,
			\label{initial2}
		\end{split}
	\end{equation}
	where $\alpha$, $\beta$, $\gamma$, $\delta$, $\eta$, $\kappa$, $\mu$ and $\nu$ are complex amplitudes satisfying $|\alpha|^{2}+|\beta|^{2}+|\gamma|^{2}+|\delta|^{2}+|\eta|^{2}+|\kappa|^{2}+|\mu|^{2}+|\nu|^{2}=1$.
	
	For the ITE of $H_{1}$, we first transfer the eigenstates of $\sigma_{4}^{x}$ to that of $\sigma_{4}^{z}$, and implement the ITE of $\sigma_{4}^{z}$. The state becomes 
	\begin{equation}
		\begin{split}
			|\phi_{1}\rangle &=\alpha|x_{1}x_{2}\bar{z}_{3}\bar{z}_{4}x_{5}x_{6}\bar{z}_{7}x_{8}x_{9}x_{10}\rangle
			+\beta|x_{1}x_{2}\bar{z}_{3}\bar{z}_{4}x_{5}x_{6}\bar{z}_{7}\bar{x}_{8}\bar{x}_{9}\bar{x}_{10}\rangle \\
			&-\gamma|x_{1}x_{2}\bar{z}_{3}\bar{z}_{4}\bar{x}_{5}\bar{x}_{6}\bar{z}_{7}x_{8}x_{9}x_{10}\rangle
			-\delta|x_{1}x_{2}\bar{z}_{3}\bar{z}_{4}\bar{x}_{5}\bar{x}_{6}\bar{z}_{7}\bar{x}_{8}\bar{x}_{9}\bar{x}_{10}\rangle \\
			&+\eta|\bar{x}_{1}\bar{x}_{2}\bar{z}_{3}\bar{z}_{4}x_{5}x_{6}\bar{z}_{7}x_{8}x_{9}x_{10}\rangle
			+\kappa|\bar{x}_{1}\bar{x}_{2}\bar{z}_{3}\bar{z}_{4}x_{5}x_{6}\bar{z}_{7}\bar{x}_{8}\bar{x}_{9}\bar{x}_{10}\rangle \\
			&-\mu|\bar{x}_{1}\bar{x}_{2}\bar{z}_{3}\bar{z}_{4}\bar{x}_{5}\bar{x}_{6}\bar{z}_{7}x_{8}x_{9}x_{10}\rangle
			-\nu|\bar{x}_{1}\bar{x}_{2}\bar{z}_{3}\bar{z}_{4}\bar{x}_{5}\bar{x}_{6}\bar{z}_{7}\bar{x}_{8}\bar{x}_{9}\bar{x}_{10}\rangle.
		\end{split}
	\end{equation}
	
	For the Hamiltonian $H_2$, we first implement the ITE of $\sigma_{8}^{z}$ and then implement the ITE of $-\sigma_{5}^{y}\sigma_{6}^{z}\sigma_{7}^{x}$. After the ITE of $\sigma_{8}^{z}$, the state becomes
	\begin{equation}
		\begin{split}
			|\phi_{21}\rangle &=\alpha|x_{1}x_{2}\bar{z}_{3}\bar{z}_{4}x_{5}x_{6}\bar{z}_{7}\bar{z}_{8}x_{9}x_{10}\rangle
			-\beta|x_{1}x_{2}\bar{z}_{3}\bar{z}_{4}x_{5}x_{6}\bar{z}_{7}\bar{z}_{8}\bar{x}_{9}\bar{x}_{10}\rangle \\
			&-\gamma|x_{1}x_{2}\bar{z}_{3}\bar{z}_{4}\bar{x}_{5}\bar{x}_{6}\bar{z}_{7}\bar{z}_{8}x_{9}x_{10}\rangle
			+\delta|x_{1}x_{2}\bar{z}_{3}\bar{z}_{4}\bar{x}_{5}\bar{x}_{6}\bar{z}_{7}\bar{z}_{8}\bar{x}_{9}\bar{x}_{10}\rangle \\
			&+\eta|\bar{x}_{1}\bar{x}_{2}\bar{z}_{3}\bar{z}_{4}x_{5}x_{6}\bar{z}_{7}\bar{z}_{8}x_{9}x_{10}\rangle
			-\kappa|\bar{x}_{1}\bar{x}_{2}\bar{z}_{3}\bar{z}_{4}x_{5}x_{6}\bar{z}_{7}\bar{z}_{8}\bar{x}_{9}\bar{x}_{10}\rangle \\
			&-\mu|\bar{x}_{1}\bar{x}_{2}\bar{z}_{3}\bar{z}_{4}\bar{x}_{5}\bar{x}_{6}\bar{z}_{7}\bar{z}_{8}x_{9}x_{10}\rangle
			+\nu|\bar{x}_{1}\bar{x}_{2}\bar{z}_{3}\bar{z}_{4}\bar{x}_{5}\bar{x}_{6}\bar{z}_{7}\bar{z}_{8}\bar{x}_{9}\bar{x}_{10}\rangle,
		\end{split}
	\end{equation}
	We then implement the ITE of $-\sigma_{5}^{y}\sigma_{6}^{z}\sigma_{7}^{x}$ and the state becomes 
	\begin{equation}
		\begin{split}
			|\phi_{2}\rangle &=\alpha(\ket{x_{1}x_{2}\bar{z}_{3}\bar{z}_{4}y_{5}z_{6}x_{7}\bar{z}_{8}x_{9}x_{10}} +i\ket{x_{1}x_{2}\bar{z}_{3}\bar{z}_{4}\bar{y}_{5}\bar{z}_{6}x_{7}\bar{z}_{8}x_{9}x_{10}} -i\ket{x_{1}x_{2}\bar{z}_{3}\bar{z}_{4}\bar{y}_{5}z_{6}\bar{x}_{7}\bar{z}_{8}x_{9}x_{10}}
			\\
			&-\ket{x_{1}x_{2}\bar{z}_{3}\bar{z}_{4}y_{5}\bar{z}_{6}\bar{x}_{7}\bar{z}_{8}x_{9}x_{10}})
			-\beta(\ket{x_{1}x_{2}\bar{z}_{3}\bar{z}_{4}y_{5}z_{6}x_{7}\bar{z}_{8}\bar{x}_{9}\bar{x}_{10}} +i\ket{x_{1}x_{2}\bar{z}_{3}\bar{z}_{4}\bar{y}_{5}\bar{z}_{6}x_{7}\bar{z}_{8}\bar{x}_{9}\bar{x}_{10}} \\
			&-i\ket{x_{1}x_{2}\bar{z}_{3}\bar{z}_{4}\bar{y}_{5}z_{6}\bar{x}_{7}\bar{z}_{8}\bar{x}_{9}\bar{x}_{10}} -\ket{x_{1}x_{2}\bar{z}_{3}\bar{z}_{4}y_{5}\bar{z}_{6}\bar{x}_{7}\bar{z}_{8}\bar{x}_{9}\bar{x}_{10}})
			-\gamma(i\ket{x_{1}x_{2}\bar{z}_{3}\bar{z}_{4}y_{5}z_{6}x_{7}\bar{z}_{8}x_{9}x_{10}}
			\\
			&-\ket{x_{1}x_{2}\bar{z}_{3}\bar{z}_{4}\bar{y}_{5}\bar{z}_{6}x_{7}\bar{z}_{8}x_{9}x_{10}} -\ket{x_{1}x_{2}\bar{z}_{3}\bar{z}_{4}\bar{y}_{5}z_{6}\bar{x}_{7}\bar{z}_{8}x_{9}x_{10}}
			+i\ket{x_{1}x_{2}\bar{z}_{3}\bar{z}_{4}y_{5}\bar{z}_{6}\bar{x}_{7}\bar{z}_{8}x_{9}x_{10}})
			\\
			&+\delta(i\ket{x_{1}x_{2}\bar{z}_{3}\bar{z}_{4}y_{5}z_{6}x_{7}\bar{z}_{8}\bar{x}_{9}\bar{x}_{10}} -\ket{x_{1}x_{2}\bar{z}_{3}\bar{z}_{4}\bar{y}_{5}\bar{z}_{6}x_{7}\bar{z}_{8}\bar{x}_{9}\bar{x}_{10}} -\ket{x_{1}x_{2}\bar{z}_{3}\bar{z}_{4}\bar{y}_{5}z_{6}\bar{x}_{7}\bar{z}_{8}\bar{x}_{9}\bar{x}_{10}} \\
			&+i\ket{x_{1}x_{2}\bar{z}_{3}\bar{z}_{4}y_{5}\bar{z}_{6}\bar{x}_{7}\bar{z}_{8}\bar{x}_{9}\bar{x}_{10}}) +\eta(\ket{\bar{x}_{1}\bar{x}_{2}\bar{z}_{3}\bar{z}_{4}y_{5}z_{6}x_{7}\bar{z}_{8}x_{9}x_{10}} +i\ket{\bar{x}_{1}\bar{x}_{2}\bar{z}_{3}\bar{z}_{4}\bar{y}_{5}\bar{z}_{6}x_{7}\bar{z}_{8}x_{9}x_{10}} \\
			&-i\ket{\bar{x}_{1}\bar{x}_{2}\bar{z}_{3}\bar{z}_{4}\bar{y}_{5}z_{6}\bar{x}_{7}\bar{z}_{8}x_{9}x_{10}} -\ket{\bar{x}_{1}\bar{x}_{2}\bar{z}_{3}\bar{z}_{4}y_{5}\bar{z}_{6}\bar{x}_{7}\bar{z}_{8}x_{9}x_{10}})
			-\kappa(\ket{\bar{x}_{1}\bar{x}_{2}\bar{z}_{3}\bar{z}_{4}y_{5}z_{6}x_{7}\bar{z}_{8}\bar{x}_{9}\bar{x}_{10}} \\
			&+i\ket{\bar{x}_{1}\bar{x}_{2}\bar{z}_{3}\bar{z}_{4}\bar{y}_{5}\bar{z}_{6}x_{7}\bar{z}_{8}\bar{x}_{9}\bar{x}_{10}} -i\ket{\bar{x}_{1}\bar{x}_{2}\bar{z}_{3}\bar{z}_{4}\bar{y}_{5}z_{6}\bar{x}_{7}\bar{z}_{8}\bar{x}_{9}\bar{x}_{10}} -\ket{\bar{x}_{1}\bar{x}_{2}\bar{z}_{3}\bar{z}_{4}y_{5}\bar{z}_{6}\bar{x}_{7}\bar{z}_{8}\bar{x}_{9}\bar{x}_{10}}) \\
			&-\mu(i\ket{\bar{x}_{1}\bar{x}_{2}\bar{z}_{3}\bar{z}_{4}y_{5}z_{6}x_{7}\bar{z}_{8}x_{9}x_{10}} -\ket{\bar{x}_{1}\bar{x}_{2}\bar{z}_{3}\bar{z}_{4}\bar{y}_{5}\bar{z}_{6}x_{7}\bar{z}_{8}x_{9}x_{10}} -\ket{\bar{x}_{1}\bar{x}_{2}\bar{z}_{3}\bar{z}_{4}\bar{y}_{5}z_{6}\bar{x}_{7}\bar{z}_{8}x_{9}x_{10}} \\
			&+i\ket{\bar{x}_{1}\bar{x}_{2}\bar{z}_{3}\bar{z}_{4}y_{5}\bar{z}_{6}\bar{x}_{7}\bar{z}_{8}x_{9}x_{10}})
			+\nu(i\ket{\bar{x}_{1}\bar{x}_{2}\bar{z}_{3}\bar{z}_{4}y_{5}z_{6}x_{7}\bar{z}_{8}\bar{x}_{9}\bar{x}_{10}} -\ket{\bar{x}_{1}\bar{x}_{2}\bar{z}_{3}\bar{z}_{4}\bar{y}_{5}\bar{z}_{6}x_{7}\bar{z}_{8}\bar{x}_{9}\bar{x}_{10}} \\
			&-\ket{\bar{x}_{1}\bar{x}_{2}\bar{z}_{3}\bar{z}_{4}\bar{y}_{5}z_{6}\bar{x}_{7}\bar{z}_{8}\bar{x}_{9}\bar{x}_{10}} +i\ket{\bar{x}_{1}\bar{x}_{2}\bar{z}_{3}\bar{z}_{4}y_{5}\bar{z}_{6}\bar{x}_{7}\bar{z}_{8}\bar{x}_{9}\bar{x}_{10}}).
		\end{split}
	\end{equation}
	It could be rewritten as
	\begin{equation}
		\begin{split}
			|\phi_{2}\rangle &=(\alpha-i\gamma)\ket{x_{1}x_{2}\bar{z}_{3}\bar{z}_{4}y_{5}z_{6}x_{7}\bar{z}_{8}x_{9}x_{10}} +(i\alpha+\gamma)\ket{x_{1}x_{2}\bar{z}_{3}\bar{z}_{4}\bar{y}_{5}\bar{z}_{6}x_{7}\bar{z}_{8}x_{9}x_{10}} \\
			&+(-i\alpha+\gamma)\ket{x_{1}x_{2}\bar{z}_{3}\bar{z}_{4}\bar{y}_{5}z_{6}\bar{x}_{7}\bar{z}_{8}x_{9}x_{10}}
			+(-\alpha-i\gamma)\ket{x_{1}x_{2}\bar{z}_{3}\bar{z}_{4}y_{5}\bar{z}_{6}\bar{x}_{7}\bar{z}_{8}x_{9}x_{10}}
			\\
			&+(-\beta+i\delta)\ket{x_{1}x_{2}\bar{z}_{3}\bar{z}_{4}y_{5}z_{6}x_{7}\bar{z}_{8}\bar{x}_{9}\bar{x}_{10}} +(-i\beta-\delta)\ket{x_{1}x_{2}\bar{z}_{3}\bar{z}_{4}\bar{y}_{5}\bar{z}_{6}x_{7}\bar{z}_{8}\bar{x}_{9}\bar{x}_{10}} \\
			&+(i\beta-\delta)\ket{x_{1}x_{2}\bar{z}_{3}\bar{z}_{4}\bar{y}_{5}z_{6}\bar{x}_{7}\bar{z}_{8}\bar{x}_{9}\bar{x}_{10}} +(\beta+i\delta)\ket{x_{1}x_{2}\bar{z}_{3}\bar{z}_{4}y_{5}\bar{z}_{6}\bar{x}_{7}\bar{z}_{8}\bar{x}_{9}\bar{x}_{10}}
			\\
			&+(\eta-i\mu)\ket{\bar{x}_{1}\bar{x}_{2}\bar{z}_{3}\bar{z}_{4}y_{5}z_{6}x_{7}\bar{z}_{8}x_{9}x_{10}}
			+(i\eta+\mu)\ket{\bar{x}_{1}\bar{x}_{2}\bar{z}_{3}\bar{z}_{4}\bar{y}_{5}\bar{z}_{6}x_{7}\bar{z}_{8}x_{9}x_{10}} \\
			&+(-i\eta+\mu)\ket{\bar{x}_{1}\bar{x}_{2}\bar{z}_{3}\bar{z}_{4}\bar{y}_{5}z_{6}\bar{x}_{7}\bar{z}_{8}x_{9}x_{10}} +(-\eta-i\mu)\ket{\bar{x}_{1}\bar{x}_{2}\bar{z}_{3}\bar{z}_{4}y_{5}\bar{z}_{6}\bar{x}_{7}\bar{z}_{8}x_{9}x_{10}}
			\\
			&+(-\kappa+i\nu)(\ket{\bar{x}_{1}\bar{x}_{2}\bar{z}_{3}\bar{z}_{4}y_{5}z_{6}x_{7}\bar{z}_{8}\bar{x}_{9}\bar{x}_{10}} +(-i\kappa-\nu)\ket{\bar{x}_{1}\bar{x}_{2}\bar{z}_{3}\bar{z}_{4}\bar{y}_{5}\bar{z}_{6}x_{7}\bar{z}_{8}\bar{x}_{9}\bar{x}_{10}} \\
			&+(i\kappa-\nu)\ket{\bar{x}_{1}\bar{x}_{2}\bar{z}_{3}\bar{z}_{4}\bar{y}_{5}z_{6}\bar{x}_{7}\bar{z}_{8}\bar{x}_{9}\bar{x}_{10}} +(\kappa+i\nu)\ket{\bar{x}_{1}\bar{x}_{2}\bar{z}_{3}\bar{z}_{4}y_{5}\bar{z}_{6}\bar{x}_{7}\bar{z}_{8}\bar{x}_{9}\bar{x}_{10}}.
		\end{split}
	\end{equation}
	
	For the $H_3$, we implement the ITE of $\sigma_{7}^{x}\sigma_{8}^{y}$. The state becomes
	\begin{equation}
		\begin{split}
			\ket{\phi_{3}} &=(\alpha-i\gamma)\ket{x_{1}x_{2}\bar{z}_{3}\bar{z}_{4}y_{5}z_{6}x_{7}\bar{y}_{8}x_{9}x_{10}} +(i\alpha+\gamma)\ket{x_{1}x_{2}\bar{z}_{3}\bar{z}_{4}\bar{y}_{5}\bar{z}_{6}x_{7}\bar{y}_{8}x_{9}x_{10}} \\
			&-(-i\alpha+\gamma)\ket{x_{1}x_{2}\bar{z}_{3}\bar{z}_{4}\bar{y}_{5}z_{6}\bar{x}_{7}y_{8}x_{9}x_{10}}
			-(-\alpha-i\gamma)\ket{x_{1}x_{2}\bar{z}_{3}\bar{z}_{4}y_{5}\bar{z}_{6}\bar{x}_{7}y_{8}x_{9}x_{10}}
			\\
			&+(-\beta+i\delta)\ket{x_{1}x_{2}\bar{z}_{3}\bar{z}_{4}y_{5}z_{6}x_{7}\bar{y}_{8}\bar{x}_{9}\bar{x}_{10}} +(-i\beta-\delta)\ket{x_{1}x_{2}\bar{z}_{3}\bar{z}_{4}\bar{y}_{5}\bar{z}_{6}x_{7}\bar{y}_{8}\bar{x}_{9}\bar{x}_{10}} \\
			&-(i\beta-\delta)\ket{x_{1}x_{2}\bar{z}_{3}\bar{z}_{4}\bar{y}_{5}z_{6}\bar{x}_{7}y_{8}\bar{x}_{9}\bar{x}_{10}} -(\beta+i\delta)\ket{x_{1}x_{2}\bar{z}_{3}\bar{z}_{4}y_{5}\bar{z}_{6}\bar{x}_{7}y_{8}\bar{x}_{9}\bar{x}_{10}}
			\\
			&+(\eta-i\mu)\ket{\bar{x}_{1}\bar{x}_{2}\bar{z}_{3}\bar{z}_{4}y_{5}z_{6}x_{7}\bar{y}_{8}x_{9}x_{10}}
			+(i\eta+\mu)\ket{\bar{x}_{1}\bar{x}_{2}\bar{z}_{3}\bar{z}_{4}\bar{y}_{5}\bar{z}_{6}x_{7}\bar{y}_{8}x_{9}x_{10}} \\
			&-(-i\eta+\mu)\ket{\bar{x}_{1}\bar{x}_{2}\bar{z}_{3}\bar{z}_{4}\bar{y}_{5}z_{6}\bar{x}_{7}y_{8}x_{9}x_{10}} -(-\eta-i\mu)\ket{\bar{x}_{1}\bar{x}_{2}\bar{z}_{3}\bar{z}_{4}y_{5}\bar{z}_{6}\bar{x}_{7}y_{8}x_{9}x_{10}}
			\\
			&+(-\kappa+i\nu)(\ket{\bar{x}_{1}\bar{x}_{2}\bar{z}_{3}\bar{z}_{4}y_{5}z_{6}x_{7}\bar{y}_{8}\bar{x}_{9}\bar{x}_{10}} +(-i\kappa-\nu)\ket{\bar{x}_{1}\bar{x}_{2}\bar{z}_{3}\bar{z}_{4}\bar{y}_{5}\bar{z}_{6}x_{7}\bar{y}_{8}\bar{x}_{9}\bar{x}_{10}} \\
			&-(i\kappa-\nu)\ket{\bar{x}_{1}\bar{x}_{2}\bar{z}_{3}\bar{z}_{4}\bar{y}_{5}z_{6}\bar{x}_{7}y_{8}\bar{x}_{9}\bar{x}_{10}} -(\kappa+i\nu)\ket{\bar{x}_{1}\bar{x}_{2}\bar{z}_{3}\bar{z}_{4}y_{5}\bar{z}_{6}\bar{x}_{7}y_{8}\bar{x}_{9}\bar{x}_{10}}.
		\end{split}
	\end{equation}
	
	For the $H_4$, we only implement the ITE of $-\sigma_{8}^{x}\sigma_{9}^{x}$. The state becomes
	\begin{equation}
		\begin{split}
			\ket{\phi_{4}} &=(\alpha-i\gamma)\ket{x_{1}x_{2}\bar{z}_{3}\bar{z}_{4}y_{5}z_{6}x_{7}x_{8}x_{9}x_{10}} +(i\alpha+\gamma)\ket{x_{1}x_{2}\bar{z}_{3}\bar{z}_{4}\bar{y}_{5}\bar{z}_{6}x_{7}x_{8}x_{9}x_{10}} \\
			&-i(-i\alpha+\gamma)\ket{x_{1}x_{2}\bar{z}_{3}\bar{z}_{4}\bar{y}_{5}z_{6}\bar{x}_{7}x_{8}x_{9}x_{10}}
			-i(-\alpha-i\gamma)\ket{x_{1}x_{2}\bar{z}_{3}\bar{z}_{4}y_{5}\bar{z}_{6}\bar{x}_{7}x_{8}x_{9}x_{10}}
			\\
			&+i(-\beta+i\delta)\ket{x_{1}x_{2}\bar{z}_{3}\bar{z}_{4}y_{5}z_{6}x_{7}\bar{x}_{8}\bar{x}_{9}\bar{x}_{10}} +i(-i\beta-\delta)\ket{x_{1}x_{2}\bar{z}_{3}\bar{z}_{4}\bar{y}_{5}\bar{z}_{6}x_{7}\bar{x}_{8}\bar{x}_{9}\bar{x}_{10}} \\
			&-(i\beta-\delta)\ket{x_{1}x_{2}\bar{z}_{3}\bar{z}_{4}\bar{y}_{5}z_{6}\bar{x}_{7}\bar{x}_{8}\bar{x}_{9}\bar{x}_{10}} -(\beta+i\delta)\ket{x_{1}x_{2}\bar{z}_{3}\bar{z}_{4}y_{5}\bar{z}_{6}\bar{x}_{7}\bar{x}_{8}\bar{x}_{9}\bar{x}_{10}}
			\\
			&+(\eta-i\mu)\ket{\bar{x}_{1}\bar{x}_{2}\bar{z}_{3}\bar{z}_{4}y_{5}z_{6}x_{7}x_{8}x_{9}x_{10}}
			+(i\eta+\mu)\ket{\bar{x}_{1}\bar{x}_{2}\bar{z}_{3}\bar{z}_{4}\bar{y}_{5}\bar{z}_{6}x_{7}x_{8}x_{9}x_{10}} \\
			&-i(-i\eta+\mu)\ket{\bar{x}_{1}\bar{x}_{2}\bar{z}_{3}\bar{z}_{4}\bar{y}_{5}z_{6}\bar{x}_{7}x_{8}x_{9}x_{10}} -i(-\eta-i\mu)\ket{\bar{x}_{1}\bar{x}_{2}\bar{z}_{3}\bar{z}_{4}y_{5}\bar{z}_{6}\bar{x}_{7}x_{8}x_{9}x_{10}}
			\\
			&+i(-\kappa+i\nu)(\ket{\bar{x}_{1}\bar{x}_{2}\bar{z}_{3}\bar{z}_{4}y_{5}z_{6}x_{7}\bar{x}_{8}\bar{x}_{9}\bar{x}_{10}} +i(-i\kappa-\nu)\ket{\bar{x}_{1}\bar{x}_{2}\bar{z}_{3}\bar{z}_{4}\bar{y}_{5}\bar{z}_{6}x_{7}\bar{x}_{8}\bar{x}_{9}\bar{x}_{10}} \\
			&-(i\kappa-\nu)\ket{\bar{x}_{1}\bar{x}_{2}\bar{z}_{3}\bar{z}_{4}\bar{y}_{5}z_{6}\bar{x}_{7}\bar{x}_{8}\bar{x}_{9}\bar{x}_{10}} -(\kappa+i\nu)\ket{\bar{x}_{1}\bar{x}_{2}\bar{z}_{3}\bar{z}_{4}y_{5}\bar{z}_{6}\bar{x}_{7}\bar{x}_{8}\bar{x}_{9}\bar{x}_{10}}.
		\end{split}
	\end{equation}
	
	For the $H_5$, we only implement the ITE of $\sigma_{7}^{z}$ from the  $-\sigma_{5}^{y}\sigma_{6}^{z}\sigma_{7}^{x}$. The state becomes
	\begin{equation}
		\begin{split}
			\ket{\phi_{5}} &=(\alpha-i\gamma)\ket{x_{1}x_{2}\bar{z}_{3}\bar{z}_{4}(\bar{x}_{5}\bar{x}_{6}+ix_{5}x_{6})\bar{z}_{7}x_{8}x_{9}x_{10}} +(i\alpha+\gamma)\ket{x_{1}x_{2}\bar{z}_{3}\bar{z}_{4}(-i\bar{x}_{5}\bar{x}_{6}+x_{5}x_{6})\bar{z}_{7}x_{8}x_{9}x_{10}} \\
			&+i(-i\alpha+\gamma)\ket{x_{1}x_{2}\bar{z}_{3}\bar{z}_{4}(i\bar{x}_{5}\bar{x}_{6}+x_{5}x_{6})\bar{z}_{7}x_{8}x_{9}x_{10}}
			+i(-\alpha-i\gamma)\ket{x_{1}x_{2}\bar{z}_{3}\bar{z}_{4}(-\bar{x}_{5}\bar{x}_{6}+ix_{5}x_{6})\bar{z}_{7}x_{8}x_{9}x_{10}}
			\\
			&+i(-\beta+i\delta)\ket{x_{1}x_{2}\bar{z}_{3}\bar{z}_{4}(\bar{x}_{5}\bar{x}_{6}+ix_{5}x_{6})\bar{z}_{7}\bar{x}_{8}\bar{x}_{9}\bar{x}_{10}} +i(-i\beta-\delta)\ket{x_{1}x_{2}\bar{z}_{3}\bar{z}_{4}(-i\bar{x}_{5}\bar{x}_{6}+x_{5}x_{6})\bar{z}_{7}\bar{x}_{8}\bar{x}_{9}\bar{x}_{10}} \\
			&+(i\beta-\delta)\ket{x_{1}x_{2}\bar{z}_{3}\bar{z}_{4}(i\bar{x}_{5}\bar{x}_{6}+x_{5}x_{6})\bar{z}_{7}\bar{x}_{8}\bar{x}_{9}\bar{x}_{10}} +(\beta+i\delta)\ket{x_{1}x_{2}\bar{z}_{3}\bar{z}_{4}(-\bar{x}_{5}\bar{x}_{6}+ix_{5}x_{6})\bar{z}_{7}\bar{x}_{8}\bar{x}_{9}\bar{x}_{10}}
			\\
			&+(\eta-i\mu)\ket{\bar{x}_{1}\bar{x}_{2}\bar{z}_{3}\bar{z}_{4}(\bar{x}_{5}\bar{x}_{6}+ix_{5}x_{6})\bar{z}_{7}x_{8}x_{9}x_{10}}
			+(i\eta+\mu)\ket{\bar{x}_{1}\bar{x}_{2}\bar{z}_{3}\bar{z}_{4}(-i\bar{x}_{5}\bar{x}_{6}+x_{5}x_{6})\bar{z}_{7}x_{8}x_{9}x_{10}} \\
			&+i(-i\eta+\mu)\ket{\bar{x}_{1}\bar{x}_{2}\bar{z}_{3}\bar{z}_{4}(i\bar{x}_{5}\bar{x}_{6}+x_{5}x_{6})\bar{z}_{7}x_{8}x_{9}x_{10}} +i(-\eta-i\mu)\ket{\bar{x}_{1}\bar{x}_{2}\bar{z}_{3}\bar{z}_{4}(-\bar{x}_{5}\bar{x}_{6}+ix_{5}x_{6})\bar{z}_{7}x_{8}x_{9}x_{10}}
			\\
			&+i(-\kappa+i\nu)(\ket{\bar{x}_{1}\bar{x}_{2}\bar{z}_{3}\bar{z}_{4}(\bar{x}_{5}\bar{x}_{6}+ix_{5}x_{6})\bar{z}_{7}\bar{x}_{8}\bar{x}_{9}\bar{x}_{10}} +i(-i\kappa-\nu)\ket{\bar{x}_{1}\bar{x}_{2}\bar{z}_{3}\bar{z}_{4}(-i\bar{x}_{5}\bar{x}_{6}+x_{5}x_{6})\bar{z}_{7}\bar{x}_{8}\bar{x}_{9}\bar{x}_{10}} \\
			&+(i\kappa-\nu)\ket{\bar{x}_{1}\bar{x}_{2}\bar{z}_{3}\bar{z}_{4}(i\bar{x}_{5}\bar{x}_{6}+x_{5}x_{6})\bar{z}_{7}\bar{x}_{8}\bar{x}_{9}\bar{x}_{10}} +(\kappa+i\nu)\ket{\bar{x}_{1}\bar{x}_{2}\bar{z}_{3}\bar{z}_{4}(-\bar{x}_{5}\bar{x}_{6}+ix_{5}x_{6})\bar{z}_{7}\bar{x}_{8}\bar{x}_{9}\bar{x}_{10}}.
		\end{split}
	\end{equation}
	And it could be rewritten as 
	\begin{equation}
		\begin{split}
			\ket{\phi_{5}} &=(\alpha+\gamma)|x_{1}x_{2}\bar{z}_{3}\bar{z}_{4}x_{5}x_{6}\bar{z}_{7}x_{8}x_{9}x_{10}\rangle
			+(\beta-\delta)|x_{1}x_{2}\bar{z}_{3}\bar{z}_{4}x_{5}x_{6}\bar{z}_{7}\bar{x}_{8}\bar{x}_{9}\bar{x}_{10}\rangle \\
			&+(\alpha-\gamma)|x_{1}x_{2}\bar{z}_{3}\bar{z}_{4}\bar{x}_{5}\bar{x}_{6}\bar{z}_{7}x_{8}x_{9}x_{10}\rangle
			-(\beta+\delta)|x_{1}x_{2}\bar{z}_{3}\bar{z}_{4}\bar{x}_{5}\bar{x}_{6}\bar{z}_{7}\bar{x}_{8}\bar{x}_{9}\bar{x}_{10}\rangle \\
			&+(\eta+\mu)|\bar{x}_{1}\bar{x}_{2}\bar{z}_{3}\bar{z}_{4}x_{5}x_{6}\bar{z}_{7}x_{8}x_{9}x_{10}\rangle
			+(\kappa-\nu)|\bar{x}_{1}\bar{x}_{2}\bar{z}_{3}\bar{z}_{4}x_{5}x_{6}\bar{z}_{7}\bar{x}_{8}\bar{x}_{9}\bar{x}_{10}\rangle \\
			&+(\eta-\mu)|\bar{x}_{1}\bar{x}_{2}\bar{z}_{3}\bar{z}_{4}\bar{x}_{5}\bar{x}_{6}\bar{z}_{7}x_{8}x_{9}x_{10}\rangle
			-(\kappa+\nu)|\bar{x}_{1}\bar{x}_{2}\bar{z}_{3}\bar{z}_{4}\bar{x}_{5}\bar{x}_{6}\bar{z}_{7}\bar{x}_{8}\bar{x}_{9}\bar{x}_{10}\rangle.
		\end{split}
	\end{equation}

	Back to the Hamiltonian $H_0$, the state becomes after the ITE of $H_0$,
	\begin{equation}
		\begin{split}
			\ket{\phi_{6}} &=(\alpha+\gamma)|x_{1}x_{2}\bar{z}_{3}x_{4}x_{5}x_{6}\bar{z}_{7}x_{8}x_{9}x_{10}\rangle
			+(\beta-\delta)|x_{1}x_{2}\bar{z}_{3}x_{4}x_{5}x_{6}\bar{z}_{7}\bar{x}_{8}\bar{x}_{9}\bar{x}_{10}\rangle \\
			&-(\alpha-\gamma)|x_{1}x_{2}\bar{z}_{3}\bar{x}_{4}\bar{x}_{5}\bar{x}_{6}\bar{z}_{7}x_{8}x_{9}x_{10}\rangle
			+(\beta+\delta)|x_{1}x_{2}\bar{z}_{3}\bar{x}_{4}\bar{x}_{5}\bar{x}_{6}\bar{z}_{7}\bar{x}_{8}\bar{x}_{9}\bar{x}_{10}\rangle \\
			&+(\eta+\mu)|\bar{x}_{1}\bar{x}_{2}\bar{z}_{3}x_{4}x_{5}x_{6}\bar{z}_{7}x_{8}x_{9}x_{10}\rangle
			+(\kappa-\nu)|\bar{x}_{1}\bar{x}_{2}\bar{z}_{3}x_{4}x_{5}x_{6}\bar{z}_{7}\bar{x}_{8}\bar{x}_{9}\bar{x}_{10}\rangle \\
			&-(\eta-\mu)|\bar{x}_{1}\bar{x}_{2}\bar{z}_{3}\bar{x}_{4}\bar{x}_{5}\bar{x}_{6}\bar{z}_{7}x_{8}x_{9}x_{10}\rangle
			+(\kappa+\nu)|\bar{x}_{1}\bar{x}_{2}\bar{z}_{3}\bar{x}_{4}\bar{x}_{5}\bar{x}_{6}\bar{z}_{7}\bar{x}_{8}\bar{x}_{9}\bar{x}_{10}\rangle.
		\end{split}
	\end{equation}
	
	It is easy to obtain the transformation of this clockwise braiding as
	\begin{equation}
		U_{\sigma_2}=\left(
		\begin{array}{cccccccc}
			1\ & 0\ & 1\ & 0\ & 0\ & 0\ & 0\ & 0\\
			0 & 1 & 0 & -1 & 0 & 0 & 0 & 0\\
			-1 & 0 & 1 & 0 & 0 & 0 & 0 & 0\\
			0 & 1 & 0 & 1 & 0 & 0 & 0 & 0\\
			0 & 0 & 0 & 0 & 1 & 0 & 1 & 0\\
			0 & 0 & 0 & 0 & 0 & 1 & 0 & -1\\
			0 & 0 & 0 & 0 & -1 & 0 & 1 & 0\\
			0 & 0 & 0 & 0 & 0 & 1 & 0 & 1\\
		\end{array}
		\right)/\sqrt{2}.
	\end{equation}
	With its transformation matrix in the logical basis
	\begin{equation}
		L_{\sigma_2}=\left(
		\begin{array}{cccccccc}
			1\ & 0\ & 0\ & 1\ & 0\ & 0\ & 0\ & 0\\
			0 & 1 & 1 & 0 & 0 & 0 & 0 & 0\\
			0 & -1 & 1 & 0 & 0 & 0 & 0 & 0\\
			-1 & 0 & 0 & 1 & 0 & 0 & 0 & 0\\
			0 & 0 & 0 & 0 & 1 & 0 & 0 & 1\\
			0 & 0 & 0 & 0 & 0 & 1 & 1 & 0\\
			0 & 0 & 0 & 0 & 0 & -1 & 1 & 0\\
			0 & 0 & 0 & 0 & -1 & 0 & 0 & 1\\
		\end{array}
		\right)/\sqrt{2}.
	\end{equation}
	
	Following the above description, similarly, we show the anticlockwise braiding of C and E in Fig. S5.
	\begin{figure}[htbp]
		\centering
		\includegraphics[width=1\columnwidth]{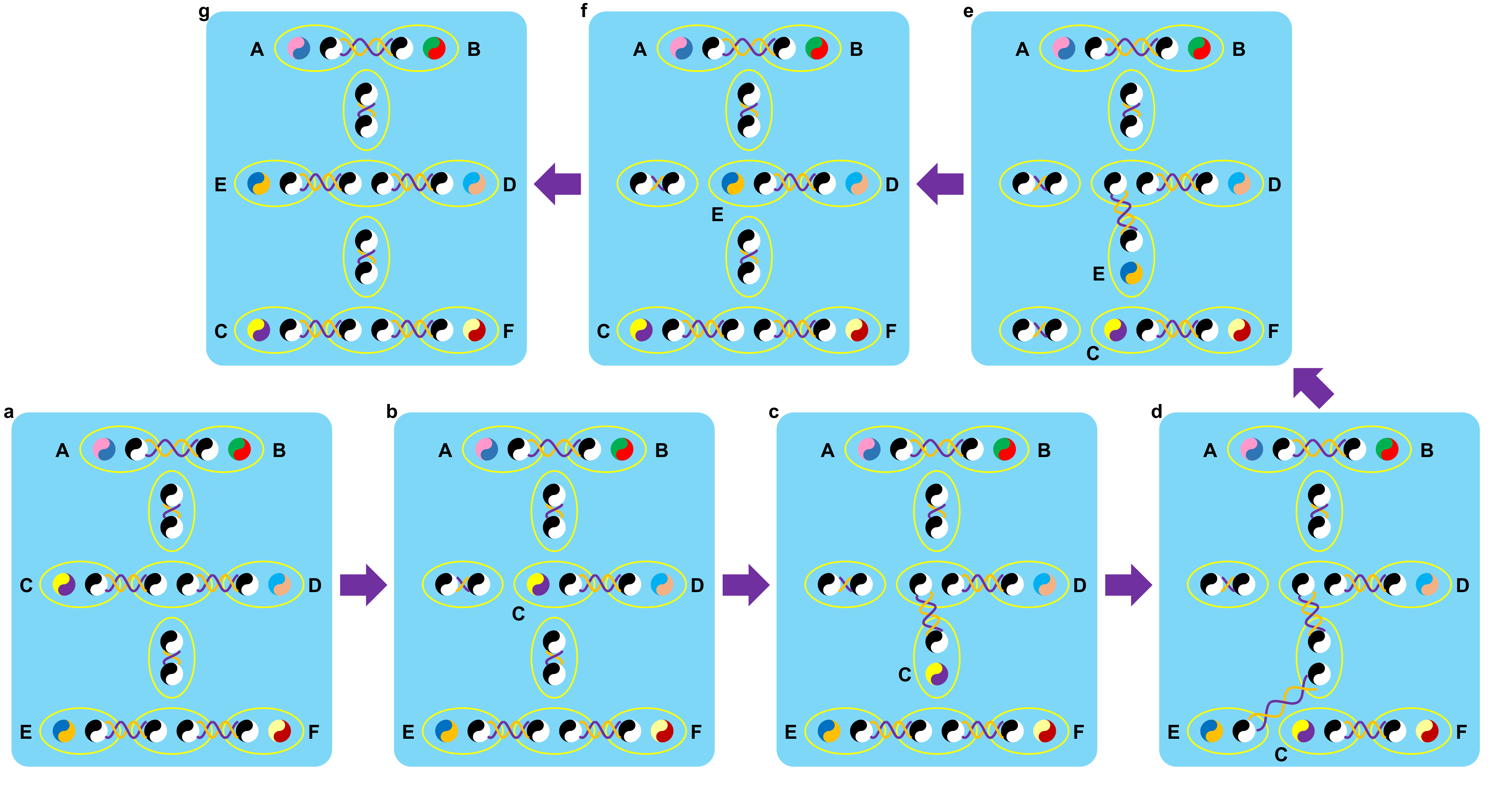}
		\small{\begin{flushleft}Fig. S5. The process of anticlockwise Majorana zero modes (MZMs) C and E. The Kitaev chain consists of ten fermions with six MZMs from A to F. Each dual-color circle represents a Majorana fermion, and a pair of Majorana fermions in the yellow circles constitutes a normal fermion. The helix lines between different Majorana fermions represent the interactions $i\gamma_{k}\gamma_{l}$ between them ($k$ and $l$= 1a, 1b, ..., 10b). $\textbf{a}, \textbf{b}, \textbf{c}, \textbf{d},  \textbf{e}, \textbf{f}$ and $\textbf{g}$ corresponds  to the Hamiltonians $H_{0}$, $H_{5}$, $H_{4}$, $H_{3}$, $H_{2}$, $H_{1}$ and $H_{0}$, respectively.\end{flushleft}}	
		
	\end{figure}

	The detailed ground states of every Hamiltonian in the anticlockwise braiding are listed as below.
	\begin{equation}
		\begin{split}
			|\psi_{1}\rangle &=\alpha|x_{1}x_{2}\bar{z}_{3}\bar{z}_{4}x_{5}x_{6}\bar{z}_{7}x_{8}x_{9}x_{10}\rangle
			+\beta|x_{1}x_{2}\bar{z}_{3}\bar{z}_{4}x_{5}x_{6}\bar{z}_{7}\bar{x}_{8}\bar{x}_{9}\bar{x}_{10}\rangle \\
			&-\gamma|x_{1}x_{2}\bar{z}_{3}\bar{z}_{4}\bar{x}_{5}\bar{x}_{6}\bar{z}_{7}x_{8}x_{9}x_{10}\rangle
			-\delta|x_{1}x_{2}\bar{z}_{3}\bar{z}_{4}\bar{x}_{5}\bar{x}_{6}\bar{z}_{7}\bar{x}_{8}\bar{x}_{9}\bar{x}_{10}\rangle \\
			&+\eta|\bar{x}_{1}\bar{x}_{2}\bar{z}_{3}\bar{z}_{4}x_{5}x_{6}\bar{z}_{7}x_{8}x_{9}x_{10}\rangle
			+\kappa|\bar{x}_{1}\bar{x}_{2}\bar{z}_{3}\bar{z}_{4}x_{5}x_{6}\bar{z}_{7}\bar{x}_{8}\bar{x}_{9}\bar{x}_{10}\rangle \\
			&-\mu|\bar{x}_{1}\bar{x}_{2}\bar{z}_{3}\bar{z}_{4}\bar{x}_{5}\bar{x}_{6}\bar{z}_{7}x_{8}x_{9}x_{10}\rangle
			-\nu|\bar{x}_{1}\bar{x}_{2}\bar{z}_{3}\bar{z}_{4}\bar{x}_{5}\bar{x}_{6}\bar{z}_{7}\bar{x}_{8}\bar{x}_{9}\bar{x}_{10}\rangle.
		\end{split}
	\end{equation}
	
	\begin{equation}
		\begin{split}
			\ket{\psi_{2}} &=(\alpha-i\gamma)\ket{x_{1}x_{2}\bar{z}_{3}\bar{z}_{4}y_{5}z_{6}x_{7}x_{8}x_{9}x_{10}} +(i\alpha+\gamma)\ket{x_{1}x_{2}\bar{z}_{3}\bar{z}_{4}\bar{y}_{5}\bar{z}_{6}x_{7}x_{8}x_{9}x_{10}} \\
			&+(-i\alpha+\gamma)\ket{x_{1}x_{2}\bar{z}_{3}\bar{z}_{4}\bar{y}_{5}z_{6}\bar{x}_{7}x_{8}x_{9}x_{10}}
			+(-\alpha-i\gamma)\ket{x_{1}x_{2}\bar{z}_{3}\bar{z}_{4}y_{5}\bar{z}_{6}\bar{x}_{7}x_{8}x_{9}x_{10}}
			\\
			&+(\beta-i\delta)\ket{x_{1}x_{2}\bar{z}_{3}\bar{z}_{4}y_{5}z_{6}x_{7}\bar{x}_{8}\bar{x}_{9}\bar{x}_{10}} +(i\beta+\delta)\ket{x_{1}x_{2}\bar{z}_{3}\bar{z}_{4}\bar{y}_{5}\bar{z}_{6}x_{7}\bar{x}_{8}\bar{x}_{9}\bar{x}_{10}} \\
			&+(-i\beta+\delta)\ket{x_{1}x_{2}\bar{z}_{3}\bar{z}_{4}\bar{y}_{5}z_{6}\bar{x}_{7}\bar{x}_{8}\bar{x}_{9}\bar{x}_{10}} +(-\beta-i\delta)\ket{x_{1}x_{2}\bar{z}_{3}\bar{z}_{4}y_{5}\bar{z}_{6}\bar{x}_{7}\bar{x}_{8}\bar{x}_{9}\bar{x}_{10}}
			\\
			&+(\eta-i\mu)\ket{\bar{x}_{1}\bar{x}_{2}\bar{z}_{3}\bar{z}_{4}y_{5}z_{6}x_{7}x_{8}x_{9}x_{10}}
			+(i\eta+\mu)\ket{\bar{x}_{1}\bar{x}_{2}\bar{z}_{3}\bar{z}_{4}\bar{y}_{5}\bar{z}_{6}x_{7}x_{8}x_{9}x_{10}} \\
			&+(-i\eta+\mu)\ket{\bar{x}_{1}\bar{x}_{2}\bar{z}_{3}\bar{z}_{4}\bar{y}_{5}z_{6}\bar{x}_{7}x_{8}x_{9}x_{10}} +(-\eta-i\mu)\ket{\bar{x}_{1}\bar{x}_{2}\bar{z}_{3}\bar{z}_{4}y_{5}\bar{z}_{6}\bar{x}_{7}x_{8}x_{9}x_{10}})
			\\
			&+(\kappa-i\nu)(\ket{\bar{x}_{1}\bar{x}_{2}\bar{z}_{3}\bar{z}_{4}y_{5}z_{6}x_{7}\bar{x}_{8}\bar{x}_{9}\bar{x}_{10}} +(i\kappa+\nu)\ket{\bar{x}_{1}\bar{x}_{2}\bar{z}_{3}\bar{z}_{4}\bar{y}_{5}\bar{z}_{6}x_{7}\bar{x}_{8}\bar{x}_{9}\bar{x}_{10}} \\
			&+(-i\kappa+\nu)\ket{\bar{x}_{1}\bar{x}_{2}\bar{z}_{3}\bar{z}_{4}\bar{y}_{5}z_{6}\bar{x}_{7}\bar{x}_{8}\bar{x}_{9}\bar{x}_{10}} +(-\kappa-i\nu)\ket{\bar{x}_{1}\bar{x}_{2}\bar{z}_{3}\bar{z}_{4}y_{5}\bar{z}_{6}\bar{x}_{7}\bar{x}_{8}\bar{x}_{9}\bar{x}_{10}}.
		\end{split}
	\end{equation}
	
	\begin{equation}
		\begin{split}
			\ket{\psi_{3}} &=i(\alpha-i\gamma)\ket{x_{1}x_{2}\bar{z}_{3}\bar{z}_{4}y_{5}z_{6}x_{7}\bar{y}_{8}x_{9}x_{10}} +i(i\alpha+\gamma)\ket{x_{1}x_{2}\bar{z}_{3}\bar{z}_{4}\bar{y}_{5}\bar{z}_{6}x_{7}\bar{y}_{8}x_{9}x_{10}} \\
			&+(-i\alpha+\gamma)\ket{x_{1}x_{2}\bar{z}_{3}\bar{z}_{4}\bar{y}_{5}z_{6}\bar{x}_{7}y_{8}x_{9}x_{10}}
			+(-\alpha-i\gamma)\ket{x_{1}x_{2}\bar{z}_{3}\bar{z}_{4}y_{5}\bar{z}_{6}\bar{x}_{7}y_{8}x_{9}x_{10}}
			\\
			&+(\beta-i\delta)\ket{x_{1}x_{2}\bar{z}_{3}\bar{z}_{4}y_{5}z_{6}x_{7}\bar{y}_{8}\bar{x}_{9}\bar{x}_{10}} +(i\beta+\delta)\ket{x_{1}x_{2}\bar{z}_{3}\bar{z}_{4}\bar{y}_{5}\bar{z}_{6}x_{7}\bar{y}_{8}\bar{x}_{9}\bar{x}_{10}} \\
			&+i(-i\beta+\delta)\ket{x_{1}x_{2}\bar{z}_{3}\bar{z}_{4}\bar{y}_{5}z_{6}\bar{x}_{7}y_{8}\bar{x}_{9}\bar{x}_{10}} +i(-\beta-i\delta)\ket{x_{1}x_{2}\bar{z}_{3}\bar{z}_{4}y_{5}\bar{z}_{6}\bar{x}_{7}y_{8}\bar{x}_{9}\bar{x}_{10}}
			\\
			&+i(\eta-i\mu)\ket{\bar{x}_{1}\bar{x}_{2}\bar{z}_{3}\bar{z}_{4}y_{5}z_{6}x_{7}\bar{y}_{8}x_{9}x_{10}}
			+i(i\eta+\mu)\ket{\bar{x}_{1}\bar{x}_{2}\bar{z}_{3}\bar{z}_{4}\bar{y}_{5}\bar{z}_{6}x_{7}\bar{y}_{8}x_{9}x_{10}} \\
			&+(-i\eta+\mu)\ket{\bar{x}_{1}\bar{x}_{2}\bar{z}_{3}\bar{z}_{4}\bar{y}_{5}z_{6}\bar{x}_{7}y_{8}x_{9}x_{10}} +(-\eta-i\mu)\ket{\bar{x}_{1}\bar{x}_{2}\bar{z}_{3}\bar{z}_{4}y_{5}\bar{z}_{6}\bar{x}_{7}y_{8}x_{9}x_{10}})
			\\
			&+(\kappa-i\nu)(\ket{\bar{x}_{1}\bar{x}_{2}\bar{z}_{3}\bar{z}_{4}y_{5}z_{6}x_{7}\bar{y}_{8}\bar{x}_{9}\bar{x}_{10}} +(i\kappa+\nu)\ket{\bar{x}_{1}\bar{x}_{2}\bar{z}_{3}\bar{z}_{4}\bar{y}_{5}\bar{z}_{6}x_{7}\bar{y}_{8}\bar{x}_{9}\bar{x}_{10}} \\
			&-i(i\kappa-\nu)\ket{\bar{x}_{1}\bar{x}_{2}\bar{z}_{3}\bar{z}_{4}\bar{y}_{5}z_{6}\bar{x}_{7}y_{8}\bar{x}_{9}\bar{x}_{10}} -i(\kappa+i\nu)\ket{\bar{x}_{1}\bar{x}_{2}\bar{z}_{3}\bar{z}_{4}y_{5}\bar{z}_{6}\bar{x}_{7}y_{8}\bar{x}_{9}\bar{x}_{10}}.
		\end{split}
	\end{equation}
	
	\begin{equation}
		\begin{split}
			|\psi_{4}\rangle &=(\alpha-i\gamma)\ket{x_{1}x_{2}\bar{z}_{3}\bar{z}_{4}y_{5}z_{6}x_{7}\bar{z}_{8}x_{9}x_{10}} +(i\alpha+\gamma)\ket{x_{1}x_{2}\bar{z}_{3}\bar{z}_{4}\bar{y}_{5}\bar{z}_{6}x_{7}\bar{z}_{8}x_{9}x_{10}} \\
			&+i(-i\alpha+\gamma)\ket{x_{1}x_{2}\bar{z}_{3}\bar{z}_{4}\bar{y}_{5}z_{6}\bar{x}_{7}\bar{z}_{8}x_{9}x_{10}}
			+i(-\alpha-i\gamma)\ket{x_{1}x_{2}\bar{z}_{3}\bar{z}_{4}y_{5}\bar{z}_{6}\bar{x}_{7}\bar{z}_{8}x_{9}x_{10}}
			\\
			&-i(\beta-i\delta)\ket{x_{1}x_{2}\bar{z}_{3}\bar{z}_{4}y_{5}z_{6}x_{7}\bar{z}_{8}\bar{x}_{9}\bar{x}_{10}} -i(i\beta+\delta)\ket{x_{1}x_{2}\bar{z}_{3}\bar{z}_{4}\bar{y}_{5}\bar{z}_{6}x_{7}\bar{z}_{8}\bar{x}_{9}\bar{x}_{10}} \\
			&+(i\beta-\delta)\ket{x_{1}x_{2}\bar{z}_{3}\bar{z}_{4}\bar{y}_{5}z_{6}\bar{x}_{7}\bar{z}_{8}\bar{x}_{9}\bar{x}_{10}} +(\beta+i\delta)\ket{x_{1}x_{2}\bar{z}_{3}\bar{z}_{4}y_{5}\bar{z}_{6}\bar{x}_{7}\bar{z}_{8}\bar{x}_{9}\bar{x}_{10}}
			\\
			&+(\eta-i\mu)\ket{\bar{x}_{1}\bar{x}_{2}\bar{z}_{3}\bar{z}_{4}y_{5}z_{6}x_{7}\bar{z}_{8}x_{9}x_{10}}
			+(i\eta+\mu)\ket{\bar{x}_{1}\bar{x}_{2}\bar{z}_{3}\bar{z}_{4}\bar{y}_{5}\bar{z}_{6}x_{7}\bar{z}_{8}x_{9}x_{10}} \\
			&+i(-i\eta+\mu)\ket{\bar{x}_{1}\bar{x}_{2}\bar{z}_{3}\bar{z}_{4}\bar{y}_{5}z_{6}\bar{x}_{7}\bar{z}_{8}x_{9}x_{10}} +i(-\eta-i\mu)\ket{\bar{x}_{1}\bar{x}_{2}\bar{z}_{3}\bar{z}_{4}y_{5}\bar{z}_{6}\bar{x}_{7}\bar{z}_{8}x_{9}x_{10}}
			\\
			&+i(-\kappa+i\nu)(\ket{\bar{x}_{1}\bar{x}_{2}\bar{z}_{3}\bar{z}_{4}y_{5}z_{6}x_{7}\bar{z}_{8}\bar{x}_{9}\bar{x}_{10}} +i(-i\kappa-\nu)\ket{\bar{x}_{1}\bar{x}_{2}\bar{z}_{3}\bar{z}_{4}\bar{y}_{5}\bar{z}_{6}x_{7}\bar{z}_{8}\bar{x}_{9}\bar{x}_{10}} \\
			&+(i\kappa-\nu)\ket{\bar{x}_{1}\bar{x}_{2}\bar{z}_{3}\bar{z}_{4}\bar{y}_{5}z_{6}\bar{x}_{7}\bar{z}_{8}\bar{x}_{9}\bar{x}_{10}} +(\kappa+i\nu)\ket{\bar{x}_{1}\bar{x}_{2}\bar{z}_{3}\bar{z}_{4}y_{5}\bar{z}_{6}\bar{x}_{7}\bar{z}_{8}\bar{x}_{9}\bar{x}_{10}}.
		\end{split}
	\end{equation}
	
	\begin{equation}
		\begin{split}
			|\psi'_{5}\rangle &=(\alpha-i\gamma)\ket{x_{1}x_{2}\bar{z}_{3}\bar{z}_{4}(ix_{5}x_{6}+\bar{x}_{5}\bar{x}_{6})\bar{z}_{7}x_{8}x_{9}x_{10}} +(i\alpha+\gamma)\ket{x_{1}x_{2}\bar{z}_{3}\bar{z}_{4}(x_{5}x_{6}-i\bar{x}_{5}\bar{x}_{6})\bar{z}_{7}x_{8}x_{9}x_{10}} \\
			&+i(i\alpha-\gamma)\ket{x_{1}x_{2}\bar{z}_{3}\bar{z}_{4}(x_{5}x_{6}+i\bar{x}_{5}\bar{x}_{6})\bar{z}_{7}x_{8}x_{9}x_{10}}
			+i(\alpha+i\gamma)\ket{x_{1}x_{2}\bar{z}_{3}\bar{z}_{4}(ix_{5}x_{6}-\bar{x}_{5}\bar{x}_{6})\bar{z}_{7}x_{8}x_{9}x_{10}}
			\\
			&+i(\beta-i\delta)\ket{x_{1}x_{2}\bar{z}_{3}\bar{z}_{4}(ix_{5}x_{6}+\bar{x}_{5}\bar{x}_{6})\bar{z}_{7}\bar{x}_{8}\bar{x}_{9}\bar{x}_{10}} +i(i\beta+\delta)\ket{x_{1}x_{2}\bar{z}_{3}\bar{z}_{4}(x_{5}x_{6}-i\bar{x}_{5}\bar{x}_{6})\bar{z}_{7}\bar{x}_{8}\bar{x}_{9}\bar{x}_{10}} \\
			&+(i\beta-\delta)\ket{x_{1}x_{2}\bar{z}_{3}\bar{z}_{4}(x_{5}x_{6}+i\bar{x}_{5}\bar{x}_{6})\bar{z}_{7}\bar{x}_{8}\bar{x}_{9}\bar{x}_{10}} +(\beta+i\delta)\ket{x_{1}x_{2}\bar{z}_{3}\bar{z}_{4}(ix_{5}x_{6}-\bar{x}_{5}\bar{x}_{6})\bar{z}_{7}\bar{x}_{8}\bar{x}_{9}\bar{x}_{10}}
			\\
			&+(\eta-i\mu)\ket{\bar{x}_{1}\bar{x}_{2}\bar{z}_{3}\bar{z}_{4}(ix_{5}x_{6}+\bar{x}_{5}\bar{x}_{6})\bar{z}_{7}x_{8}x_{9}x_{10}}
			+(i\eta+\mu)\ket{\bar{x}_{1}\bar{x}_{2}\bar{z}_{3}\bar{z}_{4}(x_{5}x_{6}-i\bar{x}_{5}\bar{x}_{6})\bar{z}_{7}x_{8}x_{9}x_{10}} \\
			&+i(i\eta-\mu)\ket{\bar{x}_{1}\bar{x}_{2}\bar{z}_{3}\bar{z}_{4}(x_{5}x_{6}+i\bar{x}_{5}\bar{x}_{6})\bar{z}_{7}x_{8}x_{9}x_{10}} +i(\eta+i\mu)\ket{\bar{x}_{1}\bar{x}_{2}\bar{z}_{3}\bar{z}_{4}(ix_{5}x_{6}-\bar{x}_{5}\bar{x}_{6})\bar{z}_{7}x_{8}x_{9}x_{10}}
			\\
			&+i(\kappa-i\nu)(\ket{\bar{x}_{1}\bar{x}_{2}\bar{z}_{3}\bar{z}_{4}(ix_{5}x_{6}+\bar{x}_{5}\bar{x}_{6})\bar{z}_{7}\bar{x}_{8}\bar{x}_{9}\bar{x}_{10}} +i(i\kappa+\nu)\ket{\bar{x}_{1}\bar{x}_{2}\bar{z}_{3}\bar{z}_{4}(x_{5}x_{6}-i\bar{x}_{5}\bar{x}_{6})\bar{z}_{7}\bar{x}_{8}\bar{x}_{9}\bar{x}_{10}} \\
			&+(i\kappa-\nu)\ket{\bar{x}_{1}\bar{x}_{2}\bar{z}_{3}\bar{z}_{4}(x_{5}x_{6}+i\bar{x}_{5}\bar{x}_{6})\bar{z}_{7}\bar{x}_{8}\bar{x}_{9}\bar{x}_{10}} +(\kappa+i\nu)\ket{\bar{x}_{1}\bar{x}_{2}\bar{z}_{3}\bar{z}_{4}(ix_{5}x_{6}-\bar{x}_{5}\bar{x}_{6})\bar{z}_{7}\bar{x}_{8}\bar{x}_{9}\bar{x}_{10}}
			\\
			&=(\alpha-\gamma)|x_{1}x_{2}\bar{z}_{3}\bar{z}_{4}x_{5}x_{6}\bar{z}_{7}x_{8}x_{9}x_{10}\rangle
			+(\beta+\delta)|x_{1}x_{2}\bar{z}_{3}\bar{z}_{4}x_{5}x_{6}\bar{z}_{7}\bar{x}_{8}\bar{x}_{9}\bar{x}_{10}\rangle \\
			&-(\alpha+\gamma)|x_{1}x_{2}\bar{z}_{3}\bar{z}_{4}\bar{x}_{5}\bar{x}_{6}\bar{z}_{7}x_{8}x_{9}x_{10}\rangle
			+(\beta-\delta)|x_{1}x_{2}\bar{z}_{3}\bar{z}_{4}\bar{x}_{5}\bar{x}_{6}\bar{z}_{7}\bar{x}_{8}\bar{x}_{9}\bar{x}_{10}\rangle \\
			&+(\eta-\mu)|\bar{x}_{1}\bar{x}_{2}\bar{z}_{3}\bar{z}_{4}x_{5}x_{6}\bar{z}_{7}x_{8}x_{9}x_{10}\rangle
			+(\kappa+\nu)|\bar{x}_{1}\bar{x}_{2}\bar{z}_{3}\bar{z}_{4}x_{5}x_{6}\bar{z}_{7}\bar{x}_{8}\bar{x}_{9}\bar{x}_{10}\rangle \\
			&-(\eta+\mu)|\bar{x}_{1}\bar{x}_{2}\bar{z}_{3}\bar{z}_{4}\bar{x}_{5}\bar{x}_{6}\bar{z}_{7}x_{8}x_{9}x_{10}\rangle
			+(\kappa-\nu)|\bar{x}_{1}\bar{x}_{2}\bar{z}_{3}\bar{z}_{4}\bar{x}_{5}\bar{x}_{6}\bar{z}_{7}\bar{x}_{8}\bar{x}_{9}\bar{x}_{10}\rangle.
		\end{split}
	\end{equation}
	
	\begin{equation}
		\begin{split}
			|\psi_{5}\rangle 
			&=(\alpha-\gamma)|x_{1}x_{2}\bar{z}_{3}{x}_{4}x_{5}x_{6}\bar{z}_{7}x_{8}x_{9}x_{10}\rangle
			+(\beta+\delta)|x_{1}x_{2}\bar{z}_{3}{x}_{4}x_{5}x_{6}\bar{z}_{7}\bar{x}_{8}\bar{x}_{9}\bar{x}_{10}\rangle \\
			&+(\alpha+\gamma)|x_{1}x_{2}\bar{z}_{3}\bar{x}_{4}\bar{x}_{5}\bar{x}_{6}\bar{z}_{7}x_{8}x_{9}x_{10}\rangle
			-(\beta-\delta)|x_{1}x_{2}\bar{z}_{3}\bar{x}_{4}\bar{x}_{5}\bar{x}_{6}\bar{z}_{7}\bar{x}_{8}\bar{x}_{9}\bar{x}_{10}\rangle \\
			&+(\eta-\mu)|\bar{x}_{1}\bar{x}_{2}\bar{z}_{3}{x}_{4}x_{5}x_{6}\bar{z}_{7}x_{8}x_{9}x_{10}\rangle
			+(\kappa+\nu)|\bar{x}_{1}\bar{x}_{2}\bar{z}_{3}{x}_{4}x_{5}x_{6}\bar{z}_{7}\bar{x}_{8}\bar{x}_{9}\bar{x}_{10}\rangle \\
			&+(\eta+\mu)|\bar{x}_{1}\bar{x}_{2}\bar{z}_{3}\bar{x}_{4}\bar{x}_{5}\bar{x}_{6}\bar{z}_{7}x_{8}x_{9}x_{10}\rangle
			-(\kappa-\nu)|\bar{x}_{1}\bar{x}_{2}\bar{z}_{3}\bar{x}_{4}\bar{x}_{5}\bar{x}_{6}\bar{z}_{7}\bar{x}_{8}\bar{x}_{9}\bar{x}_{10}\rangle.
		\end{split}
	\end{equation}
	
	And the transformation could be written as
	\begin{equation}
		U_{\sigma_2^{-1}}=\left(
		\begin{array}{cccccccc}
			1\ & 0\ & -1\ & 0\ & 0\ & 0\ & 0\ & 0\\
			0 & 1 & 0 & 1 & 0 & 0 & 0 & 0\\
			1 & 0 & 1 & 0 & 0 & 0 & 0 & 0\\
			0 & -1 & 0 & 1 & 0 & 0 & 0 & 0\\
			0 & 0 & 0 & 0 & 1 & 0 & -1 & 0\\
			0 & 0 & 0 & 0 & 0 & 1 & 0 & 1\\
			0 & 0 & 0 & 0 & 1 & 0 & 1 & 0\\
			0 & 0 & 0 & 0 & 0 & -1 & 0 & 1\\
		\end{array}
		\right)/\sqrt{2}.
	\end{equation}
	With its transformation matrix in the logical basis
	\begin{equation}
		L_{\sigma_2^{-1}}=\left(
		\begin{array}{cccccccc}
			1\ & 0\ & 0\ & -1\ & 0\ & 0\ & 0\ & 0\\
			0 & 1 & -1 & 0 & 0 & 0 & 0 & 0\\
			0 & 1 & 1 & 0 & 0 & 0 & 0 & 0\\
			1 & 0 & 0 & 1 & 0 & 0 & 0 & 0\\
			0 & 0 & 0 & 0 & 1 & 0 & 0 & -1\\
			0 & 0 & 0 & 0 & 0 & 1 & -1 & 0\\
			0 & 0 & 0 & 0 & 0 & 1 & 1 & 0\\
			0 & 0 & 0 & 0 & 1 & 0 & 0 & 1\\
		\end{array}
		\right)/\sqrt{2}.
	\end{equation}
	\clearpage
	
	\section{Spatial modes during the evolution}
	
	In the main text, we take the initial state $\ket{\phi_{0}}$ as an example and briefly give its optical spatial modes as illustration. Here, we demonstrate all the spatial modes and the corresponding two-photon correlations during the braiding operation $\sigma_{1}$ and $\sigma_{2}^{-1}$. 
	\begin{figure}[!h]
		\centering
		\includegraphics[width=1\columnwidth]{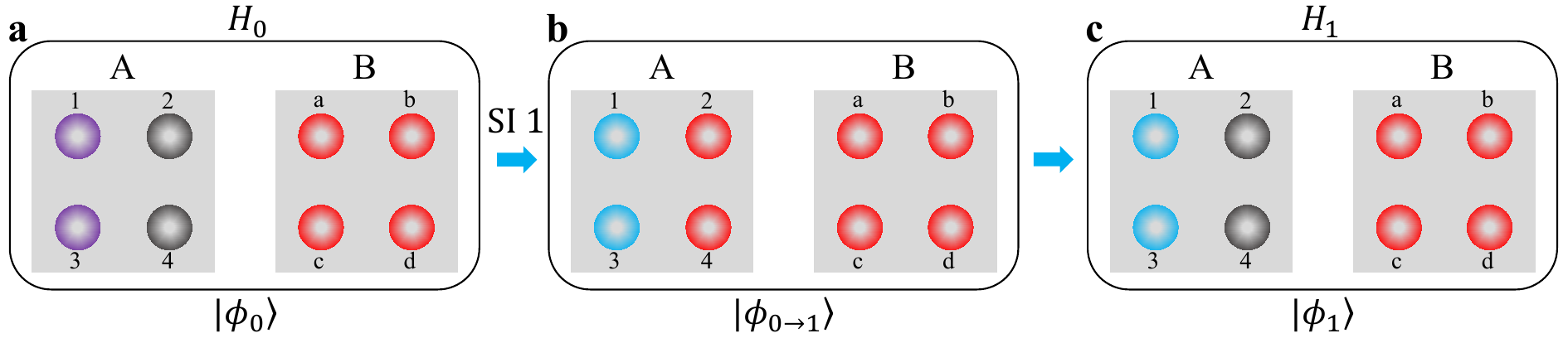}\\
		\small{\begin{flushleft}Fig. S6. The schematic for the role of the Sagnac interferometer during the non-dissipative ITE evolution.\end{flushleft} } 
	\end{figure}  
	\begin{table}[htbp]
		\tiny{TABLE. S1. Optical spatial modes of the state   from $H_{0}$ to $H_{1}$ during the braiding operation $\sigma_{1}$.\\
			\leftline{$\ket{\phi_{0}}$} }
		\begin{ruledtabular}
			\begin{tabular}{ccccc}
				&$\ket{a}_{B}$&$\ket{b}_{B}$&$\ket{c}_{B}$&$\ket{d}_{B}$\\
				\colrule
				$\ket{1}_{A}$&$\alpha|x_{1}x_{2}\bar{z}_{3}x_{4}x_{5}x_{6}\bar{z}_{7}x_{8}x_{9}x_{10}\rangle$&$\beta|x_{1}x_{2}\bar{z}_{3}x_{4}x_{5}x_{6}\bar{z}_{7}\bar{x}_{8}\bar{x}_{9}\bar{x}_{10}\rangle$&$\eta|\bar{x}_{1}\bar{x}_{2}\bar{z}_{3}x_{4}x_{5}x_{6}\bar{z}_{7}x_{8}x_{9}x_{10}\rangle$&$\kappa|\bar{x}_{1}\bar{x}_{2}\bar{z}_{3}x_{4}x_{5}x_{6}\bar{z}_{7}\bar{x}_{8}\bar{x}_{9}\bar{x}_{10}\rangle$\\
				\colrule
				$\ket{3}_{A}$&$\gamma|x_{1}x_{2}\bar{z}_{3}\bar{x}_{4}\bar{x}_{5}\bar{x}_{6}\bar{z}_{7}x_{8}x_{9}x_{10}\rangle$&
				$\delta|x_{1}x_{2}\bar{z}_{3}\bar{x}_{4}\bar{x}_{5}\bar{x}_{6}\bar{z}_{7}\bar{x}_{8}\bar{x}_{9}\bar{x}_{10}\rangle$&$\mu|\bar{x}_{1}\bar{x}_{2}\bar{z}_{3}\bar{x}_{4}\bar{x}_{5}\bar{x}_{6}\bar{z}_{7}x_{8}x_{9}x_{10}\rangle$&$\nu|\bar{x}_{1}\bar{x}_{2}\bar{z}_{3}\bar{x}_{4}\bar{x}_{5}\bar{x}_{6}\bar{z}_{7}\bar{x}_{8}\bar{x}_{9}\bar{x}_{10}\rangle$\\
			\end{tabular}
		\end{ruledtabular}
		
		\leftline{$|\phi_{0\rightarrow 1}\rangle$}
		\begin{ruledtabular}
			\begin{tabular}{ccccc}
				&$\ket{a}_{B}$&$\ket{b}_{B}$&$\ket{c}_{B}$&$\ket{d}_{B}$\\
				\colrule
				$\ket{1}_{A}$&$\alpha|x_{1}x_{2}{x}_{3}{x}_{4}{x}_{5}{x}_{6}\bar{z}_{7}x_{8}x_{9}x_{10}\rangle$&
				$\beta|x_{1}x_{2}{x}_{3}{x}_{4}{x}_{5}{x}_{6}\bar{z}_{7}\bar{x}_{8}\bar{x}_{9}\bar{x}_{10}\rangle$&$-\eta|\bar{x}_{1}\bar{x}_{2}\bar{x}_{3}{x}_{4}{x}_{5}{x}_{6}\bar{z}_{7}x_{8}x_{9}x_{10}\rangle$&$-\kappa|\bar{x}_{1}\bar{x}_{2}\bar{x}_{3}{x}_{4}{x}_{5}{x}_{6}\bar{z}_{7}\bar{x}_{8}\bar{x}_{9}\bar{x}_{10}\rangle$\\
				\colrule
				$\ket{2}_{A}$&$-\alpha|x_{1}x_{2}\bar{x}_{3}{x}_{4}{x}_{5}{x}_{6}\bar{z}_{7}x_{8}x_{9}x_{10}\rangle$&
				$-\beta|x_{1}x_{2}\bar{x}_{3}{x}_{4}{x}_{5}{x}_{6}\bar{z}_{7}\bar{x}_{8}\bar{x}_{9}\bar{x}_{10}\rangle$&$\eta|\bar{x}_{1}\bar{x}_{2}{x}_{3}{x}_{4}{x}_{5}{x}_{6}\bar{z}_{7}x_{8}x_{9}x_{10}\rangle$&$\kappa|\bar{x}_{1}\bar{x}_{2}{x}_{3}{x}_{4}{x}_{5}{x}_{6}\bar{z}_{7}\bar{x}_{8}\bar{x}_{9}\bar{x}_{10}\rangle$\\
				\colrule
				$\ket{3}_{A}$&$\gamma|x_{1}x_{2}{x}_{3}\bar{x}_{4}\bar{x}_{5}\bar{x}_{6}\bar{z}_{7}x_{8}x_{9}x_{10}\rangle$&
				$\delta|x_{1}x_{2}{x}_{3}\bar{x}_{4}\bar{x}_{5}\bar{x}_{6}\bar{z}_{7}\bar{x}_{8}\bar{x}_{9}\bar{x}_{10}\rangle$&$-\mu|\bar{x}_{1}\bar{x}_{2}\bar{x}_{3}\bar{x}_{4}\bar{x}_{5}\bar{x}_{6}\bar{z}_{7}x_{8}x_{9}x_{10}\rangle$&$-\nu|\bar{x}_{1}\bar{x}_{2}\bar{x}_{3}\bar{x}_{4}\bar{x}_{5}\bar{x}_{6}\bar{z}_{7}\bar{x}_{8}\bar{x}_{9}\bar{x}_{10}\rangle$\\
				\colrule
				$\ket{4}_{A}$&$-\gamma|x_{1}x_{2}\bar{x}_{3}\bar{x}_{4}\bar{x}_{5}\bar{x}_{6}\bar{z}_{7}x_{8}x_{9}x_{10}\rangle$&
				$-\delta|x_{1}x_{2}\bar{x}_{3}\bar{x}_{4}\bar{x}_{5}\bar{x}_{6}\bar{z}_{7}\bar{x}_{8}\bar{x}_{9}\bar{x}_{10}\rangle$&$\mu|\bar{x}_{1}\bar{x}_{2}{x}_{3}\bar{x}_{4}\bar{x}_{5}\bar{x}_{6}\bar{z}_{7}x_{8}x_{9}x_{10}\rangle$&$
				\nu|\bar{x}_{1}\bar{x}_{2}{x}_{3}\bar{x}_{4}\bar{x}_{5}\bar{x}_{6}\bar{z}_{7}\bar{x}_{8}\bar{x}_{9}\bar{x}_{10}\rangle$\\
			\end{tabular}
		\end{ruledtabular}
		
		\leftline{$|\phi_{1}\rangle$}
		\begin{ruledtabular}
			\begin{tabular}{ccccc}
				&$\ket{a}_{B}$&$\ket{b}_{B}$&$\ket{c}_{B}$&$\ket{d}_{B}$\\
				\colrule
				$\ket{1}_{A}$&$\alpha|x_{1}x_{2}{x}_{3}x_{4}x_{5}x_{6}\bar{z}_{7}x_{8}x_{9}x_{10}\rangle$&$\beta|x_{1}x_{2}{x}_{3}x_{4}x_{5}x_{6}\bar{z}_{7}\bar{x}_{8}\bar{x}_{9}\bar{x}_{10}\rangle$&$-\eta|\bar{x}_{1}\bar{x}_{2}\bar{x}_{3}x_{4}x_{5}x_{6}\bar{z}_{7}x_{8}x_{9}x_{10}\rangle$&$-\kappa|\bar{x}_{1}\bar{x}_{2}\bar{x}_{3}x_{4}x_{5}x_{6}\bar{z}_{7}\bar{x}_{8}\bar{x}_{9}\bar{x}_{10}\rangle$\\
				\colrule
				$\ket{3}_{A}$&$\gamma|x_{1}x_{2}{x}_{3}\bar{x}_{4}\bar{x}_{5}\bar{x}_{6}\bar{z}_{7}x_{8}x_{9}x_{10}\rangle$&
				$\delta|x_{1}x_{2}{x}_{3}\bar{x}_{4}\bar{x}_{5}\bar{x}_{6}\bar{z}_{7}\bar{x}_{8}\bar{x}_{9}\bar{x}_{10}\rangle$&$-\mu|\bar{x}_{1}\bar{x}_{2}\bar{x}_{3}\bar{x}_{4}\bar{x}_{5}\bar{x}_{6}\bar{z}_{7}x_{8}x_{9}x_{10}\rangle$&$-\nu|\bar{x}_{1}\bar{x}_{2}\bar{x}_{3}\bar{x}_{4}\bar{x}_{5}\bar{x}_{6}\bar{z}_{7}\bar{x}_{8}\bar{x}_{9}\bar{x}_{10}\rangle$\\
			\end{tabular}
		\end{ruledtabular}
		
	\end{table}   
	First of all, we illustrate the role of the SIs in implementing the non-dissipative ITE evolution. To show this process clearly, we take the evolution process from $H_{0}$ to $H_{1}$ in braiding operation $\sigma_{1}$ as an example. 
	After the ITE of the initial Hamiltonian
	\begin{equation}
		H_{0} =-\sigma_{1}^{x}\sigma_{2}^{x}-\sigma_{4}^{x}\sigma_{5}^{x}-\sigma_{5}^{x}\sigma_{6}^{x} -\sigma_{8}^{x}\sigma_{9}^{x}-\sigma_{9}^{x}\sigma_{10}^{x}+\sigma_{3}^{z}+\sigma_{7}^{z},
	\end{equation}
	which corresponds to creating the MZMs A, B, C, D, E and F, the state becomes
	\begin{equation}
		\begin{split}
			|\phi_{0}\rangle &=\alpha|x_{1}x_{2}\bar{z}_{3}x_{4}x_{5}x_{6}\bar{z}_{7}x_{8}x_{9}x_{10}\rangle
			+\beta|x_{1}x_{2}\bar{z}_{3}x_{4}x_{5}x_{6}\bar{z}_{7}\bar{x}_{8}\bar{x}_{9}\bar{x}_{10}\rangle \\
			&+\gamma|x_{1}x_{2}\bar{z}_{3}\bar{x}_{4}\bar{x}_{5}\bar{x}_{6}\bar{z}_{7}x_{8}x_{9}x_{10}\rangle
			+\delta|x_{1}x_{2}\bar{z}_{3}\bar{x}_{4}\bar{x}_{5}\bar{x}_{6}\bar{z}_{7}\bar{x}_{8}\bar{x}_{9}\bar{x}_{10}\rangle \\
			&+\eta|\bar{x}_{1}\bar{x}_{2}\bar{z}_{3}x_{4}x_{5}x_{6}\bar{z}_{7}x_{8}x_{9}x_{10}\rangle
			+\kappa|\bar{x}_{1}\bar{x}_{2}\bar{z}_{3}x_{4}x_{5}x_{6}\bar{z}_{7}\bar{x}_{8}\bar{x}_{9}\bar{x}_{10}\rangle \\
			&+\mu|\bar{x}_{1}\bar{x}_{2}\bar{z}_{3}\bar{x}_{4}\bar{x}_{5}\bar{x}_{6}\bar{z}_{7}x_{8}x_{9}x_{10}\rangle
			+\nu|\bar{x}_{1}\bar{x}_{2}\bar{z}_{3}\bar{x}_{4}\bar{x}_{5}\bar{x}_{6}\bar{z}_{7}\bar{x}_{8}\bar{x}_{9}\bar{x}_{10}\rangle,
			\label{initial}
		\end{split}
	\end{equation}
	where $\alpha$, $\beta$, $\gamma$, $\delta$, $\eta$, $\kappa$, $\mu$ and $\nu$ are complex amplitudes satisfying $|\alpha|^{2}+|\beta|^{2}+|\gamma|^{2}+|\delta|^{2}+|\eta|^{2}+|\kappa|^{2}+|\mu|^{2}+|\nu|^{2}=1$.

	For the ITE of $H_{1}$, we first transfer the eigenstates of $\sigma_{3}^{z}$ to that of $\sigma_{3}^{x}$, and the initial state becomes
	\begin{equation}
		\begin{split}
			|\phi_{0\rightarrow 1}\rangle &=\alpha|x_{1}x_{2}(x_{3}-\bar{x}_{3})x_{4}x_{5}x_{6}\bar{z}_{7}x_{8}x_{9}x_{10}\rangle
			+\beta|x_{1}x_{2}(x_{3}-\bar{x}_{3})x_{4}x_{5}x_{6}\bar{z}_{7}\bar{x}_{8}\bar{x}_{9}\bar{x}_{10}\rangle \\
			&+\gamma|x_{1}x_{2}(x_{3}-\bar{x}_{3})\bar{x}_{4}\bar{x}_{5}\bar{x}_{6}\bar{z}_{7}x_{8}x_{9}x_{10}\rangle
			+\delta|x_{1}x_{2}(x_{3}-\bar{x}_{3})\bar{x}_{4}\bar{x}_{5}\bar{x}_{6}\bar{z}_{7}\bar{x}_{8}\bar{x}_{9}\bar{x}_{10}\rangle \\
			&+\eta|\bar{x}_{1}\bar{x}_{2}(x_{3}-\bar{x}_{3})x_{4}x_{5}x_{6}\bar{z}_{7}x_{8}x_{9}x_{10}\rangle
			+\kappa|\bar{x}_{1}\bar{x}_{2}(x_{3}-\bar{x}_{3})x_{4}x_{5}x_{6}\bar{z}_{7}\bar{x}_{8}\bar{x}_{9}\bar{x}_{10}\rangle \\
			&+\mu|\bar{x}_{1}\bar{x}_{2}(x_{3}-\bar{x}_{3})\bar{x}_{4}\bar{x}_{5}\bar{x}_{6}\bar{z}_{7}x_{8}x_{9}x_{10}\rangle
			+\nu|\bar{x}_{1}\bar{x}_{2}(x_{3}-\bar{x}_{3})\bar{x}_{4}\bar{x}_{5}\bar{x}_{6}\bar{z}_{7}\bar{x}_{8}\bar{x}_{9}\bar{x}_{10}\rangle,
		\end{split}
	\end{equation}
	
	which corresponds to the optical modes in Fig. S6a. And then the state is sent to the SI to perform the non-dissipative ITE. The optical modes in the SI 1 are shown in the Fig. S6b, where the correlation between blue modes (red modes) located on Side A and the modes on Side B represent the ground (excited) states.
	
	We then implement the ITE of $-\sigma_{2}^{x}\sigma_{3}^{x}$. The mode 2 and mode 4 recombine at the output of the SI (as shown in Fig. S6c), which indicating that the excited states are cooled back to the ground states. Thus the state becomes
	\begin{equation}
		\begin{split}
			|\phi_{1}\rangle
			&=\alpha|x_{1}x_{2}x_{3}x_{4}x_{5}x_{6}\bar{z}_{7}x_{8}x_{9}x_{10}\rangle
			+\beta|x_{1}x_{2}x_{3}x_{4}x_{5}x_{6}\bar{z}_{7}\bar{x}_{8}\bar{x}_{9}\bar{x}_{10}\rangle \\
			&+\gamma|x_{1}x_{2}x_{3}\bar{x}_{4}\bar{x}_{5}\bar{x}_{6}\bar{z}_{7}x_{8}x_{9}x_{10}\rangle
			+\delta|x_{1}x_{2}x_{3}\bar{x}_{4}\bar{x}_{5}\bar{x}_{6}\bar{z}_{7}\bar{x}_{8}\bar{x}_{9}\bar{x}_{10}\rangle \\
			&-\eta|\bar{x}_{1}\bar{x}_{2}\bar{x}_{3}x_{4}x_{5}x_{6}\bar{z}_{7}x_{8}x_{9}x_{10}\rangle
			-\kappa|\bar{x}_{1}\bar{x}_{2}\bar{x}_{3}x_{4}x_{5}x_{6}\bar{z}_{7}\bar{x}_{8}\bar{x}_{9}\bar{x}_{10}\rangle \\
			&-\mu|\bar{x}_{1}\bar{x}_{2}\bar{x}_{3}\bar{x}_{4}\bar{x}_{5}\bar{x}_{6}\bar{z}_{7}x_{8}x_{9}x_{10}\rangle
			-\nu|\bar{x}_{1}\bar{x}_{2}\bar{x}_{3}\bar{x}_{4}\bar{x}_{5}\bar{x}_{6}\bar{z}_{7}\bar{x}_{8}\bar{x}_{9}\bar{x}_{10}\rangle.
		\end{split}
	\end{equation}     
	
	Till now, we achieve the non-dissipative ITE through the introduction of the Sagnac interferometer. The encoding during this process is shown in Table. S1. All the other non-dissipative ITE evolutions in the whole braiding operation process are similar to this.
	
	The optical spatial modes and the corresponding two-photon correlations during the braiding operation $\sigma_{1}$ are shown in Fig. S7 and Table. S2, respectively. The red dots represent the presence of the optical modes.  
	
	The optical spatial modes and the corresponding two-photon correlations during the braiding operation $\sigma_{2}^{-1}$ are shown in Fig. S8 and Table. S3, respectively. Unlike the situation in braiding operation $\sigma_{1}$, several beam displacers (BD30/BD60) are used during the evolution process. We mark the areas where beam displacers are used to achieve the quantum state evolution in Fig. S8.    
	\begin{figure}[htbp]
		\centering
		\includegraphics[width=1\columnwidth]{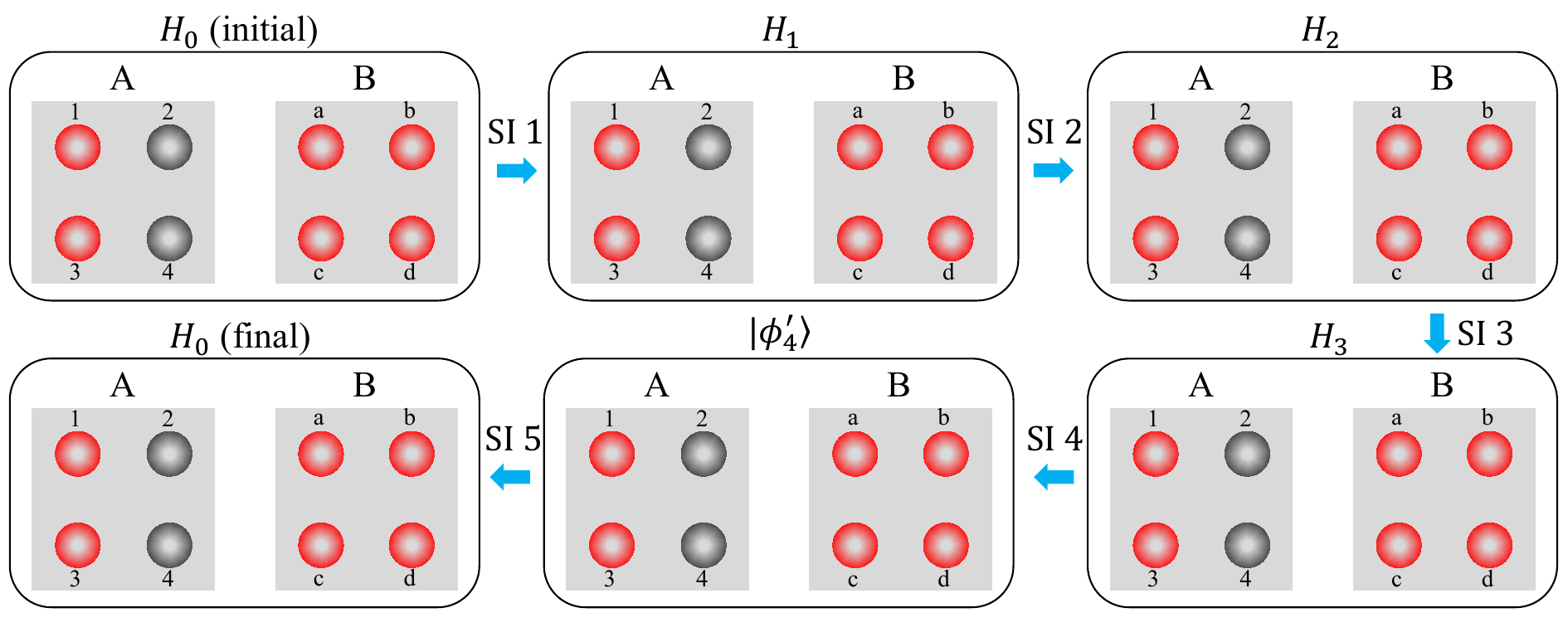}
		\small{Fig. S7. Optical spatial modes of the states during the braiding operation $\sigma_{1}$.}
	\end{figure}
	
	\begin{table}[htbp]
		\tiny{TABLE. S2. Optical spatial modes of the states during the braiding operation $\sigma_{1}$.\\
			\leftline{$H_{0}$} }
		\begin{ruledtabular}
			\begin{tabular}{ccccc}
				&$\ket{a}_{B}$&$\ket{b}_{B}$&$\ket{c}_{B}$&$\ket{d}_{B}$\\
				\colrule
				$\ket{1}_{A}$&$\alpha|x_{1}x_{2}\bar{z}_{3}x_{4}x_{5}x_{6}\bar{z}_{7}x_{8}x_{9}x_{10}\rangle$&$\beta|x_{1}x_{2}\bar{z}_{3}x_{4}x_{5}x_{6}\bar{z}_{7}\bar{x}_{8}\bar{x}_{9}\bar{x}_{10}\rangle$&$\eta|\bar{x}_{1}\bar{x}_{2}\bar{z}_{3}x_{4}x_{5}x_{6}\bar{z}_{7}x_{8}x_{9}x_{10}\rangle$&$\kappa|\bar{x}_{1}\bar{x}_{2}\bar{z}_{3}x_{4}x_{5}x_{6}\bar{z}_{7}\bar{x}_{8}\bar{x}_{9}\bar{x}_{10}\rangle$\\
				\colrule
				$\ket{2}_{A}$&$\gamma|x_{1}x_{2}\bar{z}_{3}\bar{x}_{4}\bar{x}_{5}\bar{x}_{6}\bar{z}_{7}x_{8}x_{9}x_{10}\rangle$&
				$\delta|x_{1}x_{2}\bar{z}_{3}\bar{x}_{4}\bar{x}_{5}\bar{x}_{6}\bar{z}_{7}\bar{x}_{8}\bar{x}_{9}\bar{x}_{10}\rangle$&$\mu|\bar{x}_{1}\bar{x}_{2}\bar{z}_{3}\bar{x}_{4}\bar{x}_{5}\bar{x}_{6}\bar{z}_{7}x_{8}x_{9}x_{10}\rangle$&$\nu|\bar{x}_{1}\bar{x}_{2}\bar{z}_{3}\bar{x}_{4}\bar{x}_{5}\bar{x}_{6}\bar{z}_{7}\bar{x}_{8}\bar{x}_{9}\bar{x}_{10}\rangle$\\
			\end{tabular}
		\end{ruledtabular}
		
		\tiny{\leftline{$H_{1}$}}
		\begin{ruledtabular}
			\begin{tabular}{ccccc}
				&$\ket{a}_{B}$&$\ket{b}_{B}$&$\ket{c}_{B}$&$\ket{d}_{B}$\\
				\colrule
				$\ket{1}_{A}$&$\alpha|x_{1}x_{2}{x}_{3}x_{4}x_{5}x_{6}\bar{z}_{7}x_{8}x_{9}x_{10}\rangle$&$\beta|x_{1}x_{2}{x}_{3}x_{4}x_{5}x_{6}\bar{z}_{7}\bar{x}_{8}\bar{x}_{9}\bar{x}_{10}\rangle$&$-\eta|\bar{x}_{1}\bar{x}_{2}\bar{x}_{3}x_{4}x_{5}x_{6}\bar{z}_{7}x_{8}x_{9}x_{10}\rangle$&$-\kappa|\bar{x}_{1}\bar{x}_{2}\bar{x}_{3}x_{4}x_{5}x_{6}\bar{z}_{7}\bar{x}_{8}\bar{x}_{9}\bar{x}_{10}\rangle$\\
				\colrule
				$\ket{2}_{A}$&$\gamma|x_{1}x_{2}{x}_{3}\bar{x}_{4}\bar{x}_{5}\bar{x}_{6}\bar{z}_{7}x_{8}x_{9}x_{10}\rangle$&
				$\delta|x_{1}x_{2}{x}_{3}\bar{x}_{4}\bar{x}_{5}\bar{x}_{6}\bar{z}_{7}\bar{x}_{8}\bar{x}_{9}\bar{x}_{10}\rangle$&$-\mu|\bar{x}_{1}\bar{x}_{2}\bar{x}_{3}\bar{x}_{4}\bar{x}_{5}\bar{x}_{6}\bar{z}_{7}x_{8}x_{9}x_{10}\rangle$&$-\nu|\bar{x}_{1}\bar{x}_{2}\bar{x}_{3}\bar{x}_{4}\bar{x}_{5}\bar{x}_{6}\bar{z}_{7}\bar{x}_{8}\bar{x}_{9}\bar{x}_{10}\rangle$\\
			\end{tabular}
		\end{ruledtabular}
		
		\tiny{\leftline{$H_{2}$}}
		\begin{ruledtabular}
			\begin{tabular}{ccccc}
				&$\ket{a}_{B}$&$\ket{b}_{B}$&$\ket{c}_{B}$&$\ket{d}_{B}$\\
				\colrule
				$\ket{1}_{A}$&$i\alpha|x_{1}x_{2}{x}_{3}\bar{y}_{4}x_{5}x_{6}\bar{z}_{7}x_{8}x_{9}x_{10}\rangle$&$i\beta|x_{1}x_{2}{x}_{3}\bar{y}_{4}x_{5}x_{6}\bar{z}_{7}\bar{x}_{8}\bar{x}_{9}\bar{x}_{10}\rangle$&$-\eta|\bar{x}_{1}\bar{x}_{2}\bar{x}_{3}y_{4}x_{5}x_{6}\bar{z}_{7}x_{8}x_{9}x_{10}\rangle$&$-\kappa|\bar{x}_{1}\bar{x}_{2}\bar{x}_{3}y_{4}x_{5}x_{6}\bar{z}_{7}\bar{x}_{8}\bar{x}_{9}\bar{x}_{10}\rangle$\\
				\colrule
				$\ket{2}_{A}$&$\gamma|x_{1}x_{2}{x}_{3}\bar{y}_{4}\bar{x}_{5}\bar{x}_{6}\bar{z}_{7}x_{8}x_{9}x_{10}\rangle$&
				$\delta|x_{1}x_{2}{x}_{3}\bar{y}_{4}\bar{x}_{5}\bar{x}_{6}\bar{z}_{7}\bar{x}_{8}\bar{x}_{9}\bar{x}_{10}\rangle$&$-i\mu|\bar{x}_{1}\bar{x}_{2}\bar{x}_{3}{y}_{4}\bar{x}_{5}\bar{x}_{6}\bar{z}_{7}x_{8}x_{9}x_{10}\rangle$&$-i\nu|\bar{x}_{1}\bar{x}_{2}\bar{x}_{3}{y}_{4}\bar{x}_{5}\bar{x}_{6}\bar{z}_{7}\bar{x}_{8}\bar{x}_{9}\bar{x}_{10}\rangle$\\
			\end{tabular}
		\end{ruledtabular}
		
		\tiny{\leftline{$H_{3}$}}
		\begin{ruledtabular}
			\begin{tabular}{ccccc}
				&$\ket{a}_{B}$&$\ket{b}_{B}$&$\ket{c}_{B}$&$\ket{d}_{B}$\\
				\colrule
				$\ket{1}_{A}$&$\alpha|x_{1}x_{2}{x}_{3}\bar{z}_{4}x_{5}x_{6}\bar{z}_{7}x_{8}x_{9}x_{10}\rangle$&$\beta|x_{1}x_{2}{x}_{3}\bar{z}_{4}x_{5}x_{6}\bar{z}_{7}\bar{x}_{8}\bar{x}_{9}\bar{x}_{10}\rangle$&$-i\eta|\bar{x}_{1}\bar{x}_{2}\bar{x}_{3}\bar{z}_{4}x_{5}x_{6}\bar{z}_{7}x_{8}x_{9}x_{10}\rangle$&$-i\kappa|\bar{x}_{1}\bar{x}_{2}\bar{x}_{3}\bar{z}_{4}x_{5}x_{6}\bar{z}_{7}\bar{x}_{8}\bar{x}_{9}\bar{x}_{10}\rangle$\\
				\colrule
				$\ket{2}_{A}$&$-i\gamma|x_{1}x_{2}{x}_{3}\bar{z}_{4}\bar{x}_{5}\bar{x}_{6}\bar{z}_{7}x_{8}x_{9}x_{10}\rangle$&$-i\delta|x_{1}x_{2}{x}_{3}\bar{z}_{4}\bar{x}_{5}\bar{x}_{6}\bar{z}_{7}\bar{x}_{8}\bar{x}_{9}\bar{x}_{10}\rangle$&$\mu|\bar{x}_{1}\bar{x}_{2}\bar{x}_{3}\bar{z}_{4}\bar{x}_{5}\bar{x}_{6}\bar{z}_{7}x_{8}x_{9}x_{10}\rangle$&$\nu|\bar{x}_{1}\bar{x}_{2}\bar{x}_{3}\bar{z}_{4}\bar{x}_{5}\bar{x}_{6}\bar{z}_{7}\bar{x}_{8}\bar{x}_{9}\bar{x}_{10}\rangle$\\
			\end{tabular}
		\end{ruledtabular}
		
		\tiny{\leftline{$H_{0}$}}
		\begin{ruledtabular}
			\begin{tabular}{ccccc}
				&$\ket{a}_{B}$&$\ket{b}_{B}$&$\ket{c}_{B}$&$\ket{d}_{B}$\\
				\colrule
				$\ket{1}_{A}$&$\alpha|x_{1}x_{2}\bar{z}_{3}{x}_{4}x_{5}x_{6}\bar{z}_{7}x_{8}x_{9}x_{10}\rangle$&$\beta|x_{1}x_{2}\bar{z}_{3}{x}_{4}x_{5}x_{6}\bar{z}_{7}\bar{x}_{8}\bar{x}_{9}\bar{x}_{10}\rangle$&$i\eta|\bar{x}_{1}\bar{x}_{2}\bar{z}_{3}{x}_{4}x_{5}x_{6}\bar{z}_{7}x_{8}x_{9}x_{10}\rangle$&$i\kappa|\bar{x}_{1}\bar{x}_{2}\bar{z}_{3}{x}_{4}x_{5}x_{6}\bar{z}_{7}\bar{x}_{8}\bar{x}_{9}\bar{x}_{10}\rangle$\\
				\colrule
				$\ket{2}_{A}$&$i\gamma|x_{1}x_{2}\bar{z}_{3}\bar{x}_{4}\bar{x}_{5}\bar{x}_{6}\bar{z}_{7}x_{8}x_{9}x_{10}\rangle$&$i\delta|x_{1}x_{2}\bar{z}_{3}\bar{x}_{4}\bar{x}_{5}\bar{x}_{6}\bar{z}_{7}\bar{x}_{8}\bar{x}_{9}\bar{x}_{10}\rangle$&$\mu|\bar{x}_{1}\bar{x}_{2}\bar{z}_{3}\bar{x}_{4}\bar{x}_{5}\bar{x}_{6}\bar{z}_{7}x_{8}x_{9}x_{10}\rangle$&$\nu|\bar{x}_{1}\bar{x}_{2}\bar{z}_{3}\bar{x}_{4}\bar{x}_{5}\bar{x}_{6}\bar{z}_{7}\bar{x}_{8}\bar{x}_{9}\bar{x}_{10}\rangle$\\
			\end{tabular}
		\end{ruledtabular}
	\end{table}

	\begin{figure}[htbp]
		\centering
		\includegraphics[width=1\columnwidth]{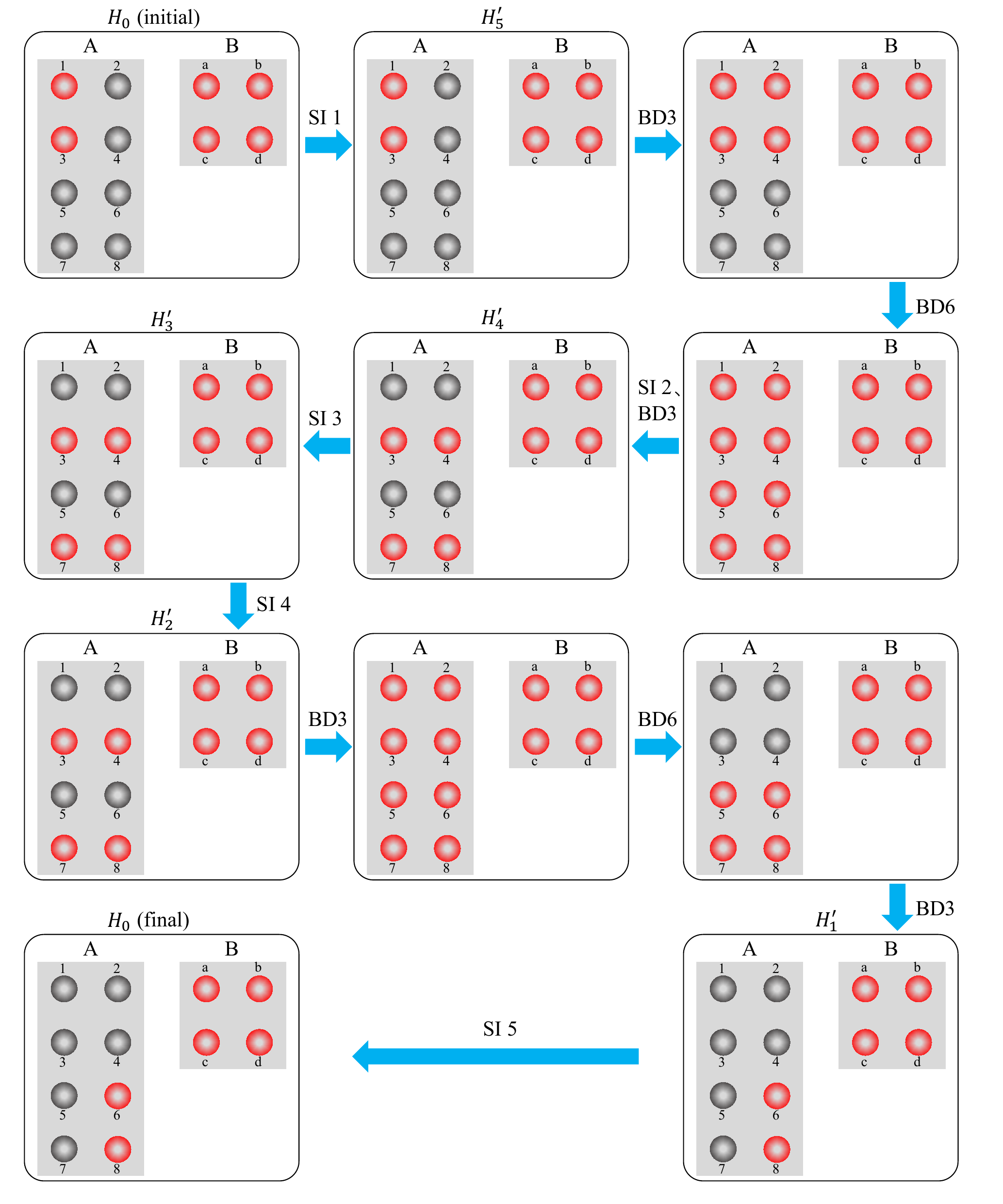}
		\small{\begin{flushleft}Fig. S8. Optical spatial modes of the states during the braiding operation $\sigma_{2}^{-1}$. BD30 and BD60 are the beam displacers with beam displacement of 3.0 mm and 6.0 mm, respectively.\end{flushleft}}
	\end{figure}
	\newpage
	
	\begin{sidewaystable}[htbp]
		\tiny{TABLE. S3. Optical spatial modes of the states during the braiding operation $\sigma_{2}^{-1}$.\\
			\leftline{$H_{0}$} }
		\begin{ruledtabular}
			\begin{tabular}{ccccc}
				&$\ket{a}_{B}$&$\ket{b}_{B}$&$\ket{c}_{B}$&$\ket{d}_{B}$\\
				\colrule
				$\ket{1}_{A}$&$\alpha|x_{1}x_{2}\bar{z}_{3}x_{4}x_{5}x_{6}\bar{z}_{7}x_{8}x_{9}x_{10}\rangle$&$\beta|x_{1}x_{2}\bar{z}_{3}x_{4}x_{5}x_{6}\bar{z}_{7}\bar{x}_{8}\bar{x}_{9}\bar{x}_{10}\rangle$&$\eta|\bar{x}_{1}\bar{x}_{2}\bar{z}_{3}x_{4}x_{5}x_{6}\bar{z}_{7}x_{8}x_{9}x_{10}\rangle$&$\kappa|\bar{x}_{1}\bar{x}_{2}\bar{z}_{3}x_{4}x_{5}x_{6}\bar{z}_{7}\bar{x}_{8}\bar{x}_{9}\bar{x}_{10}\rangle$\\
				\colrule
				$\ket{3}_{A}$&$\gamma|x_{1}x_{2}\bar{z}_{3}\bar{x}_{4}\bar{x}_{5}\bar{x}_{6}\bar{z}_{7}x_{8}x_{9}x_{10}\rangle$&
				$\delta|x_{1}x_{2}\bar{z}_{3}\bar{x}_{4}\bar{x}_{5}\bar{x}_{6}\bar{z}_{7}\bar{x}_{8}\bar{x}_{9}\bar{x}_{10}\rangle$&$\mu|\bar{x}_{1}\bar{x}_{2}\bar{z}_{3}\bar{x}_{4}\bar{x}_{5}\bar{x}_{6}\bar{z}_{7}x_{8}x_{9}x_{10}\rangle$&$\nu|\bar{x}_{1}\bar{x}_{2}\bar{z}_{3}\bar{x}_{4}\bar{x}_{5}\bar{x}_{6}\bar{z}_{7}\bar{x}_{8}\bar{x}_{9}\bar{x}_{10}\rangle$\\
			\end{tabular}
		\end{ruledtabular}
		
		{\leftline{$H_{5}$}}
		\begin{ruledtabular}
			\begin{tabular}{ccccc}
				&$\ket{a}_{B}$&$\ket{b}_{B}$&$\ket{c}_{B}$&$\ket{d}_{B}$\\
				\colrule
				$\ket{1}_{A}$&$\alpha|x_{1}x_{2}\bar{z}_{3}\bar{z}_{4}x_{5}x_{6}\bar{z}_{7}x_{8}x_{9}x_{10}\rangle$&$\beta|x_{1}x_{2}\bar{z}_{3}\bar{z}_{4}x_{5}x_{6}\bar{z}_{7}\bar{x}_{8}\bar{x}_{9}\bar{x}_{10}\rangle$&$\eta|\bar{x}_{1}\bar{x}_{2}\bar{z}_{3}\bar{z}_{4}x_{5}x_{6}\bar{z}_{7}x_{8}x_{9}x_{10}\rangle$&$\kappa|\bar{x}_{1}\bar{x}_{2}\bar{z}_{3}\bar{z}_{4}x_{5}x_{6}\bar{z}_{7}\bar{x}_{8}\bar{x}_{9}\bar{x}_{10}\rangle$\\
				\colrule
				$\ket{3}_{A}$&$-\gamma|x_{1}x_{2}\bar{z}_{3}\bar{z}_{4}\bar{x}_{5}\bar{x}_{6}\bar{z}_{7}x_{8}x_{9}x_{10}\rangle$&
				$-\delta|x_{1}x_{2}\bar{z}_{3}\bar{z}_{4}\bar{x}_{5}\bar{x}_{6}\bar{z}_{7}\bar{x}_{8}\bar{x}_{9}\bar{x}_{10}\rangle$&$-\mu|\bar{x}_{1}\bar{x}_{2}\bar{z}_{3}\bar{z}_{4}\bar{x}_{5}\bar{x}_{6}\bar{z}_{7}x_{8}x_{9}x_{10}\rangle$&$-\nu|\bar{x}_{1}\bar{x}_{2}\bar{z}_{3}\bar{z}_{4}\bar{x}_{5}\bar{x}_{6}\bar{z}_{7}\bar{x}_{8}\bar{x}_{9}\bar{x}_{10}\rangle$\\
			\end{tabular}
		\end{ruledtabular}
		
		{\leftline{$H_{4}$}}
		\begin{ruledtabular}
			\begin{tabular}{ccccc}
				&$\ket{a}_{B}$&$\ket{b}_{B}$&$\ket{c}_{B}$&$\ket{d}_{B}$\\
				\colrule
				$\ket{3}_{A}$&$(\alpha-i\gamma)|x_{1}x_{2}\bar{z}_{3}\bar{z}_{4}y_{5}z_{6}{x}_{7}x_{8}x_{9}x_{10}\rangle$&$(\beta-i\delta)|x_{1}x_{2}\bar{z}_{3}\bar{z}_{4}y_{5}z_{6}{x}_{7}\bar{x}_{8}\bar{x}_{9}\bar{x}_{10}\rangle$&$(\eta-i\mu)|\bar{x}_{1}\bar{x}_{2}\bar{z}_{3}\bar{z}_{4}y_{5}z_{6}{x}_{7}x_{8}x_{9}x_{10}\rangle$&$(\kappa-i\nu)|\bar{x}_{1}\bar{x}_{2}\bar{z}_{3}\bar{z}_{4}y_{5}z_{6}{x}_{7}\bar{x}_{8}\bar{x}_{9}\bar{x}_{10}\rangle$\\
				\colrule
				$\ket{4}_{A}$&$(i\alpha+\gamma)|x_{1}x_{2}\bar{z}_{3}\bar{z}_{4}\bar{y}_{5}\bar{z}_{6}{x}_{7}x_{8}x_{9}x_{10}\rangle$&$(i\beta+\delta)|x_{1}x_{2}\bar{z}_{3}\bar{z}_{4}\bar{y}_{5}\bar{z}_{6}{x}_{7}\bar{x}_{8}\bar{x}_{9}\bar{x}_{10}\rangle$&$(i\eta+\mu)|\bar{x}_{1}\bar{x}_{2}\bar{z}_{3}\bar{z}_{4}\bar{y}_{5}\bar{z}_{6}{x}_{7}x_{8}x_{9}x_{10}\rangle$&$(i\kappa+\nu)|\bar{x}_{1}\bar{x}_{2}\bar{z}_{3}\bar{z}_{4}\bar{y}_{5}\bar{z}_{6}{x}_{7}\bar{x}_{8}\bar{x}_{9}\bar{x}_{10}\rangle$\\
				\colrule
				$\ket{7}_{A}$&$(-i\alpha+\gamma)|x_{1}x_{2}\bar{z}_{3}\bar{z}_{4}\bar{y}_{5}{z}_{6}\bar{x}_{7}x_{8}x_{9}x_{10}\rangle$&$(-i\beta+\delta)|x_{1}x_{2}\bar{z}_{3}\bar{z}_{4}\bar{y}_{5}{z}_{6}\bar{x}_{7}\bar{x}_{8}\bar{x}_{9}\bar{x}_{10}\rangle$&$(-i\eta+\mu)|\bar{x}_{1}\bar{x}_{2}\bar{z}_{3}\bar{z}_{4}\bar{y}_{5}{z}_{6}\bar{x}_{7}x_{8}x_{9}x_{10}\rangle$&$(-i\kappa+\nu)|\bar{x}_{1}\bar{x}_{2}\bar{z}_{3}\bar{z}_{4}\bar{y}_{5}{z}_{6}\bar{x}_{7}\bar{x}_{8}\bar{x}_{9}\bar{x}_{10}\rangle$\\
				\colrule
				$\ket{8}_{A}$&$(-\alpha-i\gamma)|x_{1}x_{2}\bar{z}_{3}\bar{z}_{4}{y}_{5}\bar{z}_{6}\bar{x}_{7}x_{8}x_{9}x_{10}\rangle$&$(-\beta-i\delta)|x_{1}x_{2}\bar{z}_{3}\bar{z}_{4}{y}_{5}\bar{z}_{6}\bar{x}_{7}\bar{x}_{8}\bar{x}_{9}\bar{x}_{10}\rangle$&$(-\eta-i\mu)|\bar{x}_{1}\bar{x}_{2}\bar{z}_{3}\bar{z}_{4}{y}_{5}\bar{z}_{6}\bar{x}_{7}x_{8}x_{9}x_{10}\rangle$&$(-\kappa-i\nu)|\bar{x}_{1}\bar{x}_{2}\bar{z}_{3}\bar{z}_{4}{y}_{5}\bar{z}_{6}\bar{x}_{7}\bar{x}_{8}\bar{x}_{9}\bar{x}_{10}\rangle$\\
			\end{tabular}
		\end{ruledtabular}
		
		{\leftline{$H_{3}$}}
		\begin{ruledtabular}
			\begin{tabular}{ccccc}
				&$\ket{a}_{B}$&$\ket{b}_{B}$&$\ket{c}_{B}$&$\ket{d}_{B}$\\
				\colrule
				$\ket{3}_{A}$&$i(\alpha-i\gamma)|x_{1}x_{2}\bar{z}_{3}\bar{z}_{4}y_{5}z_{6}{x}_{7}\bar{y}_{8}x_{9}x_{10}\rangle$&$(\beta-i\delta)|x_{1}x_{2}\bar{z}_{3}\bar{z}_{4}y_{5}z_{6}{x}_{7}\bar{y}_{8}\bar{x}_{9}\bar{x}_{10}\rangle$&$i(\eta-i\mu)|\bar{x}_{1}\bar{x}_{2}\bar{z}_{3}\bar{z}_{4}y_{5}z_{6}{x}_{7}\bar{y}_{8}x_{9}x_{10}\rangle$&$(\kappa-i\nu)|\bar{x}_{1}\bar{x}_{2}\bar{z}_{3}\bar{z}_{4}y_{5}z_{6}{x}_{7}\bar{y}_{8}\bar{x}_{9}\bar{x}_{10}\rangle$\\
				\colrule
				$\ket{4}_{A}$&$i(i\alpha+\gamma)|x_{1}x_{2}\bar{z}_{3}\bar{z}_{4}\bar{y}_{5}\bar{z}_{6}{x}_{7}x_{8}x_{9}x_{10}\rangle$&$(i\beta+\delta)|x_{1}x_{2}\bar{z}_{3}\bar{z}_{4}\bar{y}_{5}\bar{z}_{6}{x}_{7}\bar{x}_{8}\bar{x}_{9}\bar{x}_{10}\rangle$&$i(i\eta+\mu)|\bar{x}_{1}\bar{x}_{2}\bar{z}_{3}\bar{z}_{4}\bar{y}_{5}\bar{z}_{6}{x}_{7}x_{8}x_{9}x_{10}\rangle$&$(i\kappa+\nu)|\bar{x}_{1}\bar{x}_{2}\bar{z}_{3}\bar{z}_{4}\bar{y}_{5}\bar{z}_{6}{x}_{7}\bar{x}_{8}\bar{x}_{9}\bar{x}_{10}\rangle$\\
				\colrule
				$\ket{7}_{A}$&$(-i\alpha+\gamma)|x_{1}x_{2}\bar{z}_{3}\bar{z}_{4}\bar{y}_{5}{z}_{6}\bar{x}_{7}y_{8}x_{9}x_{10}\rangle$&$i(-i\beta+\delta)|x_{1}x_{2}\bar{z}_{3}\bar{z}_{4}\bar{y}_{5}{z}_{6}\bar{x}_{7}y_{8}\bar{x}_{9}\bar{x}_{10}\rangle$&$(-i\eta+\mu)|\bar{x}_{1}\bar{x}_{2}\bar{z}_{3}\bar{z}_{4}\bar{y}_{5}{z}_{6}\bar{x}_{7}y_{8}x_{9}x_{10}\rangle$&$-i(i\kappa-\nu)|\bar{x}_{1}\bar{x}_{2}\bar{z}_{3}\bar{z}_{4}\bar{y}_{5}{z}_{6}\bar{x}_{7}y_{8}\bar{x}_{9}\bar{x}_{10}\rangle$\\
				\colrule
				$\ket{8}_{A}$&$(-\alpha-i\gamma)|x_{1}x_{2}\bar{z}_{3}\bar{z}_{4}{y}_{5}\bar{z}_{6}\bar{x}_{7}y_{8}x_{9}x_{10}\rangle$&$i(-\beta-i\delta)|x_{1}x_{2}\bar{z}_{3}\bar{z}_{4}{y}_{5}\bar{z}_{6}\bar{x}_{7}y_{8}\bar{x}_{9}\bar{x}_{10}\rangle$&$(-\eta-i\mu)|\bar{x}_{1}\bar{x}_{2}\bar{z}_{3}\bar{z}_{4}{y}_{5}\bar{z}_{6}\bar{x}_{7}y_{8}x_{9}x_{10}\rangle$&$-i(\kappa+i\nu)|\bar{x}_{1}\bar{x}_{2}\bar{z}_{3}\bar{z}_{4}{y}_{5}\bar{z}_{6}\bar{x}_{7}y_{8}\bar{x}_{9}\bar{x}_{10}\rangle$\\
			\end{tabular}
		\end{ruledtabular}
		
		{\leftline{$H_{2}$}}
		\begin{ruledtabular}
			\begin{tabular}{ccccc}
				&$\ket{a}_{B}$&$\ket{b}_{B}$&$\ket{c}_{B}$&$\ket{d}_{B}$\\
				\colrule
				$\ket{3}_{A}$&$(\alpha-i\gamma)|x_{1}x_{2}\bar{z}_{3}\bar{z}_{4}y_{5}z_{6}{x}_{7}\bar{z}_{8}x_{9}x_{10}\rangle$&$-i(\beta-i\delta)|x_{1}x_{2}\bar{z}_{3}\bar{z}_{4}y_{5}z_{6}{x}_{7}\bar{z}_{8}\bar{x}_{9}\bar{x}_{10}\rangle$&$(\eta-i\mu)|\bar{x}_{1}\bar{x}_{2}\bar{z}_{3}\bar{z}_{4}y_{5}z_{6}{x}_{7}\bar{z}_{8}x_{9}x_{10}\rangle$&$-i(\kappa-i\nu)|\bar{x}_{1}\bar{x}_{2}\bar{z}_{3}\bar{z}_{4}y_{5}z_{6}{x}_{7}\bar{z}_{8}\bar{x}_{9}\bar{x}_{10}\rangle$\\
				\colrule
				$\ket{4}_{A}$&$(i\alpha+\gamma)|x_{1}x_{2}\bar{z}_{3}\bar{z}_{4}\bar{y}_{5}\bar{z}_{6}{x}_{7}\bar{z}_{8}x_{9}x_{10}\rangle$&$-i(i\beta+\delta)|x_{1}x_{2}\bar{z}_{3}\bar{z}_{4}\bar{y}_{5}\bar{z}_{6}{x}_{7}\bar{z}_{8}\bar{x}_{9}\bar{x}_{10}\rangle$&$(i\eta+\mu)|\bar{x}_{1}\bar{x}_{2}\bar{z}_{3}\bar{z}_{4}\bar{y}_{5}\bar{z}_{6}{x}_{7}\bar{z}_{8}x_{9}x_{10}\rangle$&$-i(i\kappa+\nu)|\bar{x}_{1}\bar{x}_{2}\bar{z}_{3}\bar{z}_{4}\bar{y}_{5}\bar{z}_{6}{x}_{7}\bar{z}_{8}\bar{x}_{9}\bar{x}_{10}\rangle$\\
				\colrule
				$\ket{7}_{A}$&$i(-i\alpha+\gamma)|x_{1}x_{2}\bar{z}_{3}\bar{z}_{4}\bar{y}_{5}{z}_{6}\bar{x}_{7}\bar{z}_{8}x_{9}x_{10}\rangle$&$(i\beta-\delta)|x_{1}x_{2}\bar{z}_{3}\bar{z}_{4}\bar{y}_{5}{z}_{6}\bar{x}_{7}\bar{z}_{8}\bar{x}_{9}\bar{x}_{10}\rangle$&$i(-i\eta+\mu)|\bar{x}_{1}\bar{x}_{2}\bar{z}_{3}\bar{z}_{4}\bar{y}_{5}{z}_{6}\bar{x}_{7}\bar{z}_{8}x_{9}x_{10}\rangle$&$(i\kappa-\nu)|\bar{x}_{1}\bar{x}_{2}\bar{z}_{3}\bar{z}_{4}\bar{y}_{5}{z}_{6}\bar{x}_{7}\bar{z}_{8}\bar{x}_{9}\bar{x}_{10}\rangle$\\
				\colrule
				$\ket{8}_{A}$&$i(-\alpha-i\gamma)|x_{1}x_{2}\bar{z}_{3}\bar{z}_{4}{y}_{5}\bar{z}_{6}\bar{x}_{7}\bar{z}_{8}x_{9}x_{10}\rangle$&$(\beta+i\delta)|x_{1}x_{2}\bar{z}_{3}\bar{z}_{4}{y}_{5}\bar{z}_{6}\bar{x}_{7}\bar{z}_{8}\bar{x}_{9}\bar{x}_{10}\rangle$&$i(-\eta-i\mu)|\bar{x}_{1}\bar{x}_{2}\bar{z}_{3}\bar{z}_{4}{y}_{5}\bar{z}_{6}\bar{x}_{7}\bar{z}_{8}x_{9}x_{10}\rangle$&$(\kappa+i\nu)|\bar{x}_{1}\bar{x}_{2}\bar{z}_{3}\bar{z}_{4}{y}_{5}\bar{z}_{6}\bar{x}_{7}\bar{z}_{8}\bar{x}_{9}\bar{x}_{10}\rangle$\\
			\end{tabular}
		\end{ruledtabular}
		
		\leftline{$H_{1}$} 
		\begin{ruledtabular}
			\begin{tabular}{ccccc}
				&$\ket{a}_{B}$&$\ket{b}_{B}$&$\ket{c}_{B}$&$\ket{d}_{B}$\\
				\colrule
				$\ket{6}_{A}$&$(\alpha-\gamma)|x_{1}x_{2}\bar{z}_{3}\bar{z}_{4}x_{5}x_{6}\bar{z}_{7}x_{8}x_{9}x_{10}\rangle$&$(\beta+\delta)|x_{1}x_{2}\bar{z}_{3}\bar{z}_{4}x_{5}x_{6}\bar{z}_{7}\bar{x}_{8}\bar{x}_{9}\bar{x}_{10}\rangle$&$(\eta-\mu)|\bar{x}_{1}\bar{x}_{2}\bar{z}_{3}\bar{z}_{4}x_{5}x_{6}\bar{z}_{7}x_{8}x_{9}x_{10}\rangle$&$(\kappa+\nu)|\bar{x}_{1}\bar{x}_{2}\bar{z}_{3}\bar{z}_{4}x_{5}x_{6}\bar{z}_{7}\bar{x}_{8}\bar{x}_{9}\bar{x}_{10}\rangle$\\
				\colrule
				$\ket{8}_{A}$&$-(\alpha+\gamma)|x_{1}x_{2}\bar{z}_{3}\bar{z}_{4}\bar{x}_{5}\bar{x}_{6}\bar{z}_{7}x_{8}x_{9}x_{10}\rangle$&
				$(\beta-\delta)|x_{1}x_{2}\bar{z}_{3}\bar{z}_{4}\bar{x}_{5}\bar{x}_{6}\bar{z}_{7}\bar{x}_{8}\bar{x}_{9}\bar{x}_{10}\rangle$&$-(\eta+\mu)|\bar{x}_{1}\bar{x}_{2}\bar{z}_{3}\bar{z}_{4}\bar{x}_{5}\bar{x}_{6}\bar{z}_{7}x_{8}x_{9}x_{10}\rangle$&$(\kappa-\nu)|\bar{x}_{1}\bar{x}_{2}\bar{z}_{3}\bar{z}_{4}\bar{x}_{5}\bar{x}_{6}\bar{z}_{7}\bar{x}_{8}\bar{x}_{9}\bar{x}_{10}\rangle$\\
			\end{tabular}
		\end{ruledtabular}
		
		\leftline{$H_{0}$} 
		\begin{ruledtabular}
			\begin{tabular}{ccccc}
				&$\ket{a}_{B}$&$\ket{b}_{B}$&$\ket{c}_{B}$&$\ket{d}_{B}$\\
				\colrule
				$\ket{6}_{A}$&$(\alpha-\gamma)|x_{1}x_{2}\bar{z}_{3}{x}_{4}x_{5}x_{6}\bar{z}_{7}x_{8}x_{9}x_{10}\rangle$&$(\beta+\delta)|x_{1}x_{2}\bar{z}_{3}{x}_{4}x_{5}x_{6}\bar{z}_{7}\bar{x}_{8}\bar{x}_{9}\bar{x}_{10}\rangle$&$(\eta-\mu)|\bar{x}_{1}\bar{x}_{2}\bar{z}_{3}{x}_{4}x_{5}x_{6}\bar{z}_{7}x_{8}x_{9}x_{10}\rangle$&$(\kappa+\nu)|\bar{x}_{1}\bar{x}_{2}\bar{z}_{3}{x}_{4}x_{5}x_{6}\bar{z}_{7}\bar{x}_{8}\bar{x}_{9}\bar{x}_{10}\rangle$\\
				\colrule
				$\ket{8}_{A}$&$(\alpha+\gamma)|x_{1}x_{2}\bar{z}_{3}\bar{x}_{4}\bar{x}_{5}\bar{x}_{6}\bar{z}_{7}x_{8}x_{9}x_{10}\rangle$&
				$-(\beta-\delta)|x_{1}x_{2}\bar{z}_{3}\bar{x}_{4}\bar{x}_{5}\bar{x}_{6}\bar{z}_{7}\bar{x}_{8}\bar{x}_{9}\bar{x}_{10}\rangle$&$(\eta+\mu)|\bar{x}_{1}\bar{x}_{2}\bar{z}_{3}\bar{x}_{4}\bar{x}_{5}\bar{x}_{6}\bar{z}_{7}x_{8}x_{9}x_{10}\rangle$&$-(\kappa-\nu)|\bar{x}_{1}\bar{x}_{2}\bar{z}_{3}\bar{x}_{4}\bar{x}_{5}\bar{x}_{6}\bar{z}_{7}\bar{x}_{8}\bar{x}_{9}\bar{x}_{10}\rangle$\\
			\end{tabular}
		\end{ruledtabular}
	\end{sidewaystable}	
	
	\clearpage
	
	\section{Measurement}
	To reconstruct the quantum states in the braiding operation process, we need to perform the quantum state tomography. Here, under the non-local encoding framework, the readout of relative phases between different physical qubits is quite different with that in previous work. For simplicity, we use take the relative phase measurement between any two physical bits as an example. There are three cases for measurement, and we will explain them separately as follow:
	\\Case 1:  
	\begin{figure}[htbp]
		\centering
		\includegraphics[width=0.5\columnwidth]{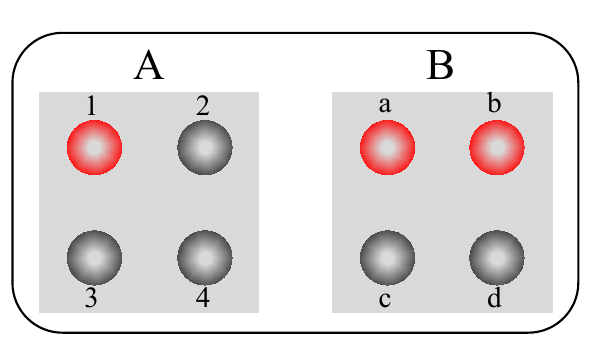}\\
		\small{Fig. S9. The case when two states to be measured contain the same optical mode in the side A.}
	\end{figure}
	
	If the two states to be measured contain the same optical mode in the side A, then the relative phase between the two states is the phase between the two modes in the side B, as shown in Fig. S9. For example, if we need to obtain the relative phase between state $\ket{1a}=\ket{1}_{A} \otimes \ket{a}_{B}$ and state $\ket{1b}=\ket{1}_{A} \otimes \ket{b}_{B}$, because they share the same optical mode $\ket{1}_{A}$ , the relative phase between them is exactly the phase between $\ket{a}_{B}$ and $\ket{b}_{B}$ . We only need to perform quantum state tomography between optical mode $\ket{a}_{B}$ and $\ket{b}_{B}$ to readout the relative phase between these two states.
	\\Case 2:  
	\begin{figure}[htbp]
		\centering
		\includegraphics[width=0.5\columnwidth]{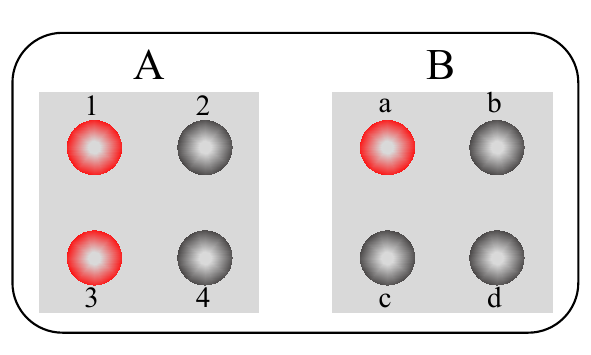}\\
		\small{Fig. S10. The case when two states to be measured contain the same optical mode in the side B.}
	\end{figure}
	
	Similar with that in Case 1, here, if the two states to be measured contain the same optical mode in the side B, then the relative phase between the two states is the phase between the two modes in the side A, as shown in Fig. S10. For example, if we need to obtain the relative phase between state $\ket{1a}=\ket{1}_{A} \otimes \ket{a}_{B}$ and state $\ket{3a}=\ket{3}_{A} \otimes \ket{a}_{B}$, because they share the same optical mode $\ket{a}_{B}$, the relative phase between them is exactly the phase between $\ket{1}_{A}$ and $\ket{3}_{A}$. We only need to perform quantum state tomography between optical mode $\ket{1}_{A}$ and $\ket{3}_{A}$ to readout the relative phase between these two states.
	\\Case 3:
	
	\begin{figure}[htbp]
		\centering
		\includegraphics[width=0.5\columnwidth]{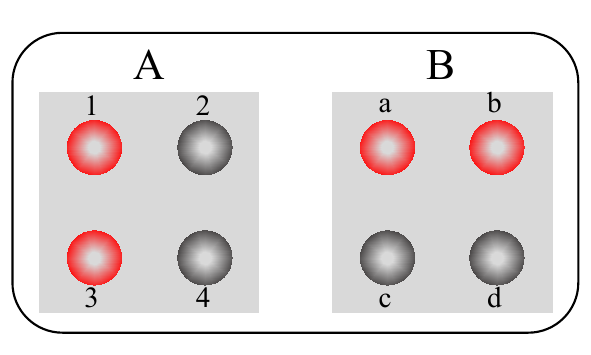}\\
		\small{Fig. S11. The case when there are no common modes of the physical qubit in any side.}
	\end{figure}
	The last and the most difficult situation is that when there are no common modes of the physical qubit in any side, as shown in Fig. S11. For example, the two states to be measured is $\ket{1a}=\ket{1}_{A} \otimes \ket{a}_{B}$ and state $\ket{3a}=\ket{3}_{A} \otimes \ket{a}_{B}$. Thus, there are actually four modes at the initial time:$\ket{1a}=\ket{0}$, $\ket{1b}=\ket{1}$, $\ket{3a}=\ket{2}$ and $\ket{3b}=\ket{3}$, and the expected measurement bases are $\ket{0}/\ket{3}$. To achieve this goal, there are two steps to be implemented in chronological order, as shown in Fig. S12.
	\begin{figure}[htbp]
		\centering
		\includegraphics[width=0.5\columnwidth]{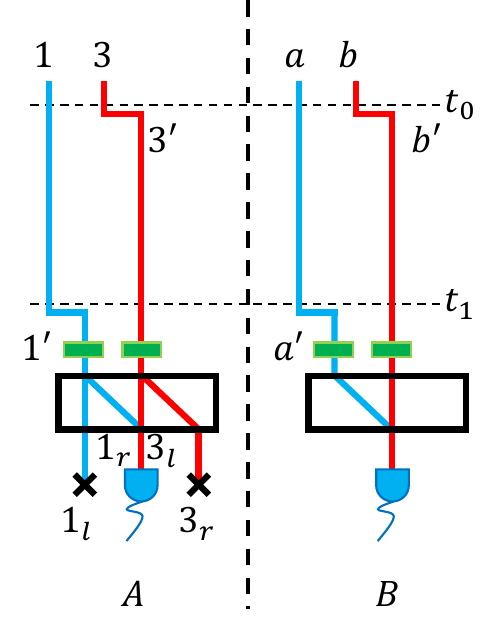}\\
		\small{\begin{flushleft}Fig. S12. The measurement setup when there are no common modes of the physical qubit in any side.\end{flushleft}}
	\end{figure} 
	Firstly, to prepare only states  $\ket{0}$ and $\ket{3}$, the same delay needs to be introduced for $\ket{3}_{A}$ and $\ket{b}_{B}$ to avoid their mutual interference at time $t_{0}$, as shown in Fig. S12. Then, mode $\ket{1}_{A}$ only has coincidence counts with mode $\ket{a}_{B}$ (displayed by blue line), and mode $\ket{3'}_{A}$ only has coincidence counts with mode  $\ket{b'}_{B}$ (displayed by red line). After this operation, only modes $\ket{1a}=\ket{0}$ and $\ket{3'b}=\ket{3}$ remained, there are no coincidence counts between states $\ket{1b'}=\ket{1}$ and $\ket{3'a}=\ket{2}$.
	
	Next, when implementing measurement at $t_{1}$, we need to combine mode $\ket{1}_{A}$ and $\ket{3'}_{A}$ together (also $\ket{a}_{B}$ and  $\ket{b'}_{B}$ together) and then perform projection measurement on each side. Here, we need to introduce the same delay as that in the first step for $\ket{3'}_{A}$ and $\ket{b'}_{B}$ to ensure the interference between $\ket{1'}_{A}$ and $\ket{3'}_{A}$ (also $\ket{a'}_{B}$ and $\ket{b'}_{B}$). Then on Side A, modes $\ket{1'}_{A}$ and $\ket{3'}_{A}$ passed through an HWP at 22.5$^\circ$ and a BD, thus were divided into $\ket{1_{l}}_{A}$ and $\ket{1_{r}}_{A}$, $\ket{3_{l}}_{A}$ and $\ket{3_{r}}_{A}$. Here, we obtained the expected modes $\ket{1_{r}a'}= \ket{0}$  and $\ket{3_{l}b'}=\ket{0}$, and modes $\ket{1_{l}b'}= \ket{1}$  and $\ket{3_{r}a'}=\ket{2}$ are discarded. Till now, we have realised the projection measurement on $\ket{0}/\ket{3}$.
	
	In the experiment, for example, when a situation where four light spots on Side B cannot meet the relative phase coordination for realising encoding of the quantum states. Based on the strategy introduced in case 3, we need to divide the original four beams into eight, and the corresponding logical encoding is shown in Fig. S13 and on Table. S4  When the relative phase coordination is met (i.e., case 1 and case 2), we still only need four beams (as shown in the optical spatial modes during braiding operation most of the time) for measurement, which can be realised easily through rotating HWPs in the optical paths.               
	
	\begin{figure}[htbp]
		\centering
		\includegraphics[width=0.6\columnwidth]{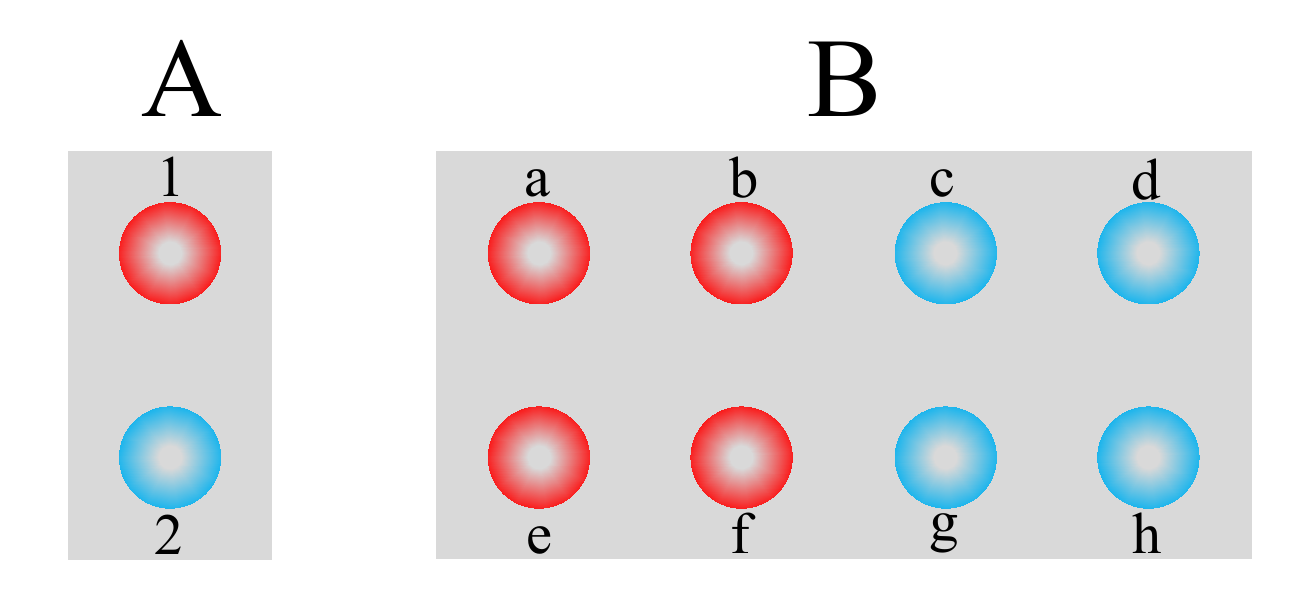}\\
		\small{\begin{flushleft}Fig. S13. The measurement case when there are no common modes of the physical qubit in any side\end{flushleft}.}
	\end{figure} 
	
	\begin{table}[htbp]
		\tiny{TABLE. S4. Optical spatial modes for case 3.\\
			\begin{ruledtabular}
				\begin{tabular}{ccccc}
					&$\ket{a}_{B}$&$\ket{b}_{B}$&$\ket{e}_{B}$&$\ket{f}_{B}$\\
					\colrule
					$\ket{1}_{A}$&$\ket{000}_{meas}$&$\ket{001}_{meas}$&$\ket{100}_{meas}$&$\ket{101}_{meas}$\\
					\colrule
					&$\ket{c}_{B}$&$\ket{d}_{B}$&$\ket{g}_{B}$&$\ket{h}_{B}$\\
					\colrule
					$\ket{2}_{A}$&$\ket{010}_{meas}$&$\ket{011}_{meas}$&$\ket{110}_{meas}$&$\ket{111}_{meas}$\\
				\end{tabular}
		\end{ruledtabular}}
	\end{table}

	\section{More experimental results}
	
	\subsection{The performance of the Sagnac interferometers.}
	\begin{figure}[!h]
		\centering
		\includegraphics[width=0.5\columnwidth]{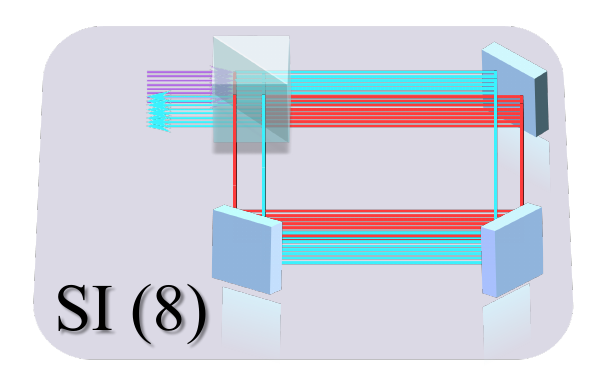}\\
		\normalsize{Fig. S14. Schematic of the Sagnac interferometer with eight input beams. 
		}
	\end{figure}

	The introduction of the Sagnac interferometer (SI) is an innovation to achieve the non-dissipative ITE in our work. The schematic of the SI with eight input beams is shown in Fig. S14. It should be mentioned that all the SIs used in the experiment are same with it except for a different amount of input beams.   Naturally, the performance of the SIs play a crucial role in the success of the quantum state evolution. Here, we provide the experimental results for all the Sagnac interferometers emerging in the experiment.  
	
	\begin{figure}[!h]
		\centering
		\includegraphics[width=1\columnwidth]{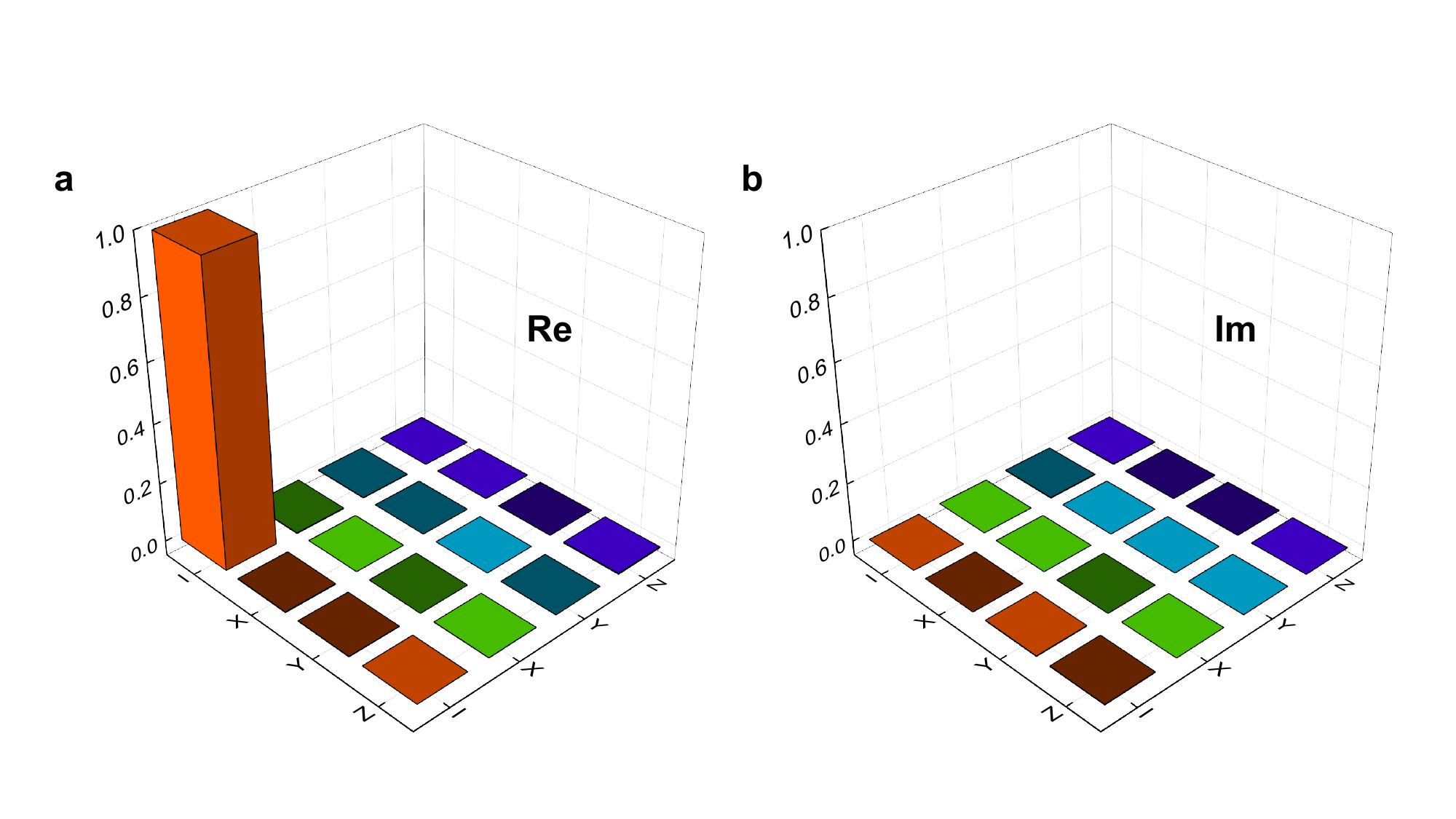}\\
		\normalsize{\begin{flushleft}Fig. S15. Experimental process density matrix $\chi_{ex}$ for the Sagnac interferometer. $\textbf{a}$. Real (Re) parts of $\chi_{ex}$. 
				$\textbf{b}$. Imaginary (Im) parts of $\chi_{ex}$.\end{flushleft}}
	\end{figure} 
	\begin{table}[!h]
		\normalsize{TABLE. S5. Interference visibility of all the Sagnac interferometers in braiding operation $\sigma_{1}$ and $\sigma_{2}^{-1}$. \\
			\begin{ruledtabular}
				\begin{tabular}{ccc}
					Sagnac interferometer&Visiblity($\sigma_{1}$)&Visiblity($\sigma_{2}^{-1}$)\\
					\colrule
					1&$0.990\pm0.003$&$0.990\pm0.003$\\
					\colrule
					2&$0.994\pm0.002$&$0.986\pm0.002$\\
					\colrule
					3&$0.996\pm0.002$&$0.987\pm0.001$\\
					\colrule
					4&$0.988\pm0.005$&$0.994\pm0.001$\\
					\colrule
					5&$0.990\pm0.003$&$0.963\pm0.003$\\
				\end{tabular}
		\end{ruledtabular}}
	\end{table}	
	Firstly, to judge the performance of the Sagnac interferometer, we take a beam in the first Sagnac interferometer used in braiding operation $\sigma_{1}$ as an example and follow the standard method described in\cite{Brien2004} to perform quantum process tomography (QPT). For any input state $\ket{\psi}=\alpha\ket{H}+e^{i\phi}\beta\ket{V}$, the output state after passing through the Sagnac interferometer should still be $\ket{\psi}$, which means the operation is an identity matrix.  The density matrix of the process matirx $\chi_{ex}$ are shown in Fig. S15. The process fidelity Tr$(\chi_{th}\chi_{ex})= 0.996 \pm 0.008$, indicating the high performance of the Sagnac interferometer.

	Next, we present the interference visibility of all the Sagnac interferometers in braiding operation $\sigma_{1}$ and $\sigma_{2}^{-1}$, as shown in Table. S5. 
	
	\subsection{Photographs of the experimental setup.}
	\begin{figure}[!h]
		\centering
		\includegraphics[width=1\columnwidth]{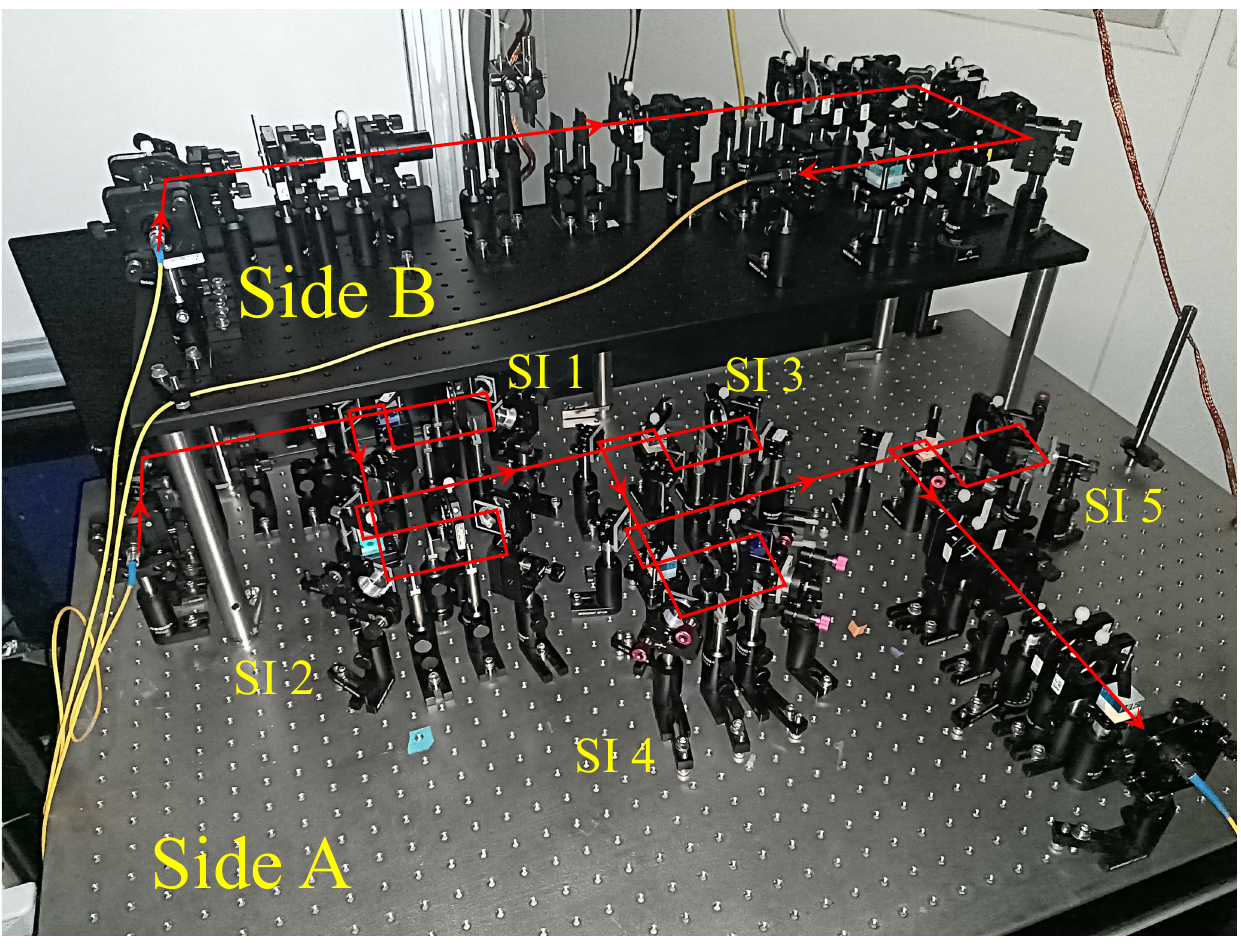}
		\normalsize{\begin{flushleft}Fig. S16. Experimental setup for implementing braiding operation $\sigma_{1}$. The red line  indicates the direction of beam propagation. \end{flushleft}}
	\end{figure}
	
	\begin{figure}[!h]
		\centering
		\includegraphics[width=1\columnwidth]{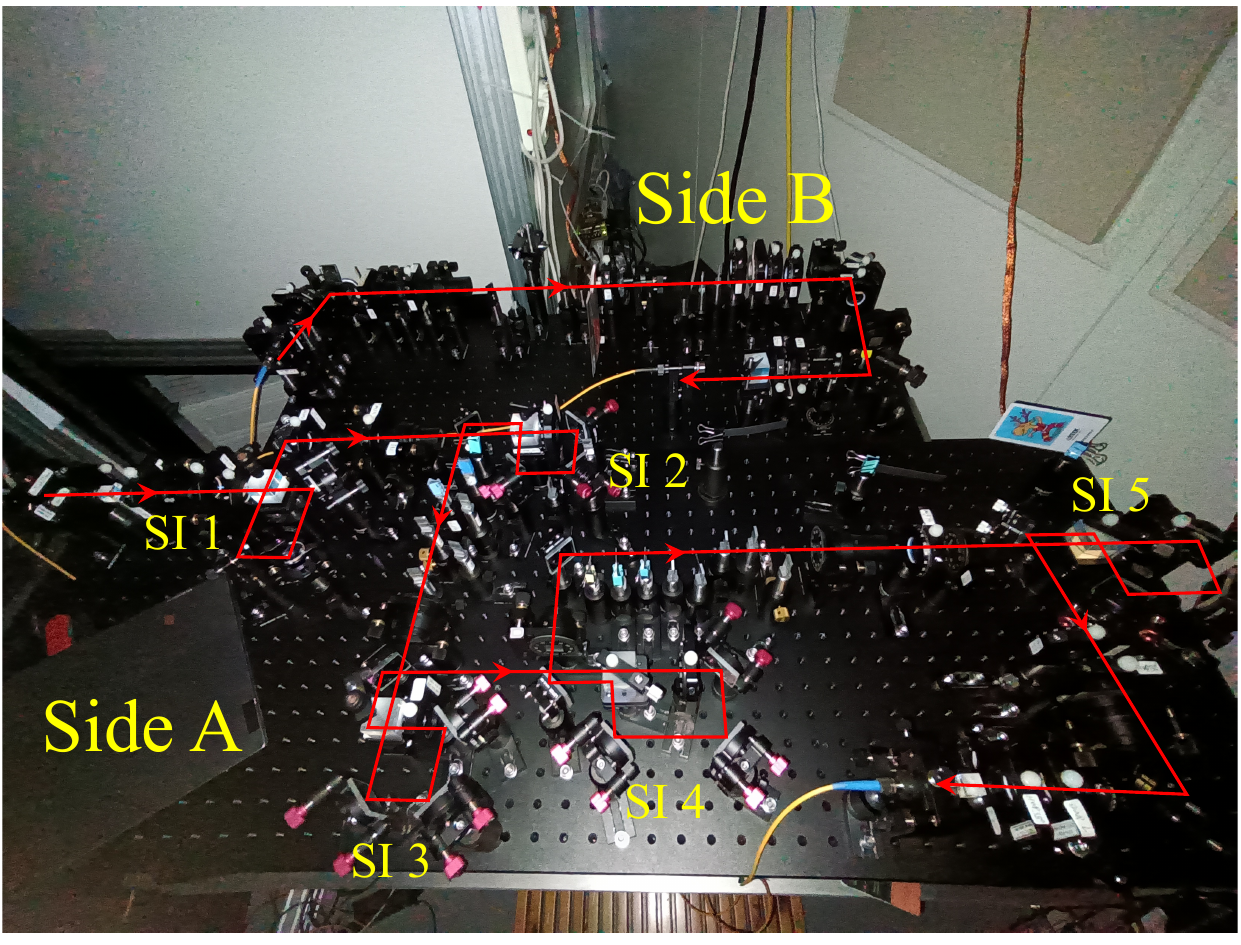}
		\normalsize{\begin{flushleft}Fig. S17. Experimental setup for implementing braiding operation $\sigma_{2}^{-1}$. The red line  indicates the direction of beam propagation.\end{flushleft}}
	\end{figure}
	\clearpage
	
	\subsection{Quantum state tomography results for all the states during the braiding operation.}
	\begin{figure}[htbp]
		\centering
		\includegraphics[width=1\columnwidth]{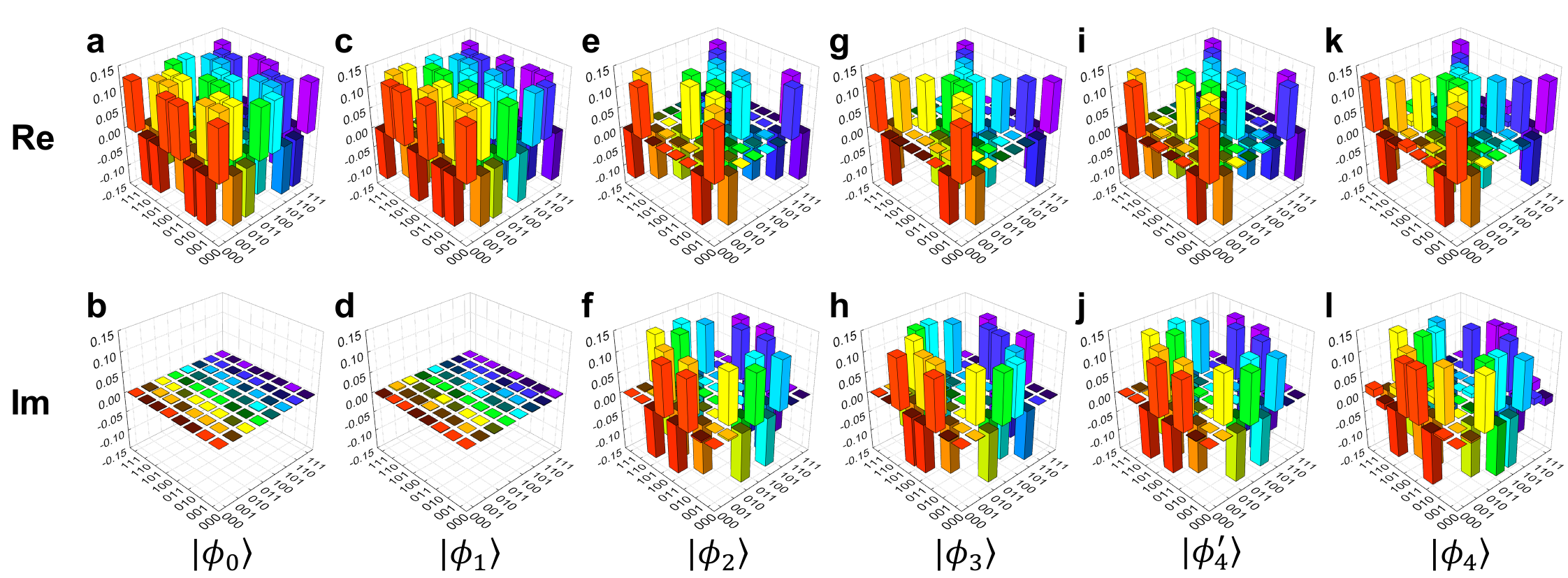}
		\normalsize{\begin{flushleft}Fig. S18. Experimental density matrices for all the states during the braiding operation $\sigma_{1}$. $\textbf{a-k}$. Real (Re) parts of the states $\ket{\phi_{0}}\textendash\ket{\phi_{4}}$, respectively. $\textbf{b-l}$. Imaginary (Im) parts of the states $\ket{\phi_{0}}\textendash\ket{\phi_{4}}$, respectively.\end{flushleft}}
	\end{figure}
	\begin{figure}[htbp]
		\centering
		\includegraphics[width=1\columnwidth]{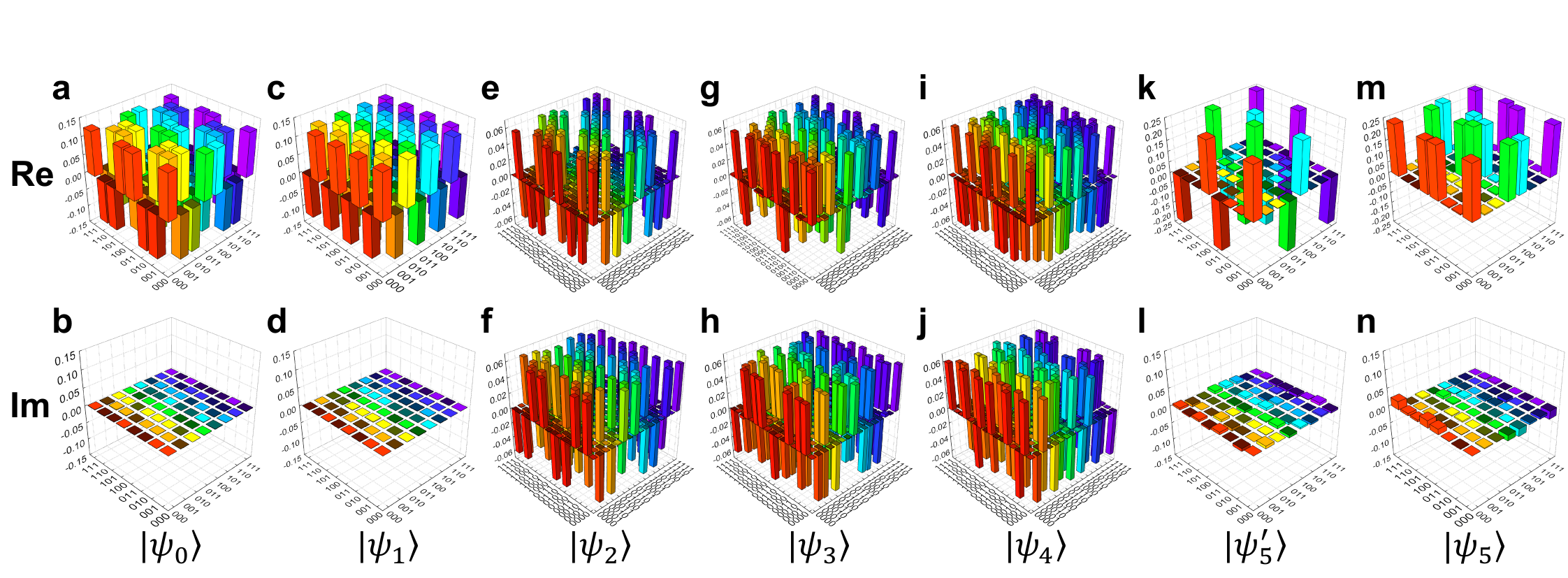}
		\normalsize{\begin{flushleft}Fig. S19. Experimental density matrices for all the states during the braiding operation $\sigma_{1}$. $\textbf{a-m}$. Real (Re) parts of the states $\ket{\psi_{0}}\textendash\ket{\psi_{5}}$, respectively. $\textbf{b-n}$. Imaginary (Im) parts of the states $\ket{\psi_{0}}\textendash\ket{\psi_{5}}$, respectively.\end{flushleft}}
		
	\end{figure}
	The experimental results for the quantum process tomography for the braiding operation $\sigma_{1}$ and  $\sigma_{2}^{-1}$ are shown in the main text. Here,  we demonstrate the quantum state tomography results for each state during the braiding operation $\sigma_{1}$ and $\sigma_{2}^{-1}$, as shown in Fig. S18, Fig. S19 and Table. S6. The obtained average fidelities for $\sigma_{1}$ and $\sigma_{2}^{-1}$ are 0.991 $\pm$ 0.005 and 0.978 $\pm$ 0.015, indicating the braiding operation is successfully achieve in each step.

	\begin{table}[!ht]
		\normalsize{TABLE. S6. Quantum state tomography fidelity of the states during braiding operation $\sigma_{1}$ and $\sigma_{2}^{-1}$. \\
			\begin{ruledtabular}
				\begin{tabular}{ccccc}
					State ($\sigma_{1}$)&Fidelity&State ($\sigma_{2}^{-1}$)&Fidelity&\\
					\colrule
					$\ket{\phi_{0}}$&0.991$\pm$0.005&$\ket{\psi_{0}}$&0.991$\pm$0.005&\\
					\colrule
					$\ket{\phi_{1}}$&0.994$\pm$0.001&$\ket{\psi_{1}}$&0.993$\pm$0.011&\\
					\colrule
					$\ket{\phi_{2}}$&0.993$\pm$0.005&$\ket{\psi_{2}}$&0.984$\pm$0.022&\\
					\colrule
					$\ket{\phi_{3}}$&0.989$\pm$0.001&$\ket{\psi_{3}}$&0.975$\pm$0.019&\\
					\colrule
					$\ket{\phi'_{4}}$&0.988$\pm$0.015&$\ket{\psi_{4}}$&0.973$\pm$0.015&\\
					\colrule
					$\ket{\phi_{4}}$&0.992$\pm$0.001&$\ket{\psi'_{5}}$&0.966$\pm$0.005&\\
					\colrule
					&&$\ket{\psi_{5}}$&0.964$\pm$0.028&\\
				\end{tabular}
		\end{ruledtabular}}
	\end{table}	
	\clearpage
	
	\subsection{Quantum process tomography results for braiding operation $\sigma_{1}$ and $\sigma_{2}^{-1}$.}
	As we mentioned in the main text,  we perform quantum process tomography of the braiding operation $\sigma_{1}$ and $\sigma_{2}^{-1}$, and present the results in the logical space. Here, we also give the matrix in the spin space. The density matrix of braiding operation $\sigma_{1}$ and $\sigma_{2}^{-1}$ in the physical qubits basis can be decomposed as:
	\begin{equation}
		U_{\sigma_1}=\left(
		\begin{array}{cccccccc}
			1\ & 0\ & 0\ & 0\ & 0\ & 0\ & 0\ & 0\\
			0 & 1 & 0 & 0 & 0 & 0 & 0 & 0\\
			0 & 0 & i & 0 & 0 & 0 & 0 & 0\\
			0 & 0 & 0 & i & 0 & 0 & 0 & 0\\
			0 & 0 & 0 & 0 & i & 0 & 0 & 0\\
			0 & 0 & 0 & 0 & 0 & i & 0 & 0\\
			0 & 0 & 0 & 0 & 0 & 0 & 1 & 0\\
			0 & 0 & 0 & 0 & 0 & 0 & 0 & 1\\
		\end{array}
		\right)/\sqrt{2}=\left(\begin{array}{cccc}
			1&0&0&0\\
			0&i&0&0\\
			0&0&i&0\\
			0&0&0&1
		\end{array}
		\right) \otimes \left(
		\begin{array}{cc}
			1&0\\
			0&1 
		\end{array}
		\right)/\sqrt{2} ,
	\end{equation}
	\begin{equation}
		U_{\sigma_2^{-1}}=\left(
		\begin{array}{cccccccc}
			1\ & 0\ & -1\ & 0\ & 0\ & 0\ & 0\ & 0\\
			0 & 1 & 0 & 1 & 0 & 0 & 0 & 0\\
			1 & 0 & 1 & 0 & 0 & 0 & 0 & 0\\
			0 & -1 & 0 & 1 & 0 & 0 & 0 & 0\\
			0 & 0 & 0 & 0 & 1 & 0 & -1 & 0\\
			0 & 0 & 0 & 0 & 0 & 1 & 0 & 1\\
			0 & 0 & 0 & 0 & 1 & 0 & 1 & 0\\
			0 & 0 & 0 & 0 & 0 & -1 & 0 & 1\\
		\end{array}
		\right)/\sqrt{2}=\left(
		\begin{array}{cc}
			1&0\\
			0&1 
		\end{array}
		\right) \otimes 
		\left(\begin{array}{cccc}
			1&0&-1&0\\
			0&1&0&1\\
			1&0&1&0\\
			0&-1&0&1
		\end{array}
		\right)/\sqrt{2}
	\end{equation}
	while in the Pauli operators $I,X,Y$ and $Z$, these two operations are written as
	
	\begin{equation}
		U_{\sigma_1}=\frac{1}{\sqrt{2}}(II-iZZ) \otimes I
	\end{equation}
	\begin{equation}
		U_{\sigma_{2}^{-1}} =\frac{1}{\sqrt{2}}I \otimes (II-iYZ)
	\end{equation}
	
	To reconstruct these 8$\times$8 matrix, one needs to prepare $4^{3}=64$ input states and a total of $64\times64=4096$ measurement bases that need to be measured, which is extremely complicated to realise in our device with such special encoding methods. In fact, the existence of the identity matrix is due to the chain that is not involved in the braiding operation. Thus, we only need to reconstruct the 4$\times$4 density matrix, which needs $4^{2}=16$ input states and $16\times16=256$ measurement bases. The output state of the process could be written as $\varepsilon(\rho)=\sum_{m,n}\chi_{mn}\hat{E}_{m}\rho\hat{E}_{n}^\dagger$ for input state $\rho$, where $\hat{E}_{m}=\{II, IX, IY, IZ, XI, XX, XY, XZ, YI, YX, YY, YZ, ZI, ZX, ZY, ZZ\}$ is in terms of the Pauli basis operators. The physical operation $\varepsilon$ is characterized by the 16$\times$16 matrix $\chi$.
	
	\subsection{Experimental results for the representative links}
	In the main text, we present the modulus values of the Jones polynomials directly. Here, we give the absolute values of the amplitudes for some typical knots, as shown in Fig. S20 and Table. S7. 
	\begin{table}[htbp]
		\normalsize{TABLE. S7. The absolute values of the amplitudes for five representative knots. \\
			\begin{ruledtabular}
				\begin{tabular}{ccc}
					&theory&experiment \\
					\colrule
					Hopf link&0&0.052$\pm$0.006\\
					\colrule
					Trefoil knot&$1/\sqrt{2}$&0.707$\pm$0.008\\
					\colrule
					Solomon link&1&0.968$\pm$0.009\\
					\colrule
					Figure Eight knot&0.5&0.488$\pm$0.021\\
					\colrule
					Borromean rings&1&0.944$\pm$0.029\\
				\end{tabular}
		\end{ruledtabular}}
	\end{table}
	\begin{figure}[htbp]
		\centering
		\includegraphics[width=0.7\columnwidth]{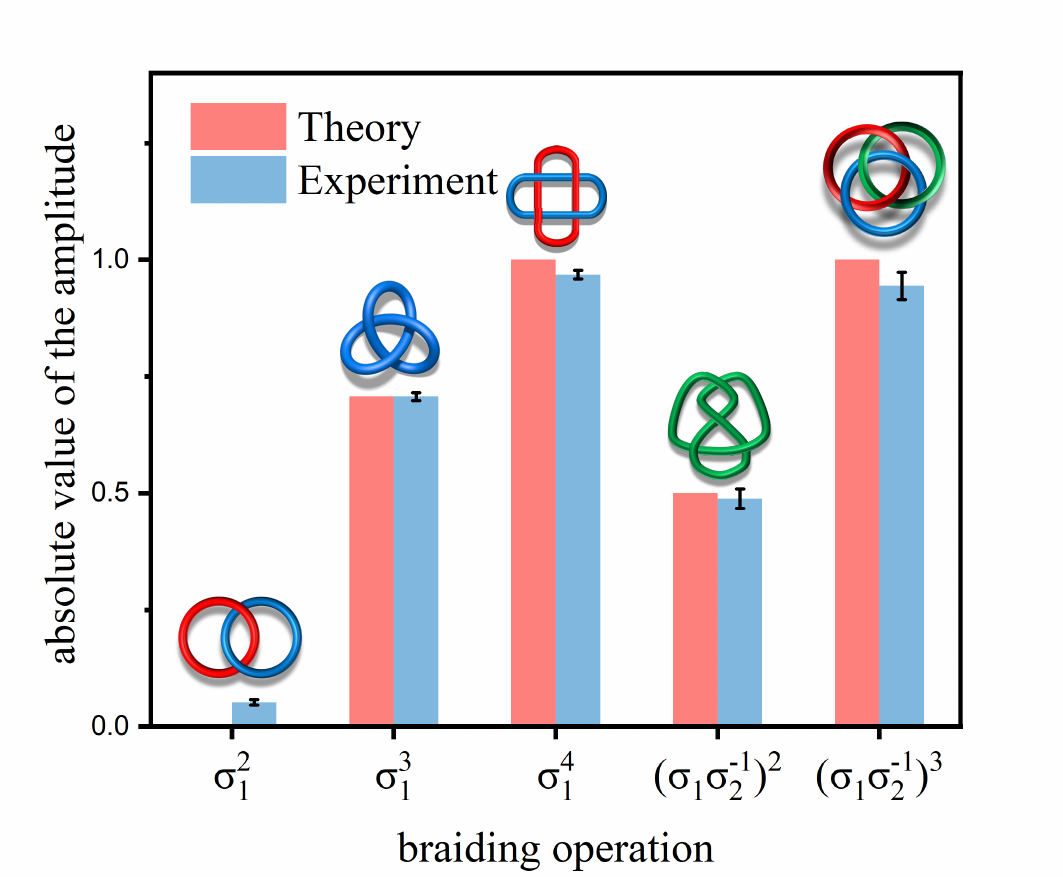}\\
		\normalsize{\begin{flushleft}Fig. S20. Theoretical (pink) and experimentally obtained (blue)  absolute values of the amplitudes for some representative knots: Hopf link, Trefoil knot, Solomon link, Figure Eight knot and Borromean rings. The error bars are estimated according to Poissonian counting statistics.\end{flushleft}
		}
	\end{figure}

	\section{Appendix: Basis rotation}
	
	\begin{equation}
		\begin{split}
			& |\bar{y}\rangle=|z\rangle-i|\bar{z}\rangle\\
			&|y\rangle=|z\rangle+i|\bar{z}\rangle\\
			&|z\rangle=|\bar{y}\rangle+|y\rangle\\
			&|\bar{z}\rangle=i(|\bar{y}\rangle-|y\rangle)\\
			&|x\rangle=|z\rangle+|\bar{z}\rangle\\
			&|\bar{x}\rangle=|z\rangle-|\bar{z}\rangle\\
			&|z\rangle=|x\rangle+|\bar{x}\rangle\\
			&|\bar{z}\rangle=|x\rangle-|\bar{x}\rangle\\
			&|\bar{y}\rangle=(1-i)|x\rangle+(1+i)|\bar{x}\rangle\\
			&|y\rangle=(1+i)|x\rangle+(1-i)|\bar{x}\rangle\\
			&|x\rangle=(1+i)|\bar{y}\rangle+(1-i)|y\rangle\\
			&|\bar{x}\rangle=(1-i)|\bar{y}\rangle+(1+i)|y\rangle\\
		\end{split}
	\end{equation}